\documentclass[prb,onecolumn,preprintnumbers,amsmath,amssymb]{revtex4}
\usepackage{graphicx}

\begin{document}

\title{  Optical Bragg, atom Bragg and  cavity QED detections of quantum phases and
 excitation spectra of ultracold atoms in bipartite and frustrated optical lattices }
\author{     Jinwu Ye $^{1,2}$, K.Y. Zhang $^{3}$, Yan Li$^{3}$, Yan Chen$^{4}$ and W.P. Zhang $^{3}$ }
\affiliation{ $^{1}$ Key Laboratory of Terahertz Optoelectronics, Ministry of Education, Department of Physics, Capital Normal University,  Beijing, 100048 China \\
$^{2}$ Department of Physics and Astronomy, Mississippi State University, P. O. Box 5167, Mississippi State, MS, 39762   \\
$^{3}$ Department of Physics, East China Normal university,Shanghai, 200062, China  \\
$^{4}$ Department of Physics, State Key Laboratory of Surface Physics and Laboratory of
Advanced Materials, Fudan University, Shanghai, 200433, China  }
\date{\today }

\begin{abstract}
  Ultracold atoms loaded on optical lattices can provide unprecedented
  experimental systems for the quantum simulations and manipulations
  of many quantum phases and quantum phase transitions between these phases.
  However, so far, how to detect these quantum phases and phase
  transitions effectively  remains an outstanding challenge.  In this paper, we will develop a systematic and unified theory of
   using the optical Bragg scattering, atomic Bragg scattering or cavity QED to detect the ground state and the excitation spectrum
   of many quantum phases of interacting bosons loaded in bipartite and frustrated optical lattices.
    The physical measurable quantities of the three
   experiments are the light scattering cross sections, the
   atom scattered clouds and the cavity leaking photons respectively.
   We show that the two photon Raman transition processes  in the three detection methods not only couple to
   the density order parameter, but also the  {\sl valence bond order } parameter due to the hopping of the bosons
   on the lattice. This valence bond order coupling is very
   sensitive to any superfluid order or any Valence bond (VB ) order
   in the quantum phases to be probed.  These quantum phases include not only the well known superfluid and
   Mott insulating phases, but also other important phases such as various kinds of charge density waves (CDW),
   valence bond solids (VBS), CDW-VBS phases with both CDW and VBS orders unique to frustrated lattices,
   and also various kinds of supersolids.
   We analyze respectively the experimental conditions of the three detection methods to
   probe these various quantum phases and their corresponding  excitation spectra.
   We also address the effects of a finite temperature and a harmonic trap.
   We contrast the three scattering methods with recent {\sl in situ } measurement inside a harmonic trap and argue that
   the two kinds of measurements are complementary  to each other.
   The combination of both kinds of detection methods could be used to
   match the combination of Scanning tunneling microscopy (STM), the Angle Resolved Photo Emission spectroscopy ( ARPES ) and
   neutron scattering in condensed matter systems,
   therefore achieve the putative goals of quantum simulations
\end{abstract}
\maketitle

\section{ Introduction }

  Various kinds of strongly correlated quantum phases of matter may
  have wide applications in quantum information processing, storage
  and communications \cite{manybody}.  It was widely believed  and also partially established that
  due to the tremendous tunability of all the parameters in this
  system, ultracold atoms loaded on optical lattices (OL) can provide an unprecedented
  experimental systems for the quantum simulations and manipulations
  of these quantum phases and quantum phase transitions between these phases.
  For example, Mott and superfluid phases \cite{boson} may have been successfully
  simulated and manipulated by ultra-cold atoms loaded in a cubic optical
  lattice \cite{bloch}. However, there are still at least two outstanding problems
  remaining. The first is to how to realize many important
  quantum phases \cite{manybody}. The second is that assuming the favorable conditions to realize these quantum phases
  are indeed achieved in experiments, how to detect them without
  ambiguities. In this paper, we will address the second question.

   Because these ultra-cold atoms are charge neutral,  so in contrast to many condensed
   matter systems,  they can not be manipulated electrically or magnetically,
   so the experimental ways to detect these quantum phases of cold atoms are
   rather limited. The earliest detection method is through
   the so called time of flight measurement \cite{manybody,bloch}
   which simply opens the trap and turn off the optical lattice and let the trapped atoms  expand
   and interfere, then take the image. This kind of measurement is
   destructive and may not be used to detect the ground state and
   excitation spectrum of quantum phases in optical lattices. There
   are also other detection methods such as the Optical Bragg
   scattering in Fig.1a and the atom  Bragg scattering in Fig.1b.
   The Optical Bragg scattering (Fig.1a) has been used previously
  to study periodic lattice structures of cold atoms loaded on
  optical lattices \cite{braggsingle}. It was also proposed as an effective method for the thermometry
  of fermions in an optical lattice \cite{braggthermal}. Recently,
  it was argued that it can be used to detect the putative anti-ferromagnetic ground state of fermions in OL \cite{braggafm}.
  There are also very recent optical Bragg scattering experimental
  data from a Mott state, a BEC and some artificial  AF  state \cite{blochlight}.
  The atom Bragg diffraction in Fig.1b ( also
  called "atom Bragg spectroscopy " ) is based on  stimulated matter waves scattering by
  two incident laser pulses \cite{braggbog,braggeng}, then take images through the time of flight measurements. There are also two kinds of "atom Bragg spectroscopies
  ". The momentum \cite{braggbog} transfer Bragg spectroscopy
  was used to detect the Bogoliubov mode inside an BEC condensate.
  The energy transfer \cite{braggeng} Bragg spectroscopy
  was used to detect the Mott gap in a Mott state in an optical lattice.
  The photon and atom Bragg diffractions are two complementary
  experimental detection methods.
 During the last several years, there have been extensive of experimental and theoretical  research combining cavity QED
 and cold atoms. For example, several experiments \cite{qedbec1,qedbec2}
 successfully achieved the strong coupling of a BEC of $ N \sim 10^{5} $ $ ^{87}Rb $ atoms to
 the photons inside an ultrahigh-finesse optical cavity.
 The super-radiant phase \cite{phased} was realized
 in a recent experiments \cite{orbital} using BEC and also in a previous experiment using thermal atoms  \cite{orbitalt}
 where the effective two "atomic" levels are the two momentum states of the cold atoms
 in the optical lattice formed by  the cavity field and the off-resonant {\sl transverse } pumping Laser.
 Since the first experimental observation of the bi-stabilities of BEC atoms at very low photon numbers inside a cavity \cite{qedbecoff},
 there have been  very active theoretical studies on the bi- or multi-instabilities of BEC spinors
 \cite{spinor}, spinless fermions \cite{fermion} or BEC with a spin-orbit coupling \cite{spinorbit} inside a cavity.
 It was also preliminarily proposed that the cavity  photons may be used to non-destructively
 detect superfluid and Mott phases of ultra-cold atoms in optical
 lattices. (  The quantitative theory will be developed in Sec.VIII and the Appendix D  ) \cite{off1,off2}.

\begin{figure}
\includegraphics[width=3cm]{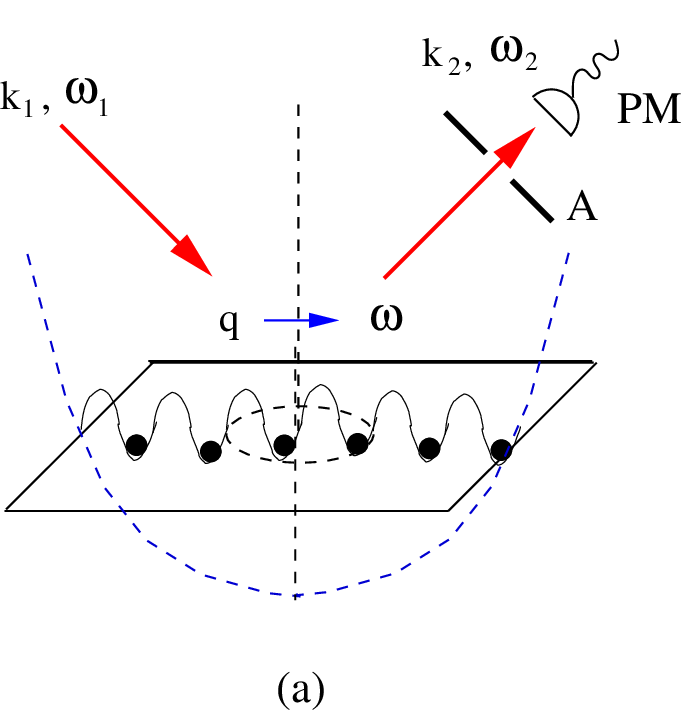}
\hspace{0.5cm}
\includegraphics[width=3cm]{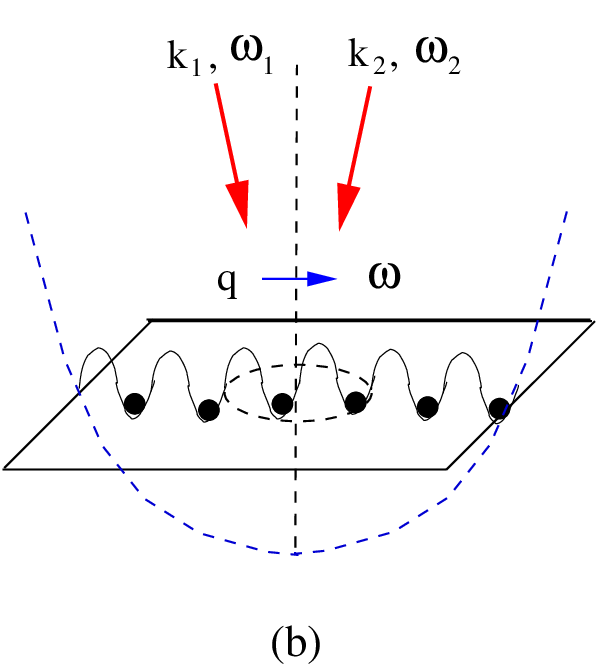}
\caption{ Optical (a) and atom (b) Bragg scattering of cold atoms
moving in 2 dimensional optical lattices. The $ \vec{q} =
\vec{k}_{1}-\vec{k}_{2}  $ and
 $ \omega = \omega_1-\omega_2 $ are momentum and energy transfer from the laser beams to the cold atoms respectively.
 Tilting and rotating the incident beam(s)
 around the dashed line can probe the ground state and the excitation spectrum of the whole 2d optical lattice.} \label{braggab}
\end{figure}

  In this paper, we will  explore the applicabilities of
  both photon Bragg diffraction and atom Bragg diffraction to detect several important quantum phases in both bipartite and frustrated lattices.
  We will also discuss in detail the possibility that the cavity QED can be used as a possible new detection
  method.  The physical measurable quantities of the three
   experiments are the light scattering cross sections, the
   atom scattered clouds and the cavity leaking photons
   respectively. All these experimental measurable quantities
   are determined  by the density-density and bond-bond  correlation functions.
   We will develop a systematic and unified theory of using the optical Bragg scattering, atomic Bragg scattering or cavity QED to detect the nature of
  quantum phases such as both the ground state and the excitation spectrum above the ground state of  interacting bosons loaded in optical lattices.

  The Extended Boson Hubbard Model (EBHM) with various  kinds of interactions, on all kinds of lattices and at different filling factors
   is described by the following Hamiltonian \cite{boson,bloch,gan,square,sca,frusqmc,pq1,mob,bipart,frus,honey}:
\begin{eqnarray}
  H_{BH} & = & -t \sum_{ \langle ij \rangle } ( b^{\dagger}_{i} b_{j} + h.c. ) -
        \mu \sum_{i} n_{i} + \frac{U}{2} \sum_{i} n_{i} ( n_{i} -1 )
                             \nonumber \\
        & +  & V_{1} \sum_{ <ij> } n_{i} n_{j}
        + V_{2} \sum_{ \langle ik \rangle } n_{i} n_{k} + \cdots
\label{boson}
\end{eqnarray}
    where $ n_{i} = b^{\dagger}_{i} b_{i} $ is the boson density, $ t $ is the nearest neighbor hopping which
    can be tuned by the depth of the optical lattice potential, the
    $ U, V_{1}, V_{2} $ are onsite, nearest neighbor (nn) and next nearest neighbor (nnn) interactions respectively,
    the $ \cdots $ may include further neighbor interactions,  dipole-dipole interaction $ V= \frac{ p^{2}}{ |\vec{r}_i- \vec{r}_j |^{3} } $
    and possible ring-exchange interactions. The filling factor $ n= N_a/N $ where $ N_a $ is the number of atoms and $ N $ is the
    number of lattice sites. The on-site interaction
    $ U $ can be tuned by the Feshbach resonance \cite{boson}.
     There are many possible ways to generate longer range interaction $ V_{1}, V_{2},....$ of
     ultra-cold atoms loaded in optical lattices.
     Being magnetically  or electrically polarized, the $ ^{52}Cr $  atoms \cite{cromium} or
     polar fermionic molecules \cite{junpolar} $ ^{40}K+ ^{87} Rb $  ( or bosonic molecules $ ^{39} K+ ^{87} Rb $ )
     interact with each other via long-rang anisotropic
     dipole-dipole interactions. Loading the $ ^{52}Cr $ or the polar bosonic molecules on a 2d optical lattice
     with the dipole moments perpendicular to the trapping plane can be mapped to
     Eqn.\ref{boson} with long-range  repulsive interactions $ \sim p^{2}/r^{3}
     $ where $ p $ is the dipole moment.
     Possible techniques to generate long-range interactions in a gas of ground state alkali atoms by
     weakly admixing excited Rydberg states  with laser light was proposed in \cite{mixture}.
     The generation of the ring exchange interaction has been discussed in \cite{qs}.

\begin{figure}
    \includegraphics[width=3.5cm]{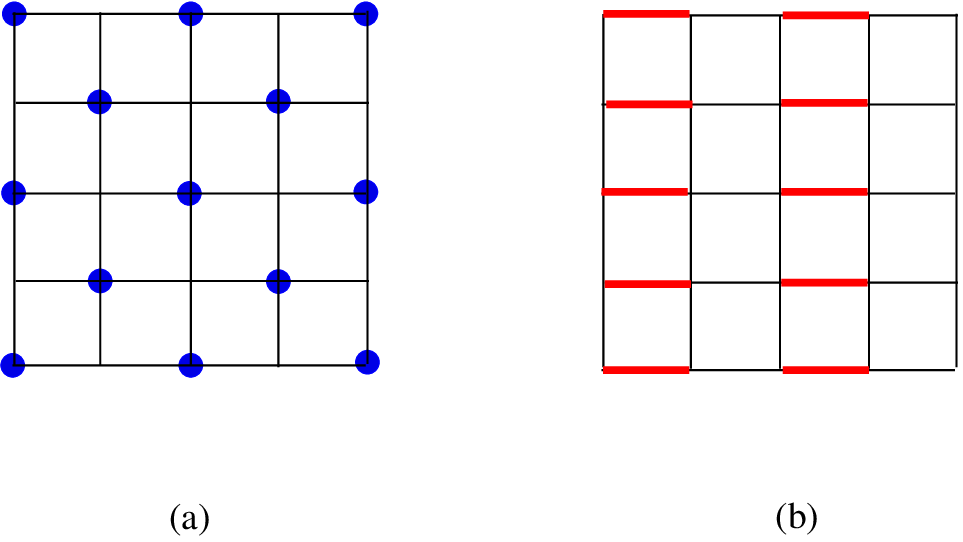}
    \hspace{0.5cm}
    \includegraphics[width=3.5cm]{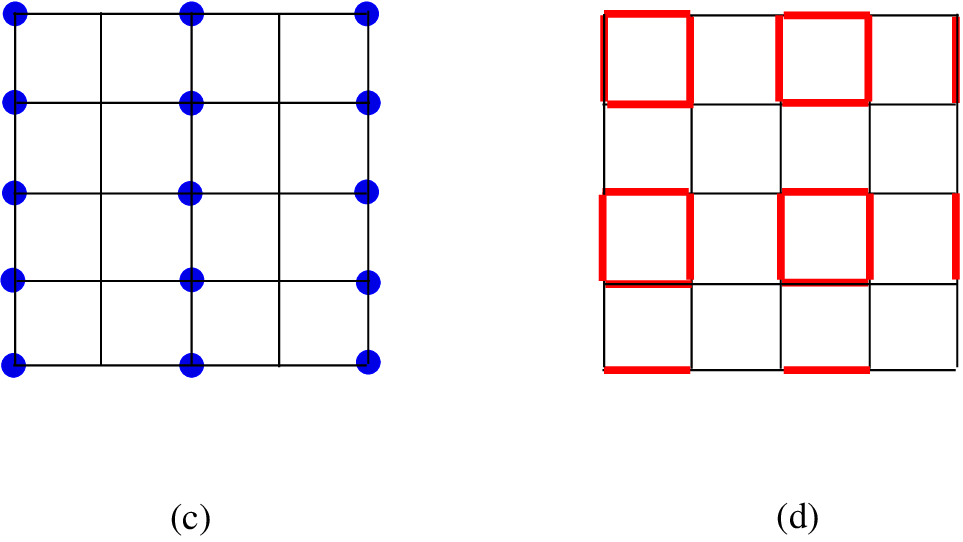}
    \caption{ Several insulating states at filling factor $ f=1/2 $ in a
    square lattice. (a) The charge density wave (CDW) phase  with
    ordering wavevector $ \vec{Q}_{n}=(\pi,\pi) $. (b) valence bond
    solid (VBS) phases with ordering wavevector $ \vec{Q}_{K}=(\pi,0) $
    where the kinetic energy $ \langle K_{ij} \rangle =  \langle
    b^{\dagger}_{i} b_{j} + h.c. \rangle $ takes a non-zero constant $ K
    $ in the two sites connected with a dimer, but $ 0 $ in the two
    sites without a dimer. (c) Stripe CDW order at  $ \vec{Q}_{n}=(\pi,
    0) $ and (d) Plaquette VBS order at $ \vec{Q}_{K1}=(\pi,0),
    \vec{Q}_{K2}=(0, \pi) $. See Refs.\cite{gan,pq1,mob,bipart,frus}. }
\label{squarephase}
\end{figure}

       Various kinds of bipartite and frustrated  optical lattices  can be realized by suitably choosing
       the geometry of the laser beams generating the optical lattices.
       For example, using three coplanar beams of equal intensity having
       the three vectors making a $ 120 ^{\circ}$ angle with each other,
       the potential wells have their minima in a honeycomb or a triangular lattice \cite{honeylattice}.
       Four beams travel along the three fold symmetry axes of a regular
       tetrahedron, the potential wells have their minima in a body-centered-cubic lattice \cite{honeylattice}.
       The authors in  \cite{kalattice} proposed to create a kagome optical lattice using superlattice techniques.
       It was known that cold atoms in a frustrated lattice show
       completely different behaviors than in a bipartite lattice.
     Some of the important phases with long range interactions in
     bipartite lattices are studied in details in \cite{square,squaresoft,frusqmc,honey,pq1,mob,bipart,frus} and
     listed in Fig.\ref{squarephase} and \ref{squarediagram}. Some quantum phases in frustrated lattices are studied in
     \cite{gan,frus} and will be briefly reviewed in the appendix A.


   In this paper, we explicitly show that the two photon Raman transition processes shown in Fig.\ref{raman} in the optical Bragg scattering,
   atomic Bragg scattering or cavity QED ( with the classical scattered light in Fig.1a replaced by the quantum cavity photon in Fig.\ref{cavityqed} ) not only couple to
   the density order parameter, but also the  {\sl valence bond order } parameter due to the hopping of the bosons
   on the lattice. This kinetic energy coupling is extremely
   sensitive to any superfluid order or any Valence bond (VB ) order
   which can be considered as a local superfluid order on a local bond at corresponding ordering wavevectors.
   By tuning the incident angle in the Fig.1a, the angle between the two laser lights in Fig.1b or the incident light
   and the cavity axis in Fig.\ref{cavityqed}a, one can detect both the ground state  and  the
   excitation spectrum
   by measuring the light scattering cross sections, the atom scattered clouds and the leaking cavity photon numbers respectively.
   The static structure function can detect not only the well known superfluid and Mott insulating phases,
   but also other interesting phases such as charge density wave (CDW), both dimer and plaquette valence bond solids (VBS),
   even some CDW-VBS phase unique in frustrated lattices at commensurate
   fillings \cite{gan,square,frusqmc,pq1,mob,bipart,frus,honey}.
   It can also detect the corresponding CDW supersolids  and VB supersolids at in-commensurate fillings in both bipartite and frustrated
   lattices. Furthermore, the dynamic structure function can detect the  excitation spectrum and the corresponding spectral weights
   in all these quantum phases.

 Recently, there are very impressive advances in probing individual atoms without and with
 optical lattices using electron and optical microscopy \cite{local0,local1,insitu0,insitu1}.
 It is a direct {\sl in situ} measurements on individual density and density fluctuations.
 For example, the superfluid phase and the Mott phase in optical lattices studied by the earliest time of flight measurements in \cite{bloch}
 are probed by the spatially resolved, {\sl in situ} imagings. As argued in Sec. IX, the {\sl in situ} measurement is an effective
 measurement of local density and density fluctuations. Under the local density approximation (LDA),
 these  {\sl in situ} imagings can be used to extract the temperature of the system and test the scaling relations of
 thermodynamic quantities across classical and quantum phase transitions \cite{trapthe0,trapthe1}, but so far, it may not be used to
 test the dynamic density-density or bond-bond scaling relations which are non-local in the space and time, their peak positions lead to the excitation spectra of the
 corresponding quantum phases ( See Fig.\ref{excitations} ). These dynamic correlation function can be precisely measured
 by the three scattering methods to be studied in this paper. However, all the three scattering methods involve large number of atoms in the system,
 so it is a momentum space probes, so it may not resolve the local information.
 So the scattering detection methods  and the {\sl in situ} measurements are complementary and dual to each other.
 Both kinds of measurement involve density or bond and their correlations, but the former focus in momentum space, the latter
 in the real space, similar to the relation between the ARPES versus STM in high temperature superconductors \cite{hightc}.
 In Sec.IX, we will compare both kinds of methods in some details.

  The rest of the paper is organized as the follows.
  In Sec.II, just from symmetry breaking points of view, we  study several very general and important
  properties of density-density, bond-bond correlation functions
  and their finite size scaling properties inside a flat trap \cite{flattrap}.
  In Sec. III, we study the off-resonant light scattering from the two level atoms hopping in an optical lattice and
  find  the off-resonant light beams not only couple to the density,
  but also the kinetic energy of the cold atoms hopping on a optical
  lattice. We also estimate the relative strengths of the two
  couplings by using the harmonic approximation \cite{boson}. We show that
  the light scattering cross section is the sum of the density-density and
  bond-bond correlation function with the corresponding form
  factors. In Sec.IV, we study the density-density correlation
  function in the superfluid, Mott and quantum critical regime at
  integer filling case from both the boson action and its dual vortex action.
  All the previous work seem focused on the order parameter
  correlation functions instead of the density-density ones which
  are measured by the three experimental detection methods.
  In Sec. V, we study the elastic scattering cross
  section in a CDW to detect the ground state, also in-elastic
  scattering cross section to detect the excitation spectrum and its scaling form near the second order
  transition from the CDW and CDW supersolid. In Sec. VI, we study the elastic scattering cross
  section in a VBS to detect the ground state, also in-elastic
  scattering cross section to detect its excitation spectrum and its
  scaling form near the second order transition from the VBS and VBS supersolid. In Sec. VII, we extend our
  discussions to frustrated lattices  and stress several new features in the density-density and bond-bond correlation functions
  unique to frustrated lattices. Then we apply the formalism to study the light scattering cross sections from the X-CDW, VBS and CDW-VBS phases
  in a triangular lattice.
  In Sec.VIII, we propose to use cavity QED
  as an effective detection method. Phase sensitive homodyne measurement and the Florescence spectrum  measurement can be used to detect
  the ground state and excitation spectrum of various quantum phases. This possible
  cavity enhanced  off-resonant light scattering
  detection method should be complementary to the light scattering
  and atom Bragg scattering discussed in the previous sections. In the Sec. IX, we discuss quantum phases and quantum phase transitions inside
  a harmonic trap. We classify the necessary conditions for the local density approximation ( LDA ) to hold and give the scaling forms for the static and equal time density-density ( or bond-bond ) correlation functions under the LDA. We compare the scattering detection methods with the {\sl in situ } local measurements.
  We also point out the possibility to experimentally observe a stable ring-structure supersolid of dipolar bosons inside a harmonic trap.
  In the final Sec.X, we contrast the three scattering experiments, also the {\sl in situ} local measurements, discuss their
  strengths and weaknesses in detecting all these quantum phases and summarize our
  results. In the appendix A, we will review several important phases in both bipartite and
  frustrated lattices. We also stress a recently
  discovered new phase which has both CDW and VBS order called CDW-VBS phase in the triangular lattice.
  This kind of CDW-VBS phase is  unique and quite common to frustrated lattices.
  In the appendix B, we discuss light Bragg scattering experiment and a possible quantum beat
  measurement to measure small energy differences in inelastic light Bragg scattering experiment.
  In the appendix C, we analyze  both momentum and energy atom Bragg scattering experiments.
  In the Appendix D, we point out some mistakes in the previous work \cite{off1} and
  show that  the light scattering cross section or cavity QED scattering at the
  classical diffraction minimum $ \vec{q}= ( \pi, \pi ) $ may be an effective measurement of the superfluid density $ \rho_s $.
  A short report on some results of this paper already appeared in \cite{braggprl}.
  {\sl In the following, we just take 2d optical lattices as examples.
  The 1d and 3d cases can be similarly discussed }.

\section{ Density-density and bond-bond correlation functions in a bipartite lattice  }

   In this section, from symmetry breaking point of view, we will study several important properties of
   the density-density correlation functions and the bond-bond correlation functions
   in these quantum phases in a bipartite lattice respectively. The
   counterparts in a frustrated lattice will be discussed in Sec.VII.
   The results achieved should
   also be useful to Quantum Monte Carlo simulations \cite{square,squaresoft,honey} on the extended
   boson Hubbard model in Eqn.1 in a finite $ N = L \times L $
   lattice.

\subsection{ Density-Density correlation function   }

  In view of the possible CDW ordering at $ \vec{q}=\vec{Q}_{N} $, one can decompose the density at site $ i $ as
 \begin{equation}
   N_{i}= n_{i} + (-1)^{i} m_{i}
 \label{nm}
 \end{equation}
    where $ i= x $ for the $ \vec{Q}_{N}=( \pi, 0 ) $ stripe CDW in Fig.\ref{squarephase}c
  and $ i= x + y $ for the $ \vec{Q}_{N}=( \pi, \pi ) $ CDW in Fig.\ref{squarephase}c. Then the density-density correlation function  is defined
  as:
 \begin{equation}
   S_{N} ( \vec{q}, t )= \frac{1}{N^2} \sum_{i,j} e^{-i \vec{k} \cdot ( \vec{r}_i -\vec{r}_{j} ) } \langle N_{i} (t) N_{j}(0) \rangle
 \label{nn}
 \end{equation}

    By substituting the decomposition of $ N_i $ into the above Eqn. we can get:
\begin{eqnarray}
  S_{N} ( \vec{q}, t ) & = &  \frac{1}{N} \sum_{i} e^{-i \vec{q} \cdot \vec{r}_i }
    \langle n (i, t) n(0,0 ) \rangle_{C}  + n^{2} \delta_{\vec{q},0}   \nonumber   \\
     & + & \frac{1}{N} \sum_{i} e^{-i (\vec{q} - \vec{Q}_{N} ) \cdot  \vec{r}_{i} }
    \langle m (i, t) m(0,0) \rangle_{C}   \nonumber  \\
    & +  &  m^{2} \delta_{\vec{q},\vec{Q}_{N}}
\label{nnt}
\end{eqnarray}
    where  we have used the translational invariance to get rid of one summation,
    $ \langle n_{i} (t) n_{j}(0) \rangle_{C}=\langle n_{i} (t) n_{j}(0) \rangle- n^{2} $ and
    $  \langle m_{i} (t) m_{j}(0) \rangle_{C}= \langle m_{i} (t) m_{j}(0) \rangle -m^{2} $ are connected
    Green functions.

    Its Fourier transform leads to the dynamic structure function:
\begin{eqnarray}
  S_{N} ( \vec{q}, \omega ) & =  & \int dt e^{-i \omega t}  S_{N} ( \vec{q}, t )
  \nonumber  \\
    & = & n^{2} \delta_{\vec{q},0} \delta (\omega) +  \frac{1}{N} S^{inel}_n (\vec{q}, \omega )      \nonumber   \\
    & + &  m^{2} \delta_{\vec{q},\vec{Q}_{N}} \delta (\omega) +
     \frac{1}{N} S^{inel}_m (\vec{q}, \omega )
\label{nnw}
\end{eqnarray}
    where the first and the third  $ \delta (\omega) $ terms denote the elastic
    scattering at $ \vec{q}=0 $ and  $ \vec{q} = \vec{Q}_{N} $
    respectively  $ S^{el}_n \delta (\omega)  = n^{2} \delta_{\vec{q},0} \delta (\omega), S^{el}_m \delta (\omega)
     = m^{2} \delta_{\vec{q},\vec{Q}_{N}} \delta (\omega) $, the second
    and the fourth terms denote the inelastic scattering near $ \vec{q}=0 $ and  $ \vec{q} = \vec{Q}_{N} $
    respectively.

\begin{figure}
\includegraphics[width=6cm]{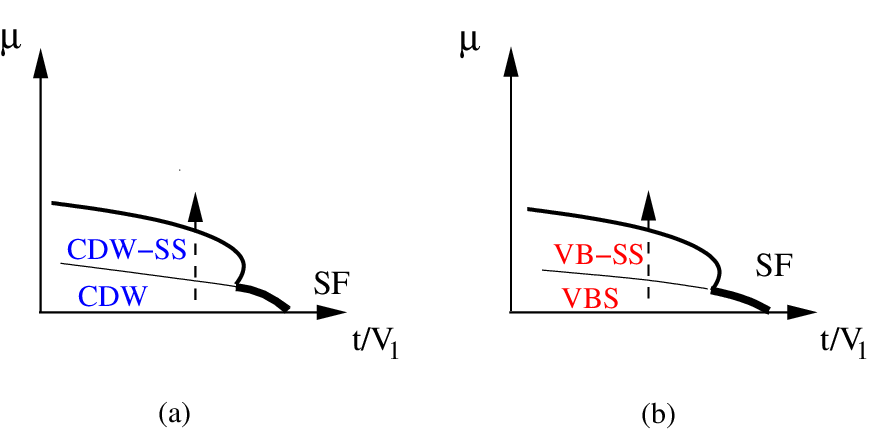}
\caption{  Zero temperature Phase diagram of the boson Hubbard model
with nearest neighbor interaction $ V_1 $ near half filling in a
bipartite lattice \cite{bipart}. The labels of the axes are given in
Eqn.1. (a) Ising limit. (b) Easy-Plane limit.  See also Fig.
\ref{sfmott23} for the finite temperature phase diagram and
Fig.\ref{cdwtrap} for the phases inside a trap. }
\label{squarediagram}
\end{figure}

  The equal time density-density correlation function is:
\begin{eqnarray}
  S_{N} ( \vec{q}, t=0 ) & =  & \frac{1}{N} \sum_{i} e^{-i \vec{q} \cdot \vec{r}_i }
   \langle n(i,0) n(0, 0) \rangle_{C}  + n^{2} \delta_{\vec{q},0}   \nonumber   \\
   & + & \frac{1}{N} \sum_{i} e^{-i (\vec{q} - \vec{Q}_{N} ) \cdot \vec{r}_i }
   \langle m (i, 0) m(0, 0) \rangle_{C}  \nonumber  \\
   & +  &  m^{2} \delta_{\vec{q},\vec{Q}_{N}}
\label{nn0}
\end{eqnarray}

    From $ S_{N} ( \vec{q}, t=0 )= \int \frac{ d \omega }{ 2 \pi }  S_{N} ( \vec{q}, \omega
    ) $, we can write the equal-time
    correlation function as the sum of the elastic one and the in-elastic one:
\begin{equation}
     S_{N} ( \vec{q}, t=0 )= S^{el}_n +  \frac{1}{N} S^{inel}_n
     (\vec{q})+ S^{el}_m + \frac{1}{N} S^{inel}_m (\vec{q} )
\end{equation}
     where $ S^{el}_n= n^{2} \delta_{\vec{q},0}, S^{el}_m  = m^{2}
     \delta_{\vec{q},\vec{Q}_{N}} $ and $ S^{inel}_n
     (\vec{q})= \int \frac{ d \omega }{ 2 \pi } S^{inel}_n (\vec{q}, \omega )
     , S^{inel}_m(\vec{q})= \int \frac{ d \omega }{ 2 \pi } S^{inel}_m (\vec{q}, \omega
     ) $.

    In the following, we discuss the density-density correlation functions in the
    superfluid, CDW and CDW supersolid respectively.

{ \sl (a) Superfluid  state near $ \vec{q}=0 $. }

     In the superfluid state,  the first term  in Eqn.\ref{nnt} stands for the gapless superfluid mode near $ \vec{q}=0 $.
     Taking the results from Ref.\cite{ssorder,sforder}, we have:
\begin{equation}
    S^{SF}_{n}( \vec{q},\omega ) = \langle \delta n (-\vec{q},-\omega_{n} )  \delta n(\vec{q},\omega_{n} ) \rangle
     =  \frac{  \rho_{s} q^{2} }{ \omega^{2}_{n} +  v^{2} q^{2} }
\label{sf}
\end{equation}
    where $ v^{2} = \rho_{s}/\kappa $ is the superfluid phonon velocity, the $ \rho_s $ is the superfluid density and the
    $ \kappa $ is the compressibility. Indeed
\begin{equation}
     \kappa_{SF}=  S^{SF}_{n}( \vec{q} \rightarrow 0 ,\omega=0 )= \rho_{s}/v^{2}
\label{cm}
\end{equation}
     as it should be. Note that the compressibility can also be directly measured by the {\sl in situ} method \cite{insitu0,insitu1}
     ( see Sec.IX for details ).  From the analytical continuation $ i \omega_{n} \rightarrow \omega+ i\delta $ and taking the imaginary part,
    we can identify the dynamic structure factor:
\begin{equation}
   S^{>}_{n}( \vec{q},\omega ) = S_{n}(q) \delta ( \omega -v q ),~~~S_{n}(q)= \rho_{s} q \pi/2v
\label{sfequal}
\end{equation}
    where the $ S_{n}(q) $ is the equal time  density-density correlation function near $ q=0 $ shown in Fig.\ref{excitations}a1.

    The second term near $ \vec{q}=\vec{Q}_{N} $ in Eqn.\ref{nnt} comes from the roton contribution
    near $ \vec{q}=\vec{Q}_{N} $ in Fig.\ref{excitations}a1. The dispersion near $ \vec{q}=\vec{Q}_{N} $ can be
    taken as $ \omega( q ) \sim  \Delta_{r} + \frac{ ( \vec{q}- \vec{Q}_{N} )^{2} }{ 2 m_r } $ where
    $ \Delta_{r} >0 $ is the roton gap. There is no CDW order yet, so $ m =0 $ in the Eqn.\ref{nnt}.
    However, as one approaches the CDW from the SF in Fig.\ref{squarediagram}a,
    there is a first order transition into the CDW driven by the collapse of the roton gap $ \Delta_r $.

{ \sl  (b)  CDW state near $ \vec{q}= \vec{Q}_{N} $. }

    In the CDW state, the $ nn $ correlator near $ \vec{q}=0 $ in Eqn.\ref{nnt} is very small, so can be dropped safely,
    so one only need to focus on the
    $ mm $ correlator near $ \vec{q}=\vec{Q}_{N} $. The discrete lattice symmetry was broken due to the non-uniform density distribution,
    so $ m \neq 0 $, there is also   a gap  $ \Delta_{CDW} $ in the CDW state, so the connected equal time correlation function decays exponentially
    in the CDW state:
\begin{equation}
     \langle m_{i} (0) m_{j}(0) \rangle_{C} \sim e^{ - | \vec{r}_i -\vec{r}_{j}|/\xi_{CDW} }
\label{cdwexp}
\end{equation}
  where $ \xi_{CDW} \sim 1/\Delta_{CDW} $ is the correlation length in the CDW.
  So the fluctuation part in Eqn.\ref{nn0} is  only at the order of $ \sim 1/N $,  we conclude:
\begin{equation}
      S_{N} ( \vec{Q}_{N}, t=0 ) = m^{2} + \frac{1}{N} S^{inel}_m (\vec{q}=\vec{Q}_{N}
      ) = m^{2}+ O(1/N)
\label{nnstatic}
\end{equation}
      Here we can see that the equal time structure factor is the sum of the elastic scattering $ S^{el}_m  =
      m^{2} $ plus a $ 1/ N $ in-elastic background.


     For $ \vec{q} \neq \vec{Q}_{N} $, but close to  $ \vec{Q}_{N} $, then we find
\begin{equation}
    S_{N} ( \vec{q}, t )  \sim    \frac{1}{N} \sum_{i} e^{-i (\vec{q} - \vec{Q}_{N} ) \cdot \vec{r}_i } \langle m (i, t) m(0, 0) \rangle_{C}
\label{nnqn}
\end{equation}
     where one can extract the excitation spectrum and spectral weight of the CDW near $ \vec{q} \neq \vec{Q}_{N} $ shown in Fig.\ref{excitations}c1.
     In fact, the excitation spectrum around $ \vec{k} - \vec{Q}_{N} =\vec{q} $ can also be extracted from the Feynman relation:
 \begin{equation}
     \omega_{CDW}( \vec{q} ) =  \frac{ \int^{\infty}_{-\infty} d \omega \omega S_{N}( \vec{q}, \omega ) }
                                       { \int^{\infty}_{-\infty} d \omega S_{N}( \vec{q}, \omega )}
\label{cdwdis}
\end{equation}
      which holds at $ \vec{q} \rightarrow 0 $, but not at $ \vec{q}=0 $.  The $ f $-sum rule
      $  \int^{\infty}_{-\infty} d \omega \omega S_{N}( \vec{q}, \omega )= \langle [ [ n( \vec{q} ), H ], n(-\vec{q}) ]
      \rangle $  at $ d=2 $ gives:
\begin{eqnarray}
        \int^{\infty}_{-\infty} d \omega \omega S_{N}( \vec{q}, \omega )
       =  - 2 t \sum_{\vec{k}} [ ( \cos q_x -1) \cos k_x  \nonumber   \\
       +  ( \cos q_y -1) \cos k_y ]
      \langle \Psi_0 | b^{\dagger}_{\vec{k}} b_{\vec{k}}  | \Psi_0
      \rangle   ~~~~~~~~~~~
\label{sumd2}
\end{eqnarray}
      where the $  | \Psi_0  \rangle $ is the ground state wavefunction.

      The $ d =1 $ version of this sum rule was used to extract the excitation spectra by QMC  in both the SF and the CDW state in Ref.\cite{sca}. Note that
      at $ d=1 $, the SF to the CDW transition along the horizontal axis in Fig.\ref{squarediagram}a is in the Kosterlitz-Thouless (KT) transition universality class instead of first order transition in $ d=2 $ and $ d=3 $.  We expect that the SF to the CDW-SS transition in Fig.\ref{squarediagram}a is also in the KT
      transition universality class.


{ \sl (c) CDW Supersolid state }

    In this case, the first term near $ \vec{q}=0 $ in Eqn.\ref{nnt} stands for the gapless superfluid mode as given by Eqn.\ref{sf}
    and shown in the lower branch in Fig.\ref{excitations}b1.

    The static order at  $ \vec{q}= \vec{Q}_{N} $ is given by
    Eqn.\ref{nnstatic} and the dynamic structure factor  close to  $ \vec{Q}_{N} $ in
    the upper branch in Fig.\ref{excitations}b1 is given by Eqn.\ref{nnqn} respectively.

\subsection{ The bond-bond correlation function }

    Inside a VBS state with a ordering wavevector $ \vec{Q}_{K} $, similar to Eqn.\ref{nm}, one can write the kinetic energy
    at a given bond $ i, \alpha, \alpha=\hat{x}, \hat{y} $ as
 \begin{equation}
   D_{i,\alpha}= B_{i,\alpha} + e^{i \vec{Q}_{K} \cdot \vec{r}_{i} }  K_{i,\alpha}
 \label{bk0}
 \end{equation}

      For example, for the $ \vec{Q}_{K}= (\pi, 0 ) $ VBS state, Eqn.\ref{bk0} becomes:
\begin{eqnarray}
     D_{i x} & = & B_{i x } + (-1)^{i_x} K_{i x }  \nonumber  \\
     D_{i y} & = & B_{i y}= B_{i x}
\label{bk}
\end{eqnarray}

      Then the bond-bond correlation function is defined as:
 \begin{equation}
      S_{D \alpha} ( \vec{q}, t )= \frac{1}{N^2} \sum_{i,j} e^{-i \vec{k} \cdot ( \vec{r}_i -\vec{r}_{j} ) } \langle D_{i \alpha} (t) D_{j \alpha}(0) \rangle
 \label{dd}
 \end{equation}

    By substituting the decomposition Eqn.\ref{bk0} into the above Eqn. we can get:
\begin{eqnarray}
  S_{D \alpha} ( \vec{q}, t ) & = &  \frac{1}{N} \sum_{i} e^{-i \vec{q} \cdot \vec{r}_i }
    \langle B_{\alpha} (i, t) B_{\alpha}(0,0 ) \rangle_{C}  + B^{2} \delta_{\vec{q},0}   \nonumber   \\
     & + & \frac{1}{N} \sum_{i} e^{-i (\vec{q} - \vec{Q}_{K} ) \cdot  \vec{r}_{i} }
    \langle K_{\alpha} (i, t) K_{\alpha}(0,0) \rangle_{C}   \nonumber  \\
    & +  &  K^{2} \delta_{\vec{q},\vec{Q}_{K}}
\label{ddt}
\end{eqnarray}
    where the  $ \langle B_{\alpha} (i, t) B_{\alpha}(0,0 ) \rangle_{C}=\langle  B_{\alpha} (i, t) B_{\alpha}(0,0 ) \rangle- B^{2} $ and
    $  \langle K_{\alpha} (i, t) K_{\alpha}(0,0) \rangle_{C}= \langle K_{\alpha} (i, t) K_{\alpha}(0,0) \rangle -K^{2} $ are connected
    Green functions.

    In a VBS or a VBS superfluid state, the density is  {\sl uniform},  then the Eqn.\ref{nnt} should be replaced by:
\begin{equation}
  S_{N} ( \vec{q}, t )  =  \frac{1}{N} \sum_{i} e^{-i \vec{q} \cdot \vec{r}_i }
    \langle n (i, t) n(0,0 ) \rangle_{C}  + n^{2} \delta_{\vec{q},0}
\label{nntvbs}
\end{equation}
    where  there is no staggered component $ m $.

    Inside a VBS state with a ordering wavevector $ \vec{Q}_{K} $, there is a big CDW gap in the connected density-density
    correlation function in Eqn.\ref{nntvbs}.  Obviously, this CDW gap is much larger than the VBS gap introduced below Eqn.\ref{kk0}.
    Furthermore,  $ S_{N} ( \vec{Q}_{K}, 0 ) =\frac{1}{N} \sum_{i} e^{-i \vec{Q}_{K} \cdot \vec{r}_i }
    \langle n (i, 0) n(0,0 ) \rangle_{C} $ is a smearing of a very small density fluctuation on a lattice scale, so it contributes to
    a very small background which is completely negligible compared to the VBS fluctuations near $ \vec{Q}_{K} $ to be discussed in the following.
    ( Note that inside a superfluid, it may still be appreciable even at the classical diffraction minimum $ \vec{q}=
    ( \pi,\pi) $ as  to be shown in the Appendix B ).

{ \sl (a) Superfluid near $ \vec{q}=0 $. }

    The gapless superfluid mode near $ \vec{q}=0 $ in Eqn.\ref{nntvbs} is still
    given by Eqn.\ref{sf}, but there is no peak  near $ \vec{q}=\vec{Q}_{K} $ in Eqn.\ref{nntvbs}.
    Instead, there is a peak near $ \vec{q}=\vec{Q}_{K} $
    in the bond-bond correlation function in Eqn.\ref{kkt}.  The excitation spectrum near $ \vec{q}=\vec{Q}_{K} $ can be
    taken as $ \omega_{v}( q ) \sim  \Delta_{rv} + \frac{ ( \vec{q}- \vec{Q}_{K} )^{2} }{ 2 m_{rv} } $ where
    $ \Delta_{rv} >0 $ is the  " valence bond roton" gap. There is no VBS order yet, so $ K =0 $ in the Eqn.\ref{kkt}.
    However, as one approaches the VBS from the SF in Fig.\ref{squarediagram}b,
    there maybe a first order transition into the VBS driven by the collapse of the gap $ \Delta_{rv} $.

{ \sl (b) VBS state }

     The lattice symmetry was broken by the non-uniform kinetic energy, so $ K \neq 0 $.
     One can neglect  the very small $ B B $ fluctuations near $ \vec{q}=0 $ in Eqn.\ref{ddt}.
     Then  Eqn.\ref{ddt} can be simplified to:
\begin{eqnarray}
  S_{K \alpha } ( \vec{q}, t ) & =  & \frac{1}{N} \sum_{i} e^{-i (\vec{q} - \vec{Q}_{K} ) \cdot \vec{r}_i }
   \langle K_{\alpha } (i, t) K_{ \alpha }(0, 0) \rangle_{C}  \nonumber  \\
   & + &  K^{2} \delta_{\vec{q},\vec{Q}_{K}}
\label{kkt}
\end{eqnarray}
     where $ \vec{Q}_{K}= ( \pi, 0 ) $ is the ordering wavevector for the VBS.

    The equal time bond-bond correlation function is:
\begin{eqnarray}
  S_{ K \alpha } ( \vec{q}, t=0 )  & = &  \frac{1}{N} \sum_{i} e^{-i (\vec{q} - \vec{Q}_{K} ) \cdot \vec{r}_i  }
   \langle K_{\alpha } (i, 0 ) K_{\alpha }(0, 0) \rangle_{C}   \nonumber  \\
   & +  & K^{2} \delta_{\vec{q},\vec{Q}_{K}}
\label{kk0}
\end{eqnarray}

    Inside the VBS, there is also  a VBS gap  $ \Delta_{VBS} $, so the connected equal time correlation function decays exponentially
    in the VBS state:
\begin{equation}
     \langle K_{i \alpha} (0) K_{j \alpha}(0) \rangle_{C} \sim e^{ - | \vec{r}_i -\vec{r}_{j}|/\xi_{VBS} }
\label{vbsexp}
\end{equation}
     where the  $ \xi_{VBS} \sim 1/\Delta_{VBS} $ is the correlation length in the VBS state. So the first term in Eqn.\ref{kk0} is  at the order of $ \sim 1/N $, so we conclude:
\begin{equation}
      S_{K \alpha } ( \vec{Q}_{K}, t=0 ) = K^{2} + O(1/N )
\label{kkstatic}
\end{equation}

    For $ \vec{q} \neq \vec{Q}_{K} $, but close to  $ \vec{Q}_{K} $, then one has
\begin{equation}
    S_{K \alpha } ( \vec{q}, t )  \sim    \frac{1}{N} \sum_{i} e^{-i (\vec{q} - \vec{Q}_{K} ) \cdot  \vec{r}_i }
     \langle K_{ \alpha } ( i, 0 ) K_{ \alpha }(0, 0) \rangle_{C}
\label{kkqk}
\end{equation}
     where one can extract the excitation spectrum and spectral weight of the VBS near $ \vec{q} \neq \vec{Q}_{K} $ shown in Fig.\ref{excitations}c2.
     The VBS excitation spectrum around $ \vec{k} - \vec{Q}_{K} =\vec{q} $ can be extracted from the Feynman relation:
 \begin{equation}
     \omega_{K \alpha}( \vec{q} ) =  \frac{ \int^{\infty}_{-\infty} d \omega \omega S_{K \alpha }( \vec{q}, \omega ) }
                                       { \int^{\infty}_{-\infty} d \omega S_{K \alpha }( \vec{q}, \omega )}
\label{vbsdis}
\end{equation}
     which holds at $ \vec{q} \rightarrow 0 $, but not at $ \vec{q}=0 $.
      The $ f $-sum rule gives
      $  \int^{\infty}_{-\infty} d \omega \omega S_{K \alpha }( \vec{q}, \omega )=
      \langle   [ [ S_{K \alpha }( \vec{q} ), H ], S_{K \alpha }(-\vec{q}) ] \rangle  $.
      Unfortunately, there is no simple expression for this double commutator which depends on the details of the Hamiltonian.

{ \sl (c) VBS supersolid state }

    In this case, the gapless superfluid mode near $ \vec{q}=0 $ in Eqn.\ref{nntvbs} is still given by Eqn.\ref{sf}.

    The static order at  $ \vec{q}= \vec{Q}_{K} $ is given by
    Eqn.\ref{kkstatic} and the dynamic structure factor  close to  $ \vec{Q}_{K} $ in
    the upper branch in Fig.\ref{excitations}b2 is given by Eqn.\ref{kkqk}.

\subsection{ Finite size scaling of the correlation functions
        near the quantum critical point in a flat trap }

    In the last two subsection, we discuss the properties of separate phases.
    We showed that well inside the phases, the mean field results dominate, the fluctuations are suppressed by $ 1/N $ due to the CDW  or VBS gap.
    However, near a 2nd order transition, the mean field theory breaks down, the fluctuations diverges.
    Because the CDW ( or VBS ) and the SF break two completely different symmetries, usually,
    there could be either a first order or second transition between them.
    Here we focus on the possible continuous quantum phase transitions
    between the phases.  Indeed, as shown in \cite{dipolarss}, the CDW-SS to the SF transition driven by the chemical potential $ \mu $ in Fig.\ref{squarediagram}
    for {\sl a dipole-dipole interaction } is a 2nd order transition in $ 3d $ Ising universality class \cite{int}.
    Then the density or the bond can be taken as the order  parameters, the density-density or bond-bond correlation  functions
    near the corresponding ordering wave vectors
    $ \vec{Q}_N $ or $ \vec{Q}_K $ will diverge at the critical point. In this section, we focus on the scalings inside a flat trap.
    The scalings inside a harmonic trap will be discussed in Sec. IX.

    At the critical point, the Eqn.\ref{cdwexp} becomes \cite{finiteqmc}.
\begin{equation}
   \langle m_{i} (0) m_{j}(0) \rangle \sim  \frac{1}{ | \vec{r}_i -\vec{r}_{j}|^{d+z-2+\eta} }
\label{cdwcr}
\end{equation}
     Substituting this equation into Eqn. \ref{nnqn}, we can see:
\begin{eqnarray}
    S_{N} ( \vec{q}, t=0)  & \sim  &  \frac{1}{Na^{d}} \int d^{d}r \frac{ e^{-i (\vec{q} - \vec{Q}_{N} ) \cdot \vec{r} } } { r^{d+z-2+\eta}}
        \nonumber   \\
         &  =  & \frac{1}{V} \frac{1}{ |\vec{q} - \vec{Q}_{N}|^{2-z-\eta}}
\label{nnqncr}
\end{eqnarray}
     where $ V= Na^{d} = L^{d} $ is the volume of the system.
     To be general, we keep the space dimension $ d $.


      In real cold atom experiments, any divergence at the critical point in the density-density
      or bond-bond correlation functions in the thermodynamic limit will be cutoff by the
    trap size $ L \sim 100 \mu m $. With the optical lattice constant $ a \sim 0.5 \mu m $, the trap
    can hold around $ N \sim 10^{5} \sim 10^{6} $ number of particles.
    In fact, this number $ N $  is comparable to the present quantum Monte-Carlo simulations.
       From the Eqn.\ref{nnqncr}, one can see that the finite size scaling form of
       the static  and equal time density-density structure factor at the ordering wavevector $ \vec{Q}_{N} $ is:
\begin{eqnarray}
        S_{n}( \vec{Q}_{N}, i \omega_{n} =0 )   =  L^{2-\eta-d} F_{ns} ( L^{1/\nu}(K-K_{c}), \beta_T/L^{z} )  \nonumber   \\
        S_{n}( \vec{Q}_{N}, \tau =0 )  =  L^{2-z-\eta-d } F_{ne} ( L^{1/\nu}(K-K_{c}), \beta_T/L^{z} )
                             \nonumber   \\
         =  L^{-2 \beta/\nu} F_{ne} ( L^{1/\nu}(K-K_{c}), \beta_T/L^{z} )  ~~~~~~~
\label{densityscaling}
\end{eqnarray}
       where  $ \beta_T=1/k_B T $ is the inverse temperature, $ L $ is size of the flat trap, $ K $ is the tuning
       parameter such as $ t/V_{1} $ or $ \mu $ in Fig.\ref{squarediagram}.
       In the last equation in \ref{densityscaling}, we used the relation between exponents $ 2 \beta/\nu= d+z-2 + \eta $.

       Similar quantities can be defined for bond-bond correlation
       functions $ S_{K, \alpha}( \vec{Q}_{K}, \tau=0 ), S_{K, \alpha}(\vec{Q}_{K}, i \omega_{n} =0
       ) $ where $ K_{ij} = b^{\dagger}_{i} b_{j} + h.c. $ and $ \alpha=\hat{x},\hat{y} $
       is the orientation of the bond $ \langle ij \rangle $.
       In principle, by doing this  finite size scaling with respect to the trap size $ L $, finite
       temperature $ T $, one can extract all the 3 exponents $ z, \nu, \eta $.
       Eqn. \ref{densityscaling} will be extended to optical lattices inside a harmonic trap in Sec.IX.


\subsection{ The prospects of realizing the quantum phases in optical lattices  }

 The  Mott and superfluid phases are already realized in the experiment \cite{bloch,insitu0}.
 There are extensive numerical evidences that
 the dipole-dipole long-range interaction is especially favorable to the formation of the CDW supersolid.
 It was argued in \cite{exss} that the dipole-dipole interaction between indirect excitons in electron-hole semi-conductor billayer
 may favor a formation of vacancy-like exciton supersolid in some intermediate distances between the bilayers.
 The QMC simulations in \cite{square} found that for hard-core bosons in a square lattice with the $ V_1 $ interaction,
 the X-CDW is not stable against a phase separation slightly away from $ 1/2 $ filling. However, the QMC simulations in \cite{dipolarss,int} found that
 with the dipole-dipole interaction, the X-CDW is stable in a large parameter regime
 slightly away from $ 1/2 $ fillings. Very similar results were found in a triangular lattice ( see Fig.\ref{triphase}a )
 \cite{dipolarsstri} and dipolar bilayer systems \cite{eplss}.   As said in the
 introduction, the $ ^{52}Cr $ atoms carry exceptionally large magnetic
 dipole moment and therefore interact with each other with
 the anisotropic long-range interaction. The dipolar bosons carry large
 electric dipole moments and provide another very important system
 with the long range dipole-dipole interaction. All kinds of CDW and CDW supersolids could be very likely realized
 in near future experiments with either  $ ^{52}Cr $ atoms \cite{cromium} or dipolar bosons \cite{junpolar} loaded in square and triangular lattices.
 It remains experimentally challenging to realize the VBS
 and VB supersolid phase. However, there is a theoretical proposal \cite{qs} that the ring exchange interaction
 can be generated in cold atomic gases subjected to an optical lattice using well-understood tools for manipulating and controlling such
 gases. If so, all the valence bond phases can be realized in the presence of the ring exchange
 interaction. The VB phase is one of the most important phases in condensed matter system which may also hold hints
 to quantum magnetisms and  high temperature superconductors, it would be necessary to quantum simulate this phase by cold atoms anyway.

\section{ Two Photon Raman scattering formalisms.}

In this section, we focus on the light scattering cross section in
Fig.1a. As shown in Sec. X, it can be straight-forwardly applied to
the atom Bragg spectroscopy experiment in Fig.1b.  As shown in Sect.
X, the cavity enhanced off-resonant scattering formalism is similar
after we take care of the physics cavity specific to a cavity QED.




\begin{figure}
\includegraphics[width=5cm]{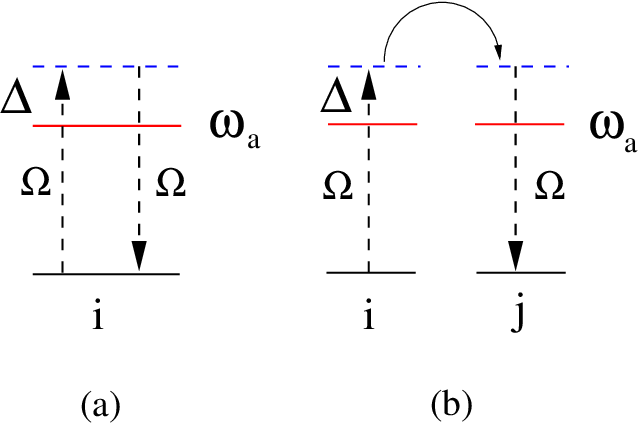}
\caption{
 The off-resonant Raman scattering processes lead to the on-site term  and the off-site term  in Eqn.\ref{laserint}.} \label{raman}
\end{figure}

   The interaction between the two laser beams in Fig.1 with the two level bosonic atoms is:
\begin{eqnarray}
  H_{int} & = & \int d^{2} \vec{r} \Psi^{\dagger}(\vec{r}) [
  \frac{\vec{p}^{2}}{ 2 m_{a}} + V_{OL}( \vec{r} ) + \frac{ \hbar \omega_{a} }{2} \sigma_{z}    \nonumber  \\
   & + & \frac{\Omega}{2} \sum_{l} ( e^{-i \omega_{l} t } \sigma^{+} u_{l}(\vec{r}) + h.c. ) ] \Psi(\vec{r})
\label{bragg}
\end{eqnarray}
   where $ \Psi(\vec{r})=( \psi_{e}, \psi_{g} ) $ is the two component boson annihilation operator,
   the incident and scattered lights in Fig.1a and the two incident lights in Fig.1b have frequencies $ \omega_{l} $ and mode functions
   $ u_{l}( \vec{r} ) = e^{i \vec{k}_{l} \cdot \vec{r} + i \phi_{l} } $ .
   The Rabi frequencies $ \Omega $ are much
   weaker than the laser beams ( not shown in Fig.1 ) which form the optical lattices.
   In the following, we develop the formalism by using  the light scattering
   geometry in Fig.1a which also applies to the atom scattering
   geometry in Fig.1b after some slight modifications.
   When it is far off the resonance, the laser light-atom detunings $ \Delta_{l}= \omega_{l}
  -\omega_{a} $ where $ \omega_a $ is the two level energy  difference are much larger than the Rabi frequency $ \Omega $ and
  the  energy transfer $ \omega= \omega_1-\omega_2 $ ( See Fig.\ref{raman} ), so $ \Delta_{1} \sim \Delta_{2} = \Delta $.
  After adiabatically eliminating the upper level $ e $ of the two level atoms,
  expanding the ground state atom field operator $ \psi_{g}(\vec{r}) = \sum _{i} b_{i} w( \vec{r} -\vec{r}_{i} )  $ in Eqn.\ref{bragg} where
  $ w( \vec{r} -\vec{r}_{i} ) $ is the localized Wannier functions of the lowest Bloch band corresponding to $ V_{OL}( \vec{r} ) $ and $  b_i $
  is the annihilation operator of an atom at the site $ i $ in the Eqn.\ref{boson}, then we get the effective interaction between the off-resonant
  laser beams and the ground level $ g $:
\begin{equation}
  H_{int} =  \hbar \frac{ \Omega^{2} }{ \Delta } e^{-i \omega t } [ \sum^{N}_{i}  J_{i,i} n_{i}
                   + \sum^{N}_{<ij>}  J_{i,j}   b^{\dagger}_{i}b_{j} ]
\label{laserint}
\end{equation}
  where the interacting matrix element is $  J_{i,j}= \int d \vec{r} w( \vec{r}-\vec{r}_{i} )
  u^{*}_{1}(\vec{r} ) u_{2}(\vec{r} ) w( \vec{r}-\vec{r}_{j} ) = J_{j,i} $.
   The first term in Eqn.\ref{laserint} is the on-site term $ \hat{D}=\sum^{N}_{i}  J_{i,i} n_{i} $ ( See Fig.\ref{raman}a ).
   The second term is the off-site term ( See Fig.\ref{raman}b ).
   Because the Wannier wavefunction $ w( \vec{r} ) $ can be taken as real in the
   lowest Bloch band \cite{justify}, the off-site term can be written as $
   \hat{K} =  \sum^{N}_{<ij>}  J_{i,j} b^{\dagger}_{i}b_{j}
    = \sum^{N}_{<ij>}  J_{i,j} ( b^{\dagger}_{i}b_{j} + h.c.) $ which
   is nothing but the off-site coupling to the nearest neighbor kinetic energy of the bosons
   $ K_{ij}=   b^{\dagger}_{i}b_{j} + h.c. $.


    It is easy to show that:
\begin{equation}
  \hat{D} ( \vec{q} ) = f_{0}( \vec{q} ) \sum^{N}_{i=1} e^{-i \vec{q} \cdot \vec{r}_{i} } n_{i}= N f_{0}( \vec{q} ) n( \vec{q}  )
\label{d}
\end{equation}
  where $ \vec{q}= \vec{k}_{1} -\vec{k}_{0} $, $ f_{0}( \vec{q} ) = \int d \vec{r} e^{-i \vec{q} \cdot \vec{r} } w^{2}( \vec{r} )$  and
  $ n( \vec{q} ) = \frac{1}{N} \sum^{N}_{i=1} e^{-i \vec{q} \cdot \vec{r}_{i} } n_{i} =
  \sum_{\vec{k}} b^{\dagger}_{\vec{k}} b_{\vec{k} + \vec{q} } $ is the Fourier transform of the  density operator at
   the  momentum $ \vec{q} $.  The wavevector is confined to $ L^{-1} < q < a^{-1} $ where the trap size
  $ L \sim 100 \mu m $ and the lattice constant $ a \sim 0.5 \mu m $ in Fig.1.
   In fact, more information is
   encoded in the off-site kinetic coupling in Eqn.\ref{laserint}.
    In a square lattice, the bonds are either oriented along the
    $ \hat{x} $ axis  $ \vec{r}_{j} - \vec{r}_{i}= \hat{x} $ or
    along the $ \hat{y} $ axis  $ \vec{r}_{j} - \vec{r}_{i}= \hat{y} $, we have:
\begin{equation}
  \hat{K}_{\Box}  =  N [  f_{x}( \vec{q} )   K_{x}( \vec{q} ) +
                f_{y}( \vec{q} ) K_{y}( \vec{q}  )]
\label{kxy}
\end{equation}
   where
   $  K_{\alpha}( \vec{q} )= \frac{1}{N} \sum^{N}_{i=1} e^{-i\vec{q} \cdot \vec{r}_{i} } K_{i,i+\alpha} =
   e^{ i q_{\alpha}/2} \sum_{\vec{k}} \cos k_{\alpha} b^{\dagger}_{\vec{k}} b_{\vec{k} + \vec{q} }   $
   are the Fourier transform  of the kinetic energy operator $ K_{ij} = b^{\dagger}_{i} b_{j} + h.c.
   $ along $ \alpha=x, y $ bonds at the  momentum $ \vec{q}  $
   and the "form" factors
   $ f_{\alpha}( \vec{q} )=f( \vec{q},  \vec{r}_{i} - \vec{r}_{j}=\alpha )=
   \int d \vec{r} e^{-i \vec{q} \cdot \vec{r} } w( \vec{r} ) w( \vec{r} + \vec{r}_{i} -\vec{r}_{j}
   ) $. Following the harmonic approximation used in \cite{boson},
   we can estimate that:
\begin{eqnarray}
    f_{0}( \pi,0  )  & \sim  & e^{ -\frac{1}{4} ( V_0/E_r)^{-1/2} }
    \nonumber   \\
   f_{x}( \pi,0  ) & \sim  & i e^{ -\frac{1}{4} ( V_0/E_r )^{-1/2} -\frac{\pi^{2}}{4} ( V_{0}/E_{r} )^{1/2} }
\end{eqnarray}
   so  $  | f_{x} ( \pi,0  )/f_{0}( \pi,0  ) | \sim e^{-\frac{\pi^{2}}{4} \sqrt{ V_{0}/E_{r}} }  $
   where $ V_{0} $ and $ E_{r} = \hbar^{2} k^{2}/2m $ are the strength of the optical lattice potential and
   the recoil energy respectively \cite{boson}. The $ f_{0}( \pi,0  ) $ is close to 1 when $ V_{0}/E_{r} > 4 $.
   It is instructive to relate this ratio to
   that of the hopping $ t $ over the onsite interaction $ U $ in the Eqn.\ref{boson}: $ | f_{x} ( \pi,0 )/f_{0}( \pi,0 ) |
   \sim \frac{t}{U} \frac{a_s}{a} $ where $ a_s $ is the zero field scattering length and $ a= \lambda/2= \pi/k $ is the lattice constant,
   using the typical values $ t/U \sim 10^{-1}, a_s/a \sim 10^{-2} $, one can estimate $ | f_{\alpha}/f_{0} | \sim 10^{-3} $.
   Note that the harmonic approximation works well only in a very deep optical lattice $ V_0 \gg E_r $, so the above value {\sl underestimates }
   the ratio, so we expect $ | f_{\alpha}/f_{0} | \ge 10^{-3} $.

   The differential scattering cross section of the light from the cold atom systems in the Fig.1 can be
   calculated by using the standard linear response theory:
\begin{eqnarray}
  \frac{ d \sigma }{ d \Omega d \omega  }  & = & ( \frac{ \Omega^{2} }{ \Delta } )^{2}
  N^{2} [ |f_{0}( \vec{q} ) |^{2} S_{n}( \vec{q}, \omega )
  \nonumber  \\
  & + &  \sum_{\alpha=\hat{x},\hat{y} } |f_{\alpha}( \vec{q} ) |^{2} S_{K_{\alpha} }( \vec{q}, \omega ) ]
\label{cross}
\end{eqnarray}
     where  $ \vec{q}=\vec{k}_{1}-\vec{k}_{0}, \omega= \omega_{1}-\omega_{2} $,
     the $ S_{n}( \vec{q}, \omega ) = \langle n(-\vec{q}, - \omega ) n(\vec{q}, \omega ) \rangle $
     is the dynamic density-density response function listed in Eqn.\ref{nn} whose Lehmann representation was listed in \cite{braggbog}.
     The $ S_{K_{\alpha}}( \vec{q}, \omega ) = \langle K_{\alpha}(-\vec{q}, - \omega ) K_{\alpha}(\vec{q}, \omega ) \rangle $
     is the  bond-bond response function whose Lehmann representation can be got from that of the $ S_{n}( \vec{q}, \omega ) $
     simply by replacing the density operator $ n( \vec{q} ) $ by the bond operator $  K_{\alpha}( \vec{q} ) $.

      The elastic scattering cross section  $  \frac{ d \sigma }{ d \Omega }|_{el} $
      is proportional to:
\begin{eqnarray}
 \frac{ d \sigma }{ d \Omega }|_{el} & =  & ( \frac{ \Omega^{2} }{ \Delta } )^{2}
  N^{2} [ |f_{0}( \vec{q} ) |^{2} S_{n}( \vec{q}, \omega=0 )
  \nonumber  \\
  & + &  \sum_{\alpha=\hat{x},\hat{y} } |f_{\alpha}( \vec{q} ) |^{2} S_{K_{\alpha} }( \vec{q}, \omega=0 ) ]
\label{cross0}
\end{eqnarray}

     The integrated differential scattering cross section over the
     final energy  $ \frac{ d \sigma }{ d \Omega } = \int d
     \omega  \frac{ d \sigma }{ d \Omega d \omega } $ is
     proportional to the {\sl equal-time} response function is
\begin{eqnarray}
      \frac{ d \sigma }{ d \Omega }  & = &   ( \frac{ \Omega^{2} }{ \Delta } )^{2} N^{2} [  |f_{0}( \vec{q} ) |^{2} S_{n}( \vec{q}
    )     \nonumber  \\
      & +  & \sum_{\alpha=\hat{x},\hat{y} } |f_{\alpha}( \vec{q} ) |^{2} S_{K_{\alpha}}( \vec{q})
    ]
\label{crossequal}
\end{eqnarray}

    In the following, we will discuss the physical implications of  Eqn.\ref{cross}
    on the CDW, VBS and corresponding supersolids summarized in the Sec.II.

\section{ Superfluid, Mott insulator and Superfluid to Mott transition at integer fillings }

   Quantum phase transitions are characterized by three critical
   exponents $ z, \nu, \eta $ called " two scale factor"
   universality. It is well known that the SF to
   Mott transition is described by the {\sl relativistic} effective action \cite{bipart}:
\begin{equation}
   {\cal S}= \int d^{2}r d \tau [ | \partial_{\tau} \phi |^{2} +
    |\nabla \phi |^{2} + r |\phi |^{2} + u |\phi |^{4} +\cdots  ]
\label{rel}
\end{equation}
   where in the Mott state $ r > 0, \langle \phi \rangle = 0 $,
   while in the SF phase $ r < 0, \langle \phi \rangle \neq 0 $.
   The SF to Mott transition described by Eqn.\ref{rel} is in the $ 3d $ XY universality class with the  critical exponents $ z=1, \nu=0.67, \eta= 0.04  $
   ( Note that for $ d=1 $, it is in $ 2d $ XY universality class, namely, KT transition ).
   In the following, we will discuss the elastics and the in-elastic scattering at the
   SF and Mott, then the quantum phase transition between the two
   phases respectively. As shown in the Sect.III, the scattering
   cross sections are determined by the density-density correlation
   functions, so we will focus on their computations.  Although the order parameter correlation functions were well
   studied theoretically from both the direct and dual pictures, they can not
   be directly measured by experiments yet. The density-density correlation functions have not been discussed theoretically so far.

\subsection{ Elastics scattering at the reciprocal lattice vector $
\vec{K} $ to detect SF and Mott states }

    We first look at the superfluid to Mott transition at integer filling factor $ n $.
    When $ \vec{q} $ is equal to the shortest reciprocal lattice vector $ \vec{K} = ( 2 \pi, 0 ) $,
    in the Mott state, $  \frac{ d \sigma^{M} }{ d
    \Omega } =  | f^{M}_{0}( 2 \pi, 0) |^{2} N^{2} n^{2} $,  in the superfluid state, $  \frac{ d \sigma^{SF} }{ d
    \Omega } =  | f^{SF}_{0}( 2 \pi, 0) |^{2} N^{2} n^{2}  + 2 | f^{SF}_{x}( 2 \pi, 0) |^{2} N^{2} B^{2} $
    where $ B $ is the average kinetic energy on a bond in the
    superfluid side. Because $ | f^{SF}_{0}( 2 \pi, 0) |^{2} \sim | f^{M}_{0}( 2 \pi, 0) |^{2} \sim 1 $ and
    $ B $ is appreciable only in the superfluid side,
    we expect an increase of the scattering cross section:
\begin{equation}
      \frac{ d \sigma^{SF} }{ d
    \Omega } -  \frac{ d \sigma^{M} }{ d \Omega } \sim  2 | f^{SF}_{x}( 2 \pi, 0) |^{2} N^{2}
    B^{2}
\label{mottsfinc}
\end{equation}
    across the  Mott to the SF transition due to the prefactor $ N^{2}$.
    The increase is most evident when moving from well inside the Mott phase to well inside the SF phase. This
    increase may be used as an effective measure of the boson  kinetic energy inside the SF.
    This prediction could be tested immediately.
    Surprisingly, there is no such optical Bragg scattering experiment in the superfluid yet.

\begin{figure}
\includegraphics[width=6cm]{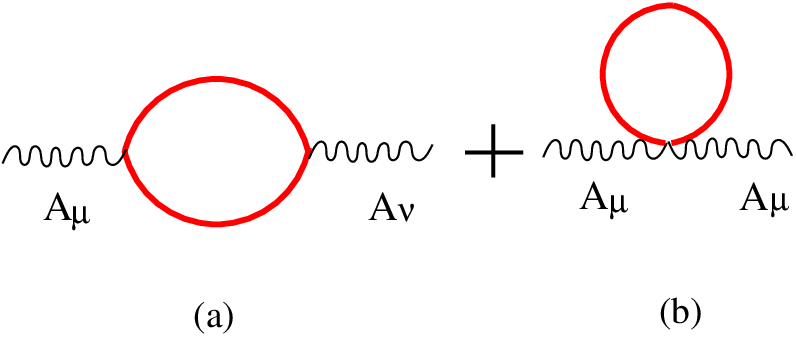}
\caption{  The current-current correlator. Its $ \mu=\nu=0 $
component leads to the boson density-density correlation function
Eqn.\ref{mott}. Although the $ D=3+1 $ case was well documented
\cite{ryder}. The  $ D=2+1 $ case seems not be calculated yet and
was done in the Sec.IV-b. } \label{cc}
\end{figure}

\subsection{ In-elastics scattering slightly away from a reciprocal
lattice vector $ \vec{K} $  to detect the excitation spectrum in SF
and Mott states }

   Now we study the in-elastic scattering away from zero ( or any
   other reciprocal lattice vector $ \vec{K} $. ). The excitation spectrum inside
   a SF was already given in Eqn.\ref{sf} and Eqn.\ref{sfequal}.
   Here, from the effective theory of both boson and dual vortex
   pictures, we will  calculate the density-density
   correlation function not only in the SF side, but also
   in the Mott side and the quantum critical point.

{\sl (1) Density-density correlation functions from a direct boson
picture }

   Observing that the density $ n= \phi^{\dagger} \partial_{\tau} \phi -
   ( \partial_{\tau} \phi^{\dagger} ) \phi $ is a conserved
   quantity. In order to calculate the density-density correlation
   function, we add a source term to Eqn.\ref{rel}
\begin{equation}
   {\cal S} [ A_{\mu}, \phi ]= \int d^{2}r d \tau [ | ( \partial_{\mu}-i A_{\mu} )\phi |^{2}  + m^{2} |\phi |^{2} + u |\phi |^{4} +\cdots  ]
\label{rela}
\end{equation}
    In the Mott side, $ m^{2} >0 $,
    Then integrating out the massive $ \phi $ field according to the
    Feymann diagram Fig.\ref{cc} paying special attentions to the
    diamagnetic term in Fig.\ref{cc}b, one gets
\begin{eqnarray}
   {\cal S} [ A_{\mu} ] & =  &\frac{1}{2} A_{\mu} ( -k ) \Pi_{\mu
   \nu}(k) A_{\nu} ( k ) +\cdots    \nonumber  \\
    \Pi_{\mu \nu}(k) &  =  & \frac{ m }{ 2 \pi } [ \frac{ k_{\mu}  k_{\nu}
    }{k^2} - \delta_{\mu \nu } ] ( I (z)-1),~~z=\frac{k^{2}}{m^{2}}
\end{eqnarray}
   where $ I(z)  =  \int^{1}_{0}  d x [ 1 + x(1-x) z ]^{1/2} = \frac{1}{2}+
    \frac{ (4 + z ) }{ 4 \sqrt{z} }  \arctan \frac{ \sqrt{z} }{2 } =
    1 + \frac{z}{12} - \frac{z^2}{240}+ \cdots $ as $ z  \rightarrow 0 $.

    Using  $  A_{\mu}= ( A_0, A_{\alpha} ) $ and $ k =( \omega, \vec{q}
    ) $ and  putting $ \mu=\nu=0 $, we can get the boson density-density correlation function:
\begin{equation}
   S^{Mott}_n(\vec{q}, \omega )= \Pi_{00}(\vec{q}, \omega )= \frac{ 5  m }{ 6 \pi }  \frac{ q^{2} }{
   \omega^{2} + q^{2} + 20 m^{2} }
\label{mott}
\end{equation}
   where one can identify the Mott gap $ \Delta^{2}_{Mott}= 20 m^{2}
   $.  The compressibility
\begin{equation}
    \kappa_{Mott} = S(\vec{q} \rightarrow 0 , \omega =0
    ) \sim q^{2} \rightarrow 0
\label{commott}
\end{equation}
     which, namely, in-compressible inside the Mott state as  expected. Again,

    One can evaluate immediately the structure factor
\begin{equation}
   S^{Mott}_n(\vec{q})= \frac{ 5 m }{ 12 \pi }  \frac{ q^{2} }{
   \sqrt{ q^{2} + 20 m^{2} }}
\end{equation}

    At the quantum critical point between the SF and the Mott  phase, $ m \rightarrow 0 $, $ z \rightarrow \infty $,
    then $ I(z) \rightarrow \sqrt{z} \int^{1}_{0} \sqrt{x(1-x)} = \frac{\pi}{8} \sqrt{z} $,  we get
\begin{equation}
    S^{QC}_n(\vec{q}, \omega ) = \frac{1}{16} \frac{  q^{2} }{ \sqrt{\omega^{2} + q^{2}} }
\label{qcdirect}
\end{equation}
    which shows that
\begin{equation}
     \kappa_{QC} = S(\vec{q} \rightarrow 0 , \omega =0 ) \sim q \rightarrow 0
\label{comqc}
\end{equation}
     which, namely, still in-compressible at the QC point.

     One can evaluate immediately the dynamic structure factor
\begin{equation}
   S^{QC}_n(\vec{q}) \sim q^{2} \log \Lambda/q
\label{qc}
\end{equation}
     where the $ \Lambda \sim 1/a $ is the ultra-violet frequency cutoff.


     Inside the superfluid $ m^{2} < 0 $, then it is convenient to
     write $ V( \phi )= \frac{1}{4} \lambda ( |\phi|^{2}-a^{2} )^{2}
     $. We write the order parameter in the polar coordinate $
     \phi= \sqrt{ a^{2} + \delta \rho } e^{i \phi } $, then Eqn.\ref{rel}
     becomes:
\begin{eqnarray}
   {\cal L} & = & (a^{2} + \delta \rho) ( \partial_\mu \theta )^{2}
   + \cdots   \nonumber \\
   & + &  \frac{1}{4} ( a^{2} + \delta \rho)^{-1} ( \partial_\mu  \delta \rho )^{2}
   + \frac{1}{4} \lambda ( \delta \rho )^{2}
\label{relrho}
\end{eqnarray}
    where one can see that there is a gapless ( Goldstone ) $ \theta $ mode and the massive Higgs magnitude $ \delta \rho $ fluctuation
    mode. It is important to stress that the density operator is different than the
    Higgs magnitude fluctuation operator. The former is a conserved quantity,
    while the later is not, although both are $ U(1) $ invariant. So they have different correlation functions.   
    Unfortunately, it is not straightforward to extract the
    density-density correlation function Eqn.\ref{sf} inside a SF from Eqn.\ref{relrho}.
    However, as shown below, it can be easily derived from the dual vortex picture.

   In a brief summary, Eqn.\ref{sf}, \ref{mott}, \ref{qcdirect} describes the density-density correlation functions
   in the SF, Mott and the QC regimes respectively. In the Mott state, when $ q \ll m $, one can see $  S^{Mott}(\vec{q})
   \sim q^{2} $ which is in sharp contrast to that inside a
   superfluid $  S^{SF}(\vec{q}) \sim q $. While at the QC,  $  S^{QC}(\vec{q}) \sim q^{2} \log \Lambda/q $.
   Note that all the compressibilities at Mott, SF and QC can also be directly measured by the {\sl in situ} method
   at different positions inside a trap \cite{insitu0,insitu1}.

   The scattering cross section at the classical diffraction
    minimum $ \vec{q}= ( \pi, \pi ) $ will be computed in the
    Appendix B.

{\sl (2) The density-density correlation functions from a dual
vortex picture }

   Alternatively, one can calculate the density-density correlation
   from the dual vortex action \cite{bipart,hightc}. It is well known that the boson
   action Eqn.\ref{rel} is dual to the vortex action:
\begin{eqnarray}
   {\cal S}_{d} [ a_{\mu}, \psi ] & = & \int d^{3} x [ | (\partial_{\mu} - i  a_{\mu} ) \psi_v |^{2}  + r_d |\psi_v |^{2} + u_d |\psi_v |^{4} +\cdots
       \nonumber   \\
   & +  & \frac{1}{ 4 e^{2} } f^{2}_{\mu \nu} ]
\label{dual}
\end{eqnarray}
   where in the Mott state $ r_d < 0, \langle \psi_v \rangle \neq 0 $,
   while in the SF phase $ r_d > 0, \langle \psi_v \rangle = 0 $  with also the $ 3d $ XY  critical exponents $ z=1, \nu=0.67, \eta= 0.04  $.

   In the Mott state, the vortex condensation  $ \langle \psi_v \rangle \neq 0
   $ leads to  a mass term for the gauge field $ \frac{1}{2}
    |\langle \psi_v \rangle|^{2}( a^{t}_{\mu} )^{2} $ where  $ a^{t}_{\mu} $ is the transverse
    component of the gauge field. In the Landau gauge $ \partial_{\mu} a_{\mu} =0 $, the effective action for the gauge
    field is:
\begin{eqnarray}
   {\cal S}_{d} [ a_{\mu} ] & = & \frac{1}{2} a_{\mu} (-k) [  \frac{ k^{2} }{e^{2}} ( \delta_{\mu \nu}
   -\frac{ k_{\mu} k_{\nu} }{ k^2 } ) + \frac{1}{\alpha} k_{\mu}
   k_{\nu}   \nonumber  \\
   & + & |\langle \psi \rangle|^{2}  ( \delta_{\mu \nu} -\frac{ k_{\mu} k_{\nu} }{ k^2 }  ) ] a_{\nu} (k)
\label{amu}
\end{eqnarray}
   where $ \alpha \rightarrow 0 $ indicates the Landau gauge.

   Using $ a_{\mu} = ( a_0, \vec{a} ) $ and  $ k =( \omega, \vec{q}  )
   $, we can find the density-density correlation inside the Mott phase:
\begin{equation}
   S^{Mott}_n(\vec{q}, \omega )= \langle (\nabla \times \vec{a})( \nabla \times \vec{a} ) \rangle =  \frac{ e^{2}  q^{2} }{ \omega^{2} + q^{2} +  m^{2}_{d} }
\label{mottdual}
\end{equation}
    where $ m^{2}_{d}= e^{2} |\langle \psi_v \rangle|^{2} $. We can
    identify the quasi-particle spectral weight $ {\cal A}=  e^2 $.
    It is identical to Eqn.\ref{mott} after we identify $
    m^{2}_{d}= 20 m^{2} $.

    Inside the superfluid state $ r_d > 0 $, the mass term for the gauge field is absent, integrating out the massive
    vortex fluctuation leads to the density-density correlation inside the superfluid phase:
\begin{equation}
    S^{SF}_n(\vec{q}, \omega )= \langle (\nabla \times \vec{a})( \nabla \times \vec{a} ) \rangle =  \frac{ e^{2}  q^{2} }{ \omega^{2} + q^{2} }
\label{sfdual}
\end{equation}
     which is identical to Eqn.\ref{sf} after putting back the
     corresponding the superfluid density $ \rho_s $ and the phonon
     velocity $ v^{2} $.

    At the quantum critical point between the SF and the Mott  phase, integrating out the
    massless vortex fluctuation leads to the density-density correlation at the QC:
\begin{equation}
    S^{QC}_n(\vec{q}, \omega ) \sim \langle (\nabla \times \vec{a})( \nabla \times \vec{a} ) \rangle =  \frac{  q^{2} }{ \sqrt{\omega^{2} + q^{2}} }
\label{qcdual}
\end{equation}
     which is identical to Eqn.\ref{qcdirect} from the direct boson
     picture. Eqn.\ref{qc} follows.

\subsection{ The scaling functions for the density-density correlation functions across the SF to Mott transition }

\begin{figure}
\includegraphics[width=6cm]{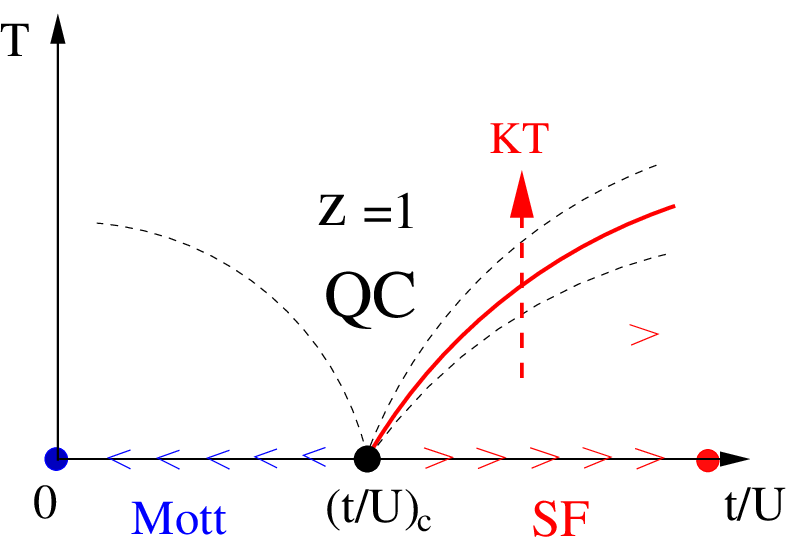}
\caption{ Finite temperature phase diagram of the SF to Mott
transition.  There are 3 fixed points: the Mott fixed point at $
t/U=0 $, the SF fixed point at $ t/U =\infty $ and the quantum
critical point at  $ ( t/U )_c $. The QC is the quantum Critical
regime. The KT stands for the Kosterlize-Thouless transition. See
Fig.\ref{motttrap} for the zero temperature phase diagram and the phases inside a trap. } \label{sfmott}
\end{figure}

   So far, we discussed the properties of different quantum phases where mean field theory works well. It is important to study
   quantum fluctuations near the quantum critical point
   between different quantum phases.  Ref.\cite{boson0} focused on the scaling form of the single particle
  Green function $ \langle \phi^{\dagger}(\vec{x}, \tau) \phi(0,0) \rangle $. Here, motivated by the light scattering
  experiments, we are studying the scaling form of the density-density correlation
  function $ \langle \delta n(\vec{x}, \tau) \delta n(0, 0) \rangle $ where $ \delta n =  \phi^{\dagger} \partial_{\tau} \phi -
   ( \partial_{\tau} \phi^{\dagger} ) \phi$.
   Following the scaling theory developed in \cite{scaling} and observing $ \delta n $ is a conserved quantity, we can
   write down the scaling functions near the SF to Mott transition in Fig.\ref{sfmott}:
\begin{equation}
   S^{SF}_n ( \vec{q}, \omega, k_{B} T ) = | C |^{2}
    \frac{k_{B} T}{v^{2} } \Phi_{SF}( \frac{\hbar v q}{k_{B} T }, \frac{\hbar \omega }{k_{B} T
    }, \frac{ k_{B} T}{ 2 \pi \rho_{s} } )
\label{relscaling}
\end{equation}
    In the Mott state $ 2 \pi \rho_{s} $ should be replaced by the Mott gap $ \Delta
    $. At the QC,  $ \Phi_{SF}( \frac{\hbar v q}{k_{B} T }, \frac{\hbar \omega }{k_{B} T
    }, \infty ) = \Phi_{Mott}( \frac{\hbar v q}{k_{B} T }, \frac{\hbar \omega }{k_{B} T
    }, \infty ) $.  Note that due to $ z=1 $, the spin wave velocity $ v $ remains not critical across the
    SF to Mott transition in Fig.\ref{sfmott}, while $ \rho_{s} \sim ( t-t_{c} )^{(d+z-2) \nu } \sim ( t-t_{c} )^{\nu} $
    with $ z=1, d=2 $ and $ \Delta \sim ( t_{c}-t )^{\nu}, \nu \sim 0.67 $.

    Note that Eqn.\ref{sf} in the SF and Eqn.\ref{mott} in the Mott side only work deep inside the two phases
    controlled by the SF fixed point and the Mott fixed point in Fig.\ref{sfmott} respectively,
    but will break down near to the QCP.   The equal time correlation function follow from $ \int \frac{d
    \omega }{ 2 \pi} $ over Eqn.\ref{relscaling}.

    At $ T=0 $, Eqn.\ref{relscaling} simplifies to
\begin{equation}
   S^{SF}_n ( \vec{q}, \omega, T=0 ) = | C |^{2}
    \frac{ 2 \pi \rho_s }{v^{2} } \Phi_{SF}( \frac{\hbar v q}{2 \pi \rho_{s}}, \frac{\hbar \omega
    }{ 2 \pi \rho_{s} } )
\label{relscaling0}
\end{equation}
    In the Mott state $ 2 \pi \rho_{s} $ should be replaced by the Mott gap $
    \Delta $. From Eqn.\ref{sf}, we can determine that $  \Phi_{SF}( x,y)
    = \frac{x^{2}}{x^{2}+y^{2}} $.  From Eqn.\ref{mott}, we can determine that $  \Phi_{Mott}( x,y)
    = \frac{x^{2}}{x^{2}+y^{2} + 1 } $.

    At the QC $ \rho_s= \Delta =0 $,  Eqn.\ref{relscaling0} simplifies further to
\begin{equation}
   S^{QC}_n ( \vec{q}, \omega, T=0 ) = | C |^{2}
    \frac{ \hbar v q }{v^{2} } \Phi_{QC}( \frac{\hbar \omega }{ \hbar v q } )
\label{relscalingqc}
\end{equation}
     From Eqn.\ref{qcdirect}, we can determine that $ \Phi_{QC}(x)=
     \frac{1}{\sqrt{x^{2} + 1}} $.

\section{ CDW and CDW supersolid at and near half fillings }

\begin{figure}
\includegraphics[width=7cm]{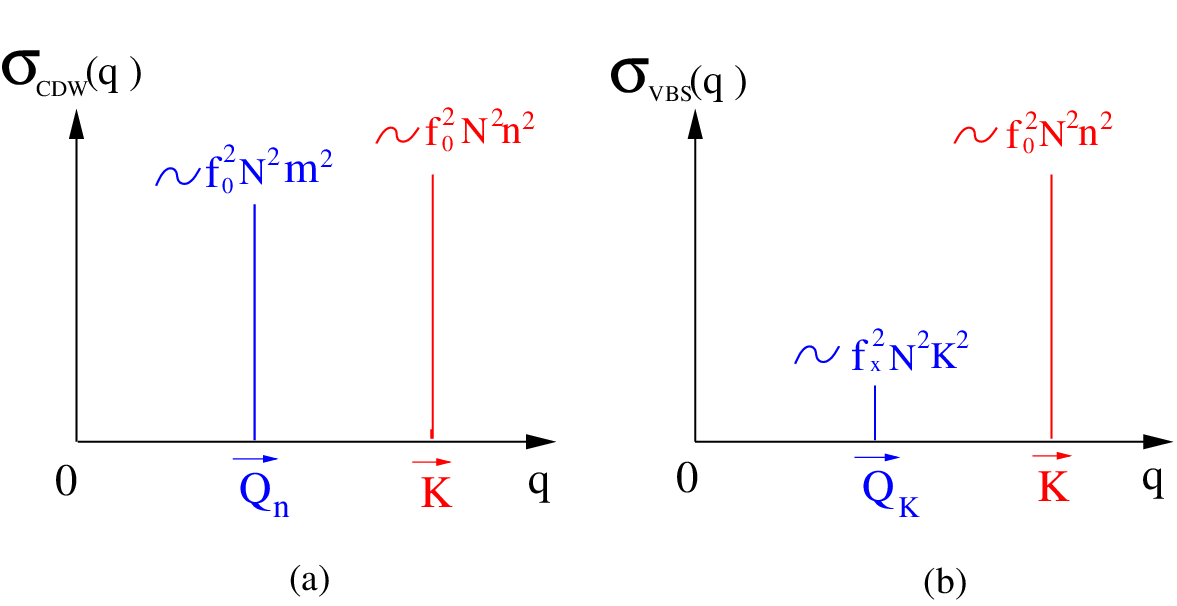}
\caption{ The characteristics of  optical scattering cross section
in a square lattice.(a) CDW, the ratio of the peak at $ \vec{Q}_{n}
$ over that at $ \vec{K} $ is $ \sim m^{2}/n^{2} \sim 1 $. (b) VBS
state, the ratio of the peak at $ \vec{Q}_{K} $ over that at $
\vec{K} $ is $ \sim |f_{x}/f_{0}|^{2} K^{2}/n^{2} \sim 10^{-5} $. It
should be visible in the current optical Bragg scattering
experiments. } \label{squareoptical}
\end{figure}

At the $ 1/2 $ filling along the horizontal axis in
Fig.\ref{squarediagram}a, inside the SF state near the CDW,
   there is a peak of $ S_{n}( \vec{q} ) $ near $ \vec{Q}_{n} = (\pi,\pi ) $, so the
   SF to the CDW transition is a first order one driven by the instability of the peak.
   Due to the lack of VBS order on both sides, the second term in Eqn.\ref{cross} can be neglected, so that
\begin{equation}
  \frac{ d \sigma }{ d \Omega dE }|_{CDW} \sim ( \frac{ \Omega^{2} }{ \Delta } )^{2} N^{2} |f_{0}( \vec{q} ) |^{2} S_{N}( \vec{q}, \omega )
\label{linearcdw}
\end{equation}

{\sl (a) Elastic scattering at the CDW and CDW-SS ordering vector $
\vec{Q}_{n} $. }

   When one gets into the  CDW state at $ \vec{Q}_{n}  $ in Fig.\ref{squarediagram}b in \cite{bipart}, then the
   $ S_{n}( \vec{q} ) $ should show a peak at $ \vec{q}= \vec{Q}_{n} $
   whose amplitude scales as the {\sl square} of the number of atoms inside the trap
   $ S_{CDW}( \vec{Q}_{n} ) \sim  |f_{0}(\pi,\pi ) |^{2} N^{2} m^{2}  $ where $ m= n_A-n_B $ is the CDW order parameter \cite{bipart}.
   When $ \vec{q} =\vec{K} $, then  $ S_{CDW}( \vec{K} ) \sim  |f_{0}(2 \pi, 0 ) |^{2} N^{2} n^{2} $
   where $ f_{0}(2 \pi, 0 ) \sim f^{2}_{0}(\pi,\pi ) $ ( Fig.\ref{squareoptical}a ).
   So the ratio of the two peaks in Fig.\ref{squareoptical}a is $ S_{CDW}( \vec{Q}_{n} )/S_{CDW}( \vec{K} ) \sim  m^{2}/n^{2} $ if one
   neglects the small difference of the two form factors.
   Slightly away from $ 1/2 $ filling, the CDW in Fig.\ref{squarediagram}a may turn into the
   CDW supersolid ( CDW-SS ) phase through a second order phase transition described by Eqn.\ref{non}. Then we have $
   \langle n(\vec{q}) \rangle = m  \delta_{\vec{q},\vec{Q}_{n}} +  n \delta_{\vec{q},0}  $ where $ n= n_A+n_B = 1/2 + \delta n $.
   The superfluid density $ \rho_s \sim \delta n= n -1/2 $.
   The scattering cross section inside the CDW-SS at $ \vec{Q}_n $ : $ S_{CDW-SS}( \vec{Q}_n ) \sim |f_{0}( \pi, \pi ) |^{2} N^{2}
   m^{2} $ stays more or less the same as that inside the CDW, but at $ \vec{K}=(2 \pi, 0 ) $:
   $ S_{CDW-SS}( \vec{K} ) \sim  |f_{0}(2 \pi, 0 ) |^{2} N^{2}
   n^{2} + 2 |f_{x}(2 \pi, 0 ) |^{2} N^{2} (\delta n)^{2} B^{2} $ will increase where $ n = 1/2 + \delta n $. The $ B $ is
   the average bond strength due to very small superfluid component  $ \rho_s \sim \delta n= n -1/2 $
   flowing through the whole lattice. So we expect the right peak in
   Fig.\ref{squareoptical}a will increase due to the increase of the total density and the superfluid component inside the CDW-SS phase.
   Of course, the small  superfluid component at $ \vec{k}=0 $ can also be detected by
   the TOF with the peak strength near $ \vec{k}=0 $  proportional to $ \rho_s \sim \delta n $.


{\sl (b) In-elastics scattering to detect the excitation spectrum in
         CDW and CDW-SS }

\begin{figure}
\hspace{-0.5cm}
\includegraphics[width=6cm]{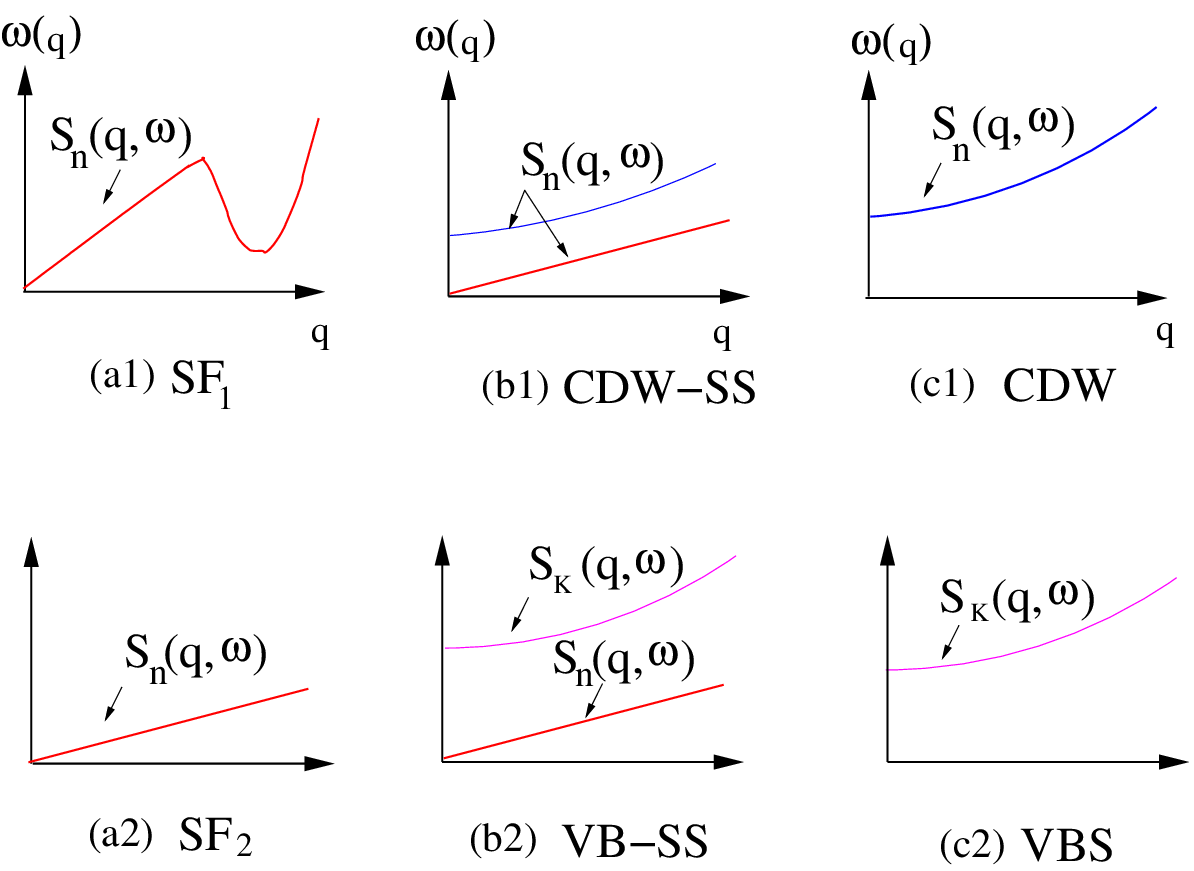}
\caption{The excitations spectrum in the CDW, VBS, SF, CDW-SS and
VB-SS states which are the peak positions of the corresponding
dynamic response functions shown with arrows. In the (b1) and (c1)
cases, the starting wavevector is $ \vec{Q}_{n} $ in the upper CDW
branch. In the (b2) and (c2) cases, the starting wavevector is $
\vec{Q}_{K} $ in the upper VBS branch.
The corresponding spectral weights are worked out in the text.}
\label{excitations}
\end{figure}

  So far, we only discussed the ground state properties of various quantum states.
  The elementary excitation spectrum above these ground states can be determined  from
  the peak positions of the corresponding dynamic density-density or bond-bond response functions.
  Eqn.\ref{cross} shows that the response function of the cold atom system is the sum of the two
  response functions with the corresponding spectral weight $ \sim |f_{0} |^{2} $ and $  \sim |f_{\alpha} |^{2} $.
  The  $ \langle \delta n  \delta n \rangle $ correlation function inside a SF was studied in several different physical systems in \cite{ssorder,sforder}
  and was listed in Eqn.\ref{sf} and \ref{sfequal}.
  In the SF {\sl near } to the CDW, there is a roton minimum near $ \vec{q}= \vec{Q}_{n} $,
  the $ S_{n}( \vec{q} ) $ shows a peak near $  \vec{Q}_{n} $ in Fig.\ref{excitations}a1. The superfluid mode in the Fig.\ref{excitations}a1
  has a spectral weight $ \sim N | f_{0} |^{2} \times \rho_{s} $ where $ \rho_{s} \sim  n  $.
  Inside the CDW-SS, the roton minimum disappears and is replaced by the upper branch
  with a CDW gap $ \Delta_{CDW} \sim  U $ and a spectral weight $ \sim  N | f_{0} |^{2} \times 1/2 $ in the Fig.\ref{excitations}b1,
  the lower superfluid branch in the Fig.\ref{excitations}b1
  has a spectral weight $ \sim N | f_{0} |^{2} \times \rho_{s} $ where $ \rho_{s} \sim  \delta n = n - 1/2 $ is the superfluid density inside the CDW-SS.
  Inside the CDW, the superfluid lower branch disappears, the upper CDW branch in Fig.\ref{excitations}c1
  has the spectral weight $ \sim  N | f_{0} |^{2} \times 1/2 $.

{\sl (c) Scaling function across the CDW to CDW-SS transition. }

  Slightly away from $ 1/2 $ filling, the transition from the CDW to the
  CDW-SS along the vertical axis in Fig.\ref{squarediagram}b  is described by the {\sl non-relativistic} effective action \cite{bipart}:
\begin{equation}
   {\cal S}_{non}  =  \int d^{2}r d \tau [ \phi^{\dagger} \partial_{\tau} \phi
   +  \frac{ \hbar^{2} }{ 2 m_a } |\nabla \phi
   |^{2}  -  \mu |\phi |^{2} + u |\phi |^{4} +\cdots  ]
\label{non}
\end{equation}
  with the critical exponents $ z=2, \nu=1/2, \eta=0 $  with logarithmic corrections \cite{bipart}.

  It is the chemical potential $ \mu $ tuning the zero density
  transition from the CDW at $ \mu < 0 $ with $ \langle \phi \rangle
  =0 $ to the CDW-SS with  $ \mu > 0 $ with $ \langle \phi \rangle \neq 0
  $, so the critical chemical potential $ \mu_c=0 $.  Well inside the CDW-SS phase, one can set $  \langle \phi \rangle = |  \langle \phi
  \rangle | e^{i \theta } $, then the linear derivative term in
  Eqn.\ref{non} becomes irrelevant, the density-density correlation
  function  of the SF component reduces to Eqn.\ref{sf} with $ \rho_s \sim \frac{ \hbar^{2} |  \langle \phi
  \rangle |^{2} }{ 2 m } $. It leads to the lower branch in Fig.\ref{excitations}b1.

  Ref.\cite{z2} focused on the scaling form of the single particle
  Green function $ \langle \phi^{\dagger}(\vec{x}, \tau) \phi(0,0) \rangle $. Here, motivated by the light scattering
  experiments, we are studying the scaling form of the density-density correlation
  function $ \langle \delta n(\vec{x}, \tau) \delta n(0, 0) \rangle $ where $ \delta n/a^2 = \phi^{\dagger} \phi
  $ is the density above the CDW background. Near the QCP,
  we have $  \langle n(\vec{q}) \rangle = m
\delta_{\vec{q},\vec{Q}_{n}} +  n \delta_{\vec{q},0}  $ where $ n=
n_A+n_B = 1/2 + \delta n $.
  Because $ \delta n/a^2 = \phi^{\dagger} \phi $ is a conserved quantity, so it has no anomalous dimension.
  From the scaling analysis in \cite{z2},  one can show that at $ T=0 $
\begin{equation}
   \delta n= \frac{  m_{a} \mu a^{2} }{ 4 \pi \hbar^{2}} \ln [ \frac{\hbar^{2}}{ 2 m_{a} \mu a^{2} }]
\label{deltanz2}
\end{equation}
  where $ a $ is the lattice constant  and $ m_a $ is the effective atom mass in Eqn.\ref{non} and \ref{bragg},
  the logarithmic factor is also explicitly written.

  The superfluid density inside the CDW-SS is $ \rho_{s} \sim ( \mu-\mu_{c} )^{(d+z-2) \nu } \sim  \mu-\mu_{c}  $
  with $ d=2, z=2, \mu=1/2, \eta=0 $ upto a logarithmic correction. From Eqn.\ref{deltanz2}, we expect that
\begin{equation}
   \rho_{s} \sim \frac{  m_{a} \mu a^{2} }{ 4 \pi \hbar^{2}} \ln [ \frac{\hbar^{2}}{ 2 m_{a} \mu a^{2} }]
\label{rhos}
\end{equation}
   where we set $ \mu_c=0 $.

  In the scaling limit $ q \ll 1/a, \hbar \omega \ll \frac{\hbar^{2}}{ 2 m_a a^{2} }
  $, the density-density correlation function should take the
  following scaling form:
\begin{equation}
   S_{n}( \vec{q}, \omega )=  \frac{ 2 m_{a} a^{4} }{ \hbar }  \Phi_{n}( \frac{ \hbar \omega}{ k_{B} T }, \frac{ \hbar q }{ \sqrt{ 2 m_{a} k_{B} T } },
  \frac{\mu}{ k_{B} T } )
\label{scalingz2}
\end{equation}
  where $ \Phi_{n} $ is a universal function independent of the atom-atom interactions in Eqn.\ref{boson}.
  Inside the CDW-SS phase in Fig.\ref{squarediagram}b, there should also
  be a $ S_{n}( \vec{q} ) $ peak $  \sim |f_{0}|^{2} N^{2} m^{2} $ at $ \vec{q}= \vec{Q}_{n} $  signaling its CDW order.

\begin{figure}
\includegraphics[width=3.5cm]{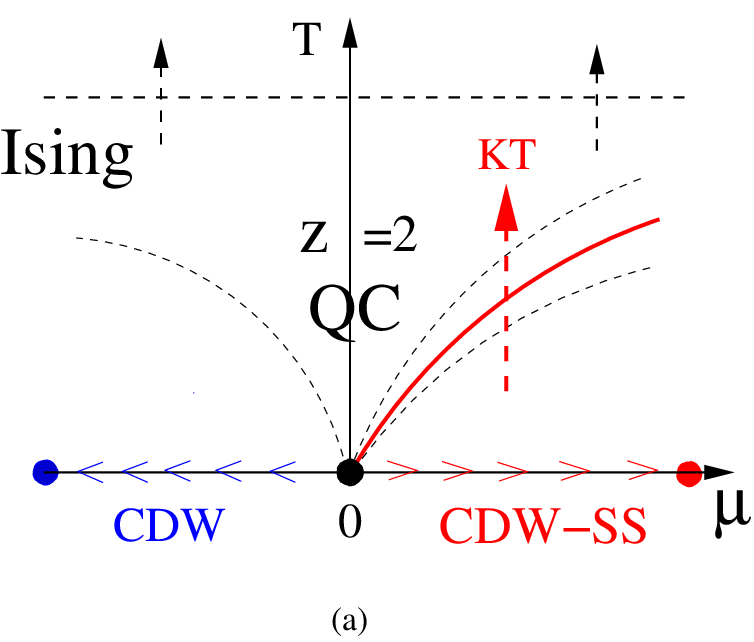}
\hspace{0.5cm}
\includegraphics[width=3.5cm]{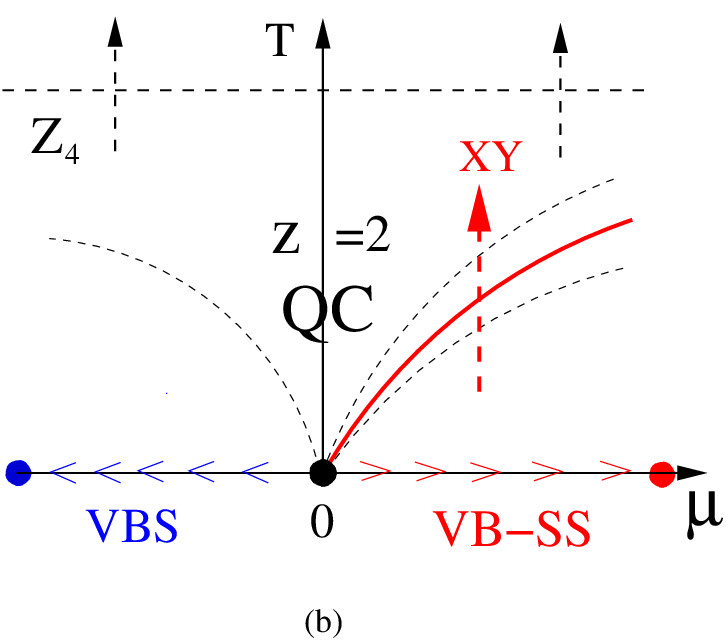}
\caption{ (a) The CDW to CDW-SS driven by the chemical potential $
\mu $. The dashed line is a finite temperature Ising melting
transition. (b) The VBS to VB-SS driven by the chemical potential $
\mu $. The dashed line is a finite temperature $ Z_4 $ Clock melting
transition. See also Fig.\ref{squarediagram} for the zero
temperature phase diagram and Fig.\ref{cdwtrap} for the phases inside a
trap. } \label{sfmott23}
\end{figure}

   In fact, Eqns.\ref{deltanz2},\ref{rhos}, \ref{scalingz2} should also apply to the the Mott to SF transition and the CDW to the CDW-SS inside an harmonic trap
   driven by a local chemical potential ( or equivalently the distance from the center to the boundary ) under the LDA and will be discussed
   in Sec. XI.

\section{ VBS and VB supersolid at and near half fillings}

   At the $ 1/2 $ filling along the horizontal axis in Fig.\ref{squarediagram}b, inside the SF state near the VBS,
  there is a peak of $ S_{K}( \vec{q} ) $ near $ \vec{Q}_{K} = (\pi, 0 ) $, so the SF to the VBS
  transition is a very weak first order one.

{\sl (a) Elastic scattering at the VBS and VB-SS ground state
 at its ordering wavevector $ \vec{Q}_K $. }

  When  $ \vec{q}=\vec{K} $, due to the {\sl uniform } distribution of the density in the VBS,
  the  second term in Eqn.\ref{cross} can be neglected, so there is a diffraction peak ( Fig.\ref{squareoptical}b ) whose amplitude scales as the
  {\sl square} of the number of atoms inside the trap
  $ S_{VBS}( \vec{K} )\sim |f_{0}(2\pi,0) |^{2} N^{2} n^{2} $ where $ f_{0}(2 \pi, 0 ) \sim  f^{4}_{0}(\pi, 0 ) $ and $ n=1/2 $ is the uniform
  density in the VBS state.

   However, when one tunes $ \vec{q} $ near
  $ \vec{Q}_{K} $, the first term in Eqn.\ref{cross} can be neglected, then
\begin{equation}
\frac{ d \sigma }{ d \Omega dE }|_{VBS}  \sim ( \frac{ \Omega^{2} }{
\Delta } )^{2} N^{2}
   \sum_{\alpha=\hat{x},\hat{y} } |f_{\alpha}( \vec{q} ) |^{2} S_{K_{\alpha} }( \vec{q}, \omega )
\label{linearvbs}
\end{equation}
  It should show a peak at $ \vec{q} = \vec{Q}_{K} $ signifying the VBS ordering at $ \vec{Q}_{K} $
  whose amplitude scales also as the {\sl square} of the number of atoms inside the trap
  $ S_{VBS}( \vec{Q}_K )\sim |f_{x}(\pi,0) |^{2} N^{2} K^{2} $ where $ K=K_{x}-K_{y} $ is the
  VBS order parameter \cite{bipart}.
  So the ratio of the VBS peak at $ \vec{q}= \vec{Q}_{K} $ over the uniform density
  peak at $ \vec{q}= \vec{K} $ is $ S_{VBS}( \vec{Q}_K )/S_{VBS}( \vec{K} )  \sim |f_{x}(\pi,0)/f_{0}(2\pi,0) |^{2} \geq 10^{-5} $.
  However, the smallness of $ |f_{x} |^{2} $ is compensated by the large number of atoms $ N \sim  10^{6} $,
  $  |f_{x} |^{2} N^{2} =  ( |f_{x} |^{2} N ) \times N  \sim N \sim 10^{6} $.
  Therefore, the Bragg scattering cross section from the VBS order is $ \geq 10^{-5} $ smaller than that
  at $ \vec{q} =\vec{K} $ at the same incident energy $ I_{in} $ ( Fig.\ref{squareoptical}b ), but still $ \sim 10^{6} $ above the background,
  so very much visible in the current optical Bragg scattering experiments.
  Slightly away from $ 1/2 $ filling, the VBS may turn into VB Supersolid (VB-SS) through a second order transition \cite{bipart}.
  We have $ \langle K_{x}(\vec{q}) \rangle =  B \delta_{\vec{q},0} + K \delta_{\vec{q},\vec{Q}_{K}}  $
  and $ \langle n(\vec{q}) \rangle = ( \delta n + 1/2 ) \delta_{\vec{q},0}  $.
  The superfluid density $ \rho_s \sim \delta n= n -1/2 $.
   The scattering cross section inside VB-SS: $ S_{VB-SS}( \vec{Q}_K ) \sim |f_{x}( \pi, 0 ) |^{2} N^{2}
   K^{2} $ stays more or less the same as that inside the VBS, but $ S_{VB-SS}( \vec{K} ) \sim  |f_{0}(2 \pi, 0 ) |^{2} N^{2}
   n^{2} +   |f_{x}(2 \pi, 0 ) |^{2} N^{2} (\delta n )^{2}  B^{2}_{x} + |f_{y}(2 \pi, 0 ) |^{2} N^{2} (\delta n )^{2} B^{2}_{y} $
   where $ n = 1/2 + \delta n $ and the $ B_x, B_y $
   are  the average bond strengths along $ x $ and $ y $ due to very small superfluid component  $ \rho_s \sim \delta n= n -1/2 $
   flowing through the whole lattice. So we expect the right peak in
   Fig.\ref{squareoptical}b will increase due to the increase of the total density and the superfluid component inside the VB-SS phase.
   Of course, the  superfluid component at $ \vec{k}=0 $ can also be detected by
  the TOF with the peak strength near $ \vec{k}=0 $  proportional to $ \rho_s \sim \delta n $.
  Very similarly, one can discuss the VBS order at $   \vec{q}= \vec{Q}_{K}= ( 0, \pi ) $.
  For the plaquette VBS order in Fig.\ref{squarephase} which has both $ (\pi, 0 $ and $ (0, \pi) $ order, then one should be able to
  see the $ S_{K}( \vec{q} ) $ peaks at both $ ( \pi, 0 ) $ and $ (0,\pi) $. So the dimer VBS and the plaquette VBS can also be distinguished by
  the optical Bragg scattering.



{\sl (b) In-elastics scattering  to detect the excitation spectrum
in VBS and VB-SS }

  In the SF near to the VB, there is no peak in $ S_{n}( \vec{q} ) $.
  The superfluid mode in the Fig.\ref{excitations}a2 has a spectral weight $ \sim N | f_{0} |^{2} \times \rho_{s} $ where $ \rho_{s} \sim  n  $.
  However,  $ S_{VBS}( \vec{q} ) $ shows a peak near $  \vec{Q}_{K} $ which is suppressed by a factor $ | f_{x} |^{2} $ as compared to Fig.\ref{excitations}a1.
  Inside the VB-SS, the SF order of the VB-SS  is
  the same as the SF inside the CDW-SS, so its $ \langle \delta n  \delta n \rangle $ correlation function is also given by Eqn.\ref{sf}.
  So there is a  upper branch with a VBS gap
  $ \Delta_{VBS} \sim t^{2}/U $ and a spectral weight $ \sim  N | f_{x} |^{2} \times 1/2 $ in the Fig.\ref{excitations}b2,
  also a lower superfluid branch in the Fig.\ref{excitations}b2 with the spectral
  weight $ \sim N | f_{0} |^{2} \times \rho_s $ where $ \rho_s \sim \delta n =n-1/2 $ is the superfluid density inside the VB-SS.
  Inside the VBS, the superfluid lower branch disappears,
  the upper VBS branch in Fig.\ref{excitations}c2 has the spectral weight $  \sim  N | f_{x} |^{2} \times 1/2 $.

{\sl (c) The transition from the VBS to the VB-SS }

  Slightly away from $ 1/2 $ filling, the transition from the VBS to the VB-SS along the vertical axis in Fig.\ref{squarediagram}b
  near $ \vec{q}=(0,0) $ is also described by Eqn.\ref{non}, so it is also in the same universality class of SF to
  Mott transition with the critical exponents $ z=2, \nu=1/2, \eta=0 $
  upto a logarithmic correction \cite{bipart}. Eqn.\ref{deltanz2}  and
  Eqn.\ref{scalingz2} also hold with $ \delta n = \phi^{\dagger}
  \phi $ as the boson density above the VBS background.


\section{ Detection of quantum phases in frustrated lattices }

       The procedures discussed in bipartite lattices in the previous sections can be generalized to frustrated lattices such
       as triangular and kagome lattices. There are several
       new features due to the frustrations (1) The ordering
       wavevector $ \vec{Q}_{n} \neq - \vec{Q}_{n} $, while for a
       bi-partite lattice  $ \vec{Q}_{n} = - \vec{Q}_{n} $ upto a
       reciprocal lattice. (2) The corresponding fluctuation near
       the ordering wavevector $ \vec{Q}_{n} $ will be a complex
       order parameter, in contrast to that on a bipartite lattice
       which is just a real ( or Ising ) order parameter. The results achieved should
  also be useful to Quantum Monte Carlo simulations on the extended
  boson Hubbard model in  a frustrated  finite $ N = L \times L $
  lattice \cite{frusqmc}. The analysis in the direct picture in this
  section can be contrasted to that by the dual vortex method in the
  dual picture\cite{frus}. Several important CDW, VBS and CDW-VBS phases are reviewed in the appendix A.

\subsection{ Density-Density and bond-bond correlation functions in a frustrated lattice   }

 In a frustrated lattice, in general, one can write the density at site $ i $ as
\begin{equation}
  N ( \vec{r}_i, t )= n( \vec{r}_i, t )  + Re \sum^{P}_{\alpha=1} \phi_{\alpha} e^{
  i \vec{Q}_{\alpha} \cdot \vec{r}_i }
\label{nmp}
\end{equation}
   where $ \vec{Q}_{\alpha}, \alpha=1,2,\cdots, P $ are the ordering
   wavevectors, the $ \phi_{\alpha} $ is the complex CDW order parameter
   near the $ \vec{Q}_{\alpha} $.
   For the X-CDW in Fig.\ref{triphase}a, $ P=1 $, for the CDW-VBS in Fig.\ref{tricdwvbs}, $ P=3 $.
   Then the density-density correlation function in Eqn.\ref{nn} can
   be written as:
\begin{eqnarray}
  S_{N} ( \vec{q}, t ) & = &  \frac{1}{N} \sum_{i} e^{-i \vec{q} \cdot \vec{r}_i }
    \langle n (i, t) n(0,0 ) \rangle_{C}  + n^{2} \delta_{\vec{q},0}   \nonumber   \\
     & + & \sum^{P}_{\alpha=1} [\frac{1}{N} \sum_{i} e^{-i (\vec{q} - \vec{Q}_{\alpha} ) \cdot  \vec{r}_{i} }
    \langle \phi_{\alpha} (i, t) \phi^{*}_{\alpha} (0,0) \rangle_{C}
     \nonumber  \\
    &  + &  \frac{1}{N} \sum_{i} e^{-i (\vec{q} + \vec{Q}_{\alpha} ) \cdot  \vec{r}_{i} }
    \langle \phi^{*}_{\alpha} (i, t) \phi_{\alpha} (0,0) \rangle_{C} ]    \nonumber  \\
    & +  &  \sum^{P}_{\alpha=1} | \langle \phi_{\alpha} \rangle | ^{2} (
    \delta_{\vec{q},\vec{Q}_{\alpha}}+
    \delta_{\vec{q},-\vec{Q}_{\alpha}} )
\label{nntp}
\end{eqnarray}
    where  we have used the translational invariance to get rid of one
    summation and
    $ \langle n_{i} (t) n_{j}(0) \rangle_{C}=\langle n_{i} (t) n_{j}(0) \rangle- n^{2} $ and
    $ \langle \phi^{*}_{\alpha} (i, t) \phi_{\alpha} (0,0) \rangle_{C}
    = \langle \phi^{*}_{\alpha} (i, t) \phi_{\alpha} (0,0) \rangle - | \langle \phi_{\alpha} \rangle | ^{2}   $ are connected  Green functions.
    The $  \langle \phi_{\alpha} \rangle = m_{\alpha} e^{ i
    \theta_{\alpha} } $ is the expectation value of the CDW order
    parameter.  The translational invariance also dictates $ \langle \phi^{*}_{\alpha} (i, t) \phi_{\beta} (0,0)
    \rangle = 0 $ for $ \alpha \neq \beta $.

    Its Fourier transform leads to the dynamic structure function:
\begin{eqnarray}
  S_{N} ( \vec{q}, \omega ) & =  & \int dt e^{-i \omega t}  S_{N} ( \vec{q}, t )
  \nonumber  \\
    & = & n^{2} \delta_{\vec{q},0} \delta (\omega) + \sum^{P}_{\alpha=1} | \langle \phi_{\alpha} \rangle | ^{2} (
    \delta_{\vec{q},\vec{Q}_{\alpha}}+
    \delta_{\vec{q},-\vec{Q}_{\alpha}} ) \delta (\omega)   \nonumber   \\
    & + & \frac{1}{N} S^{inel}_n (\vec{q}, \omega )  +  \frac{1}{N} S^{inel}_{\phi} (\vec{q}, \omega )
\label{nnwp}
\end{eqnarray}
    where the first and second $ \delta (\omega) $ terms denote the elastic
    scatterings at $ \vec{q}=0 $ and  $ \vec{q} = \vec{Q}_{\alpha} $
    respectively, the third and the fourth  term denote the inelastic scatterings near $ \vec{q}=0 $ and  $ \vec{q} = \vec{Q}_{\alpha} $
    respectively.

    From $ S_{N} ( \vec{q}, t=0 )= \int \frac{ d \omega }{ 2 \pi }  S_{N} ( \vec{q}, \omega
    ) $, we can see the equal-time
    correlation function is the sum of the elastic one and the
    in-elastic one:
\begin{equation}
     S_{N} ( \vec{q}, t=0 )= S^{el}_n +  \frac{1}{N} S^{inel}_n
     (\vec{q})+ S^{el}_{\phi} + \frac{1}{N} S^{inel}_{\phi} (\vec{q} )
\end{equation}
     where $ S^{el}_n= n^{2} \delta_{\vec{q},0}, S^{el}_{\phi}  = \sum^{P}_{\alpha=1} | \langle \phi_{\alpha} \rangle | ^{2} (
    \delta_{\vec{q},\vec{Q}_{\alpha}}+
    \delta_{\vec{q},-\vec{Q}_{\alpha}} ) $ and $ S^{inel}_n
     (\vec{q})= \int \frac{ d \omega }{ 2 \pi } S^{inel}_n (\vec{q}, \omega )
     , S^{inel}_{\phi} (\vec{q})= \int \frac{ d \omega }{ 2 \pi } S^{inel}_{\phi} (\vec{q}, \omega   ) $.

    In the following, we discuss the density-density correlation functions in the
    superfluid, CDW and CDW supersolid respectively.

{ \sl (a) Superfluid  state near the CDW or the CDW-VBS state }

     In the superfluid state,  the first term  in Eqn.\ref{nntp} stands for the gapless superfluid mode near $ \vec{q}=0 $ and is also given by
     Eqn.\ref{sf}.  The second term near $ \vec{q}= \pm \vec{Q}_{\alpha} $ in Eqn.\ref{nntp} comes from the roton contribution
    near $ \vec{q}= \pm \vec{Q}_{\alpha} $ in Fig.\ref{excitations}a1. The dispersion near $ \vec{q}= \pm \vec{Q}_{\alpha} $ can be
    taken as $ \omega( q ) \sim  \Delta_{r\alpha} + \frac{ ( \vec{q} \mp \vec{Q}_{\alpha} )^{2} }{ 2 m_{r \alpha }} $ where
    $ \Delta_{r \alpha } >0 $ is the roton gap near $ \pm \vec{Q}_{\alpha} $. There is no CDW order yet, so $  \langle \phi_{\alpha} \rangle=0 $
    in the Eqn.\ref{nntp}.
    However, as one approaches the CDW from the SF in Fig.\ref{squarediagram}a,
    there is a first order transition into the CDW driven by the collapse of all the roton gaps $ \Delta_{r \alpha} $.

{ \sl  (b)  CDW or CDW-VBS state }

    In the CDW state, the $ nn $ correlator near $ \vec{q}=0 $ in Eqn.\ref{nntp} is very small, so can be dropped safely,
    so one only need to focus on the
    $ \phi^{*}_{\alpha} \phi_{\alpha}   $ correlator near $ \vec{q}= \pm \vec{Q}_{\alpha} $.
    The discrete lattice symmetry was broken due to the non-uniform density distribution,
    so $  \langle \phi_{\alpha} \rangle \neq 0 $, there is also a gap  $ \Delta_{CDW,\alpha} $ in the CDW state,
    so the connected equal time correlation function decays exponentially
    in the CDW state:  $ \langle \phi^{*}_{\alpha}(i, 0)  \phi_{\alpha} ( j,0) \rangle_{C} \sim e^{ - | \vec{r}_i -\vec{r}_{j}|/\xi_{CDW,\alpha} } $ with
    $ \xi_{CDW,\alpha} \sim 1/\Delta_{CDW,\alpha} $, so the in-coherent term in Eqn.\ref{nntp} is  only at the order of $ \sim 1/N $,  we conclude:
\begin{eqnarray}
      S_{N} ( \pm \vec{Q}_{\alpha}, t=0 ) & = & | \langle \phi_{\alpha} \rangle | ^{2} + \frac{1}{N} S^{inel}_\alpha (\vec{q}= \pm \vec{Q}_{\alpha}
      )    \nonumber  \\
      & = & | \langle \phi_{\alpha} \rangle | ^{2} + O(1/N)
\label{nnstaticp}
\end{eqnarray}
      Here we can see that the equal time structure factor is the sum of the elastic scattering $ S^{el}_{\alpha}  =
      | \langle \phi_{\alpha} \rangle | ^{2} $ plus a $ 1/ N $ in-elastic background.

     For $ \vec{q} \neq  \vec{Q}_{\alpha} $, but close to  $ \vec{Q}_{\alpha} $, then we find
\begin{equation}
    S_{N} ( \vec{q}, t )  \sim    \frac{1}{N} \sum_{i} e^{-i (\vec{q} - \vec{Q}_{\alpha} ) \cdot \vec{r}_i
    } \langle \phi_{\alpha} (i, t) \phi^{*}_{\alpha} (0,0) \rangle_{C}
\label{nnqnp}
\end{equation}
     where one can extract the excitation spectrum and spectral weight of the CDW near $ \vec{q} \neq \pm \vec{Q}_{N} $ shown in Fig.\ref{excitations}c1.
     Again, the excitation spectrum around $ \vec{k} - \vec{Q}_{N} =\vec{q} $ can also be extracted from the Feynman relation
     Eqn.\ref{cdwdis} which holds at $ \vec{q} \rightarrow 0 $, but not at $ \vec{q}=0 $.  The $ f $-sum rule
      $ I(\vec{q})= \int^{\infty}_{-\infty} d \omega \omega S_{N}( \vec{q}, \omega )= \langle [ [ n( \vec{q} ), H ], n(-\vec{q}) ]
      \rangle $ also holds independent of the underlying lattices.
      For example,  the Eqn.\ref{sumd2}  at a $ d=2 $ square lattice can be extended to a triangular lattice:
\begin{eqnarray}
      I(\vec{q})   =  - 2 t \sum_{\vec{k}} [ ( \cos q_x -1) \cos k_x + ( \cos q_y -1) \cos
      k_y       \nonumber   \\
       +  ( \cos(  q_x + q_y ) - 1) \cos ( k_x + k_y ) ]
      \langle \Psi_0 | b^{\dagger}_{\vec{k}} b_{\vec{k}}  | \Psi_0
      \rangle   ~~~~~~~~~~~
\label{sumd2tri}
\end{eqnarray}
     which can be used to extract the excitation spectra by QMC  in both the SF and the CDW state
     in a triangular lattice. The explicit forms for the $ f $ sum
     rule for other frustrated lattices can be similarly derived.

{ \sl (c) CDW Supersolid or CDW-VBS Supersolid state }

    In this case, the first term near $ \vec{q}=0 $ in Eqn.\ref{nntp} stands for the gapless superfluid mode as given by Eqn.\ref{sf}
    and shown in the lower branch in Fig.\ref{excitations}b1.

    The static order at  $ \vec{q}= \pm \vec{Q}_{\alpha} $ is given by
    Eqn.\ref{nnstaticp} and the dynamic structure factor  close to  $ \pm \vec{Q}_{\alpha} $ in
    the upper branch in Fig.\ref{excitations}b1 is given by Eqn.\ref{nnqnp} respectively.

    Combining the density-density discussions above in a frustrated lattice  with the bond-bond correlation
    functions in a bipartite lattice presented in Sec.II-B, we can
    similarly  discuss the bond-bond correlation functions in a frustrated lattice

\subsection{ Applications to CDW, VBS and CDW-VBS phases in a triangular lattice }

       From Eqn.\ref{xcdw}, one can see that the scattering cross section
       for the X-CDW  in \ref{triphase}a is similar to Fig.\ref{squareoptical}a with $  \pm \vec{Q}_{n}= \pm 2\pi/3( 1, 1 )
       $ and  $ \vec{K} =( 2\pi, 0 )  $ as the shortest reciprocal lattice vector of the underlying triangular OL.
       The CDW order parameter $ m^{2} $ replaced by $  | \langle \phi_{c}
       \rangle |^{2}  $ where $ \phi_c= A $.

       For a triangular lattice, there are three different orientations of
       bonds aligned along $ \hat{a}_1, \hat{a}_d, \hat{a}_2 $. From Eqn.\ref{trivbs}, one can see that  Eqn.\ref{kxy} should be generalized to:
\begin{equation}
  \hat{K}_{\triangle}  =    f_{a_1} ( \vec{q} ) K_{a_1}( \vec{q} ) + f_{a_d} ( \vec{q} )  K_{a_d}( \vec{q} )
                +  f_{a_2}( \vec{q} )  K_{a_2}( \vec{q} )
\label{kabc}
\end{equation}
       where $ \vec{q}= \vec{k}_{1}-\vec{k}_{0} $.
       For the $ \pm \vec{Q}_{K}=  \pm 2\pi/3 ( 1, 1 ) $ ordering of the Triangular valence bond in
       Fig.\ref{triphase}b, $ S_{K1}= S_{Kd}=S_{K2} $ with the VBS order parameter $ \phi_{v1}= \phi_{vd}= K_1, \phi_{v2}= K_1 e^{-i 2 \pi/3}
       $. So  the scattering cross section
       for the VBS in Fig.\ref{triphase}b is similar to Fig.\ref{squareoptical}b with $  \pm \vec{Q}_{K}= \pm 2\pi/3( 1, 1 )
       $ and the  $ \vec{K}=( 2\pi, 0 )  $ as the shortest reciprocal lattice vector of the underlying triangular OL.
       The VBS order parameter $ K^{2} $ replaced by $  | \langle \phi_{v}
       \rangle |^{2} = 3 K^{2}_{1} $.

\begin{figure}
\includegraphics[width=7cm]{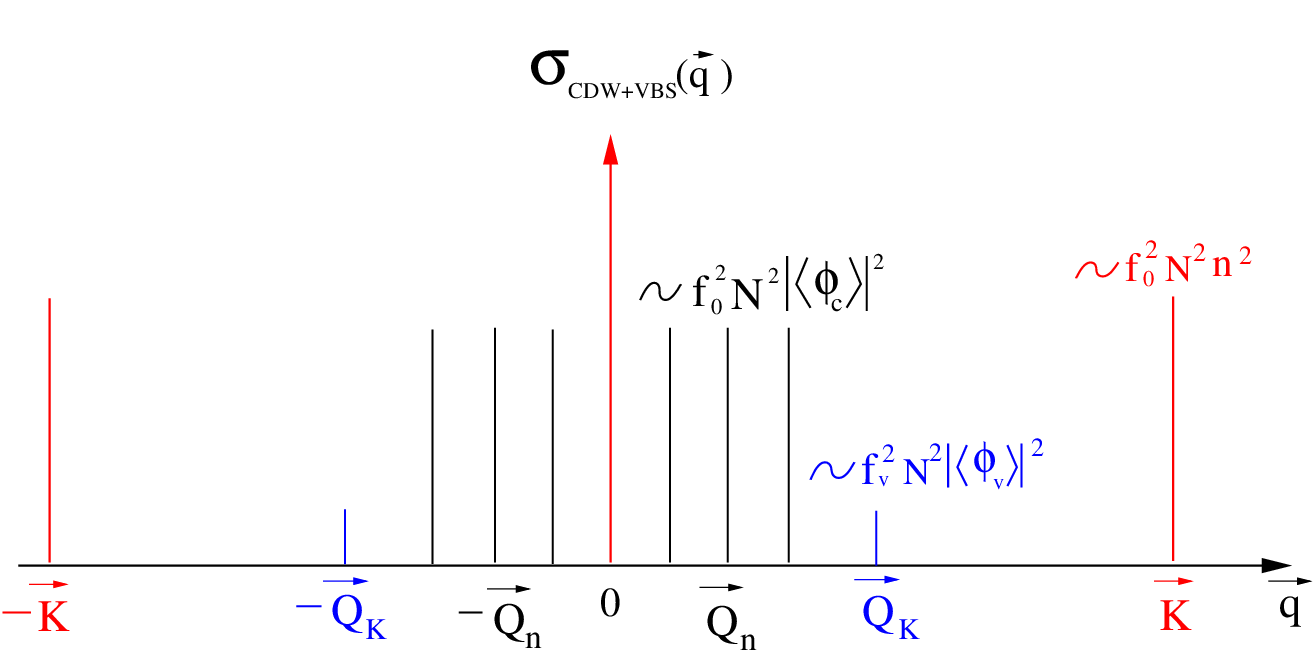}
\caption{ The characteristics of  optical scattering cross section
in the CDW+VBS phase in a triangular lattice. The $ \vec{Q}_n $
stands for the three CDW ordering wavevectors $ \vec{Q}_{\alpha}=
2\pi/3(1,0), 2\pi/3(0,1), 2\pi/3(1,-1),\alpha=1,2,3 $. Its magnitude
is proportional to the CDW order parameter $ | \langle \phi_{c}
\rangle | ^{2} $ in Eqn.\ref{nnstaticp} and Eqn.\ref{directtricv}.
The $ \vec{Q}_K $ stands for the extra VBS ordering wavevector $
\vec{Q}_{4}= 2\pi/3(1,1) $. Its magnitude is proportional to the VBS
order parameter $ | \langle \phi_{v} \rangle | ^{2} $ in
Eqn.\ref{directtricvb}.  The ratio of the peak at $ \vec{Q}_{n} $
over that at $ \vec{K} $ is $ \sim | \langle \phi_{c}
\rangle^{2}/n^{2} \sim 1 $, while the ratio of the peak at $
\vec{Q}_{K} $ over that at $ \vec{K} $ is $ \sim |f_{a}/f_{0}|^{2} |
\langle \phi_{v} \rangle |^{2}/n^{2}\sim 10^{-5} $. It should be
visible in the current optical Bragg scattering experiments. }
\label{trioptical}
\end{figure}

       For the CDW-VBS phase in Fig.\ref{tricdwvbs} and Eqn.\ref{directtricv}, Eqn.\ref{directtricvb}, there are 3 CDW ordering wave
       vectors $ \vec{Q}_{\alpha}= 2\pi/3(1,0), 2\pi/3(0,1), 2\pi/3(1,-1),\alpha=1,2,3 $ withe the corresponding CDW order parameters
       $ \phi_1= - \delta e^{i 5 \pi/18}, \phi_2=\delta e^{i \pi/18}, \phi_3=\delta e^{i \pi/18} $.
       In addition to the same 3 ordering wave vectors, the VBS order
       has its own new ordering wavevector $ \vec{Q}_{4}= 2\pi/3(1,1) $ with the VBS order parameter
       $ \phi_{v1}= \phi_{vd}= c \delta (I_1+I_2), \phi_{v2}= c \delta (I_1+I_2) e^{-i 2 \pi/3} $,
       so although  the scattering cross section  $ \sim  | \langle \phi_{c} \rangle |^{2} \sim  \delta^{2} $ at the ordering
       vectors $ \vec{Q}_{\alpha}, \alpha=1,2,3 $ comes from the sum of the contributions  from both the CDW and the VBS,
       it's value $ \sim |f_{a}(\vec{Q}_{4}) |^{2} \times  | \langle \phi_{v} \rangle |^{2} \sim |f_{a}(\vec{Q}_{4}) |^{2} \times  c^{2} \delta^{2} $ at the
       ordering wavevector $ \vec{Q}_{4}= 2\pi/3(1,1) $ is solely due to the VBS at this ordering wavevector. So the small scattering cross section peak
       at $ \vec{Q}_{4}= 2\pi/3(1,1) $  in Fig.\ref{trioptical} can be used to
       determine the VBS order in the CDW-VBS phase in Fig.\ref{tricdwvbs}.

       The SF to Mott transition in a triangular lattice can be
       similarly discussed as in Sec.V.  The CDW ( CDW-VBS)  to the CDW-SS ( CDW-VB-SS ) transition in a triangular
       lattice can also be similarly discussed  as in Sec. V.
       However, it was shown in \cite{frus} that there is no VB-SS
       in a frustrated lattice, the transition from a VBS to the SF
       can only be a direct first order, there is no an intervening
       VB-SS, so the Sect.VI does not apply to a frustrated
       lattice.

       The generalization to non-Bravais lattices such the
       honeycomb, Kagome, Sutherland-Shastry  and checkboard
       lattices \cite{frus,frusrev} can also be worked out similarly.

\section{  Cavity QED detection method }

Very recently, a non-destructive method \cite{off1} was proposed to
detect Mott and superfluid phases by using cavity enhanced
off-resonant light scattering from ultra-cold atoms loaded on
optical lattices. In the off-resonant scattering, the atom-field
detuning $ \Delta $ is much larger than the atom-field coupling $ g
$ which, in turn, so the upper level of the atoms can be
adiabatically eliminated. However, on the experimental side, the two
cavities used in \cite{off1} are very hard to implement, on the
theory side, all the concepts and calculations on quantum phases and
phase transitions in \cite{off1} are not correct.

\begin{figure}
\includegraphics[width=3cm]{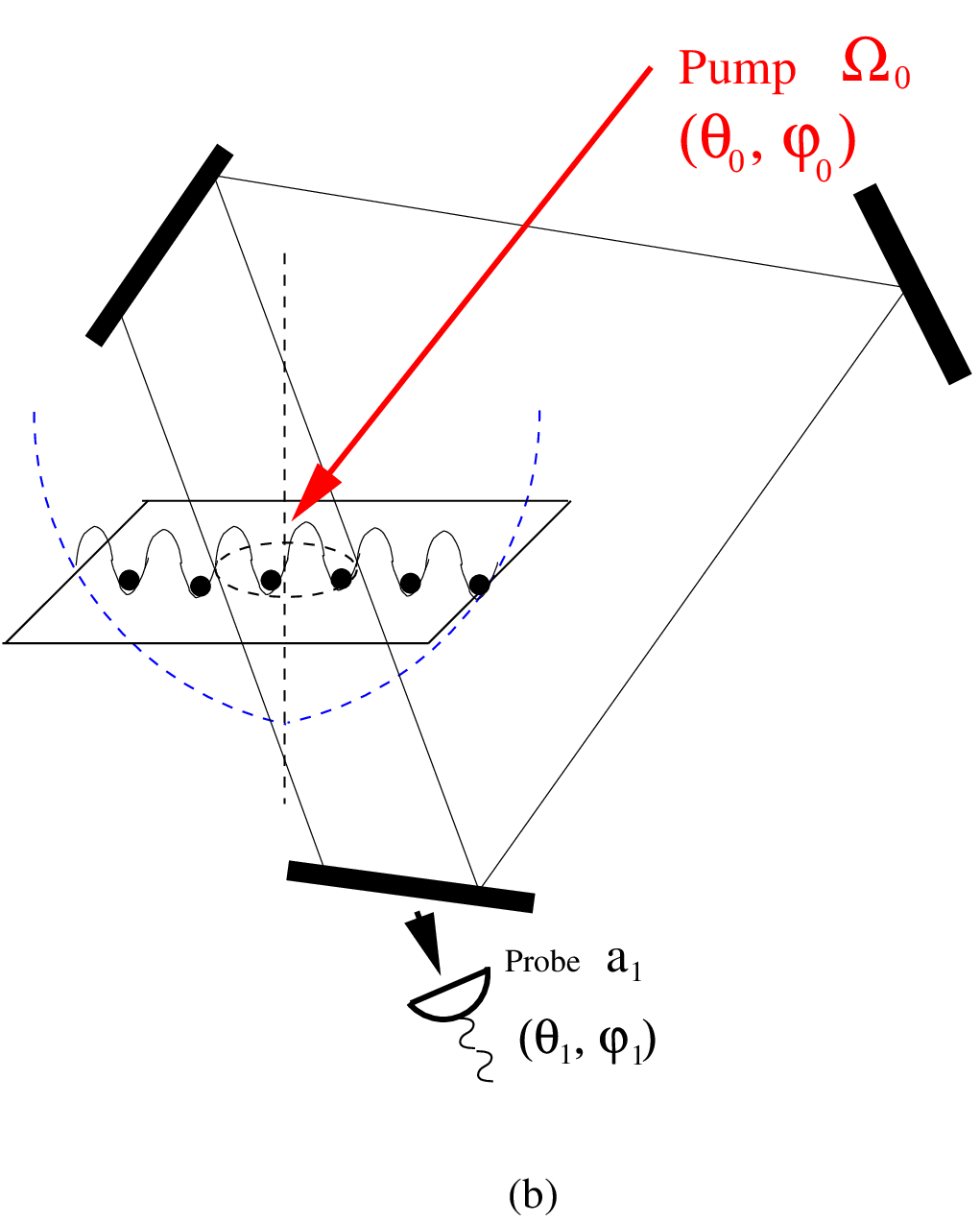}
\hspace{0.5cm}
\includegraphics[width=3cm]{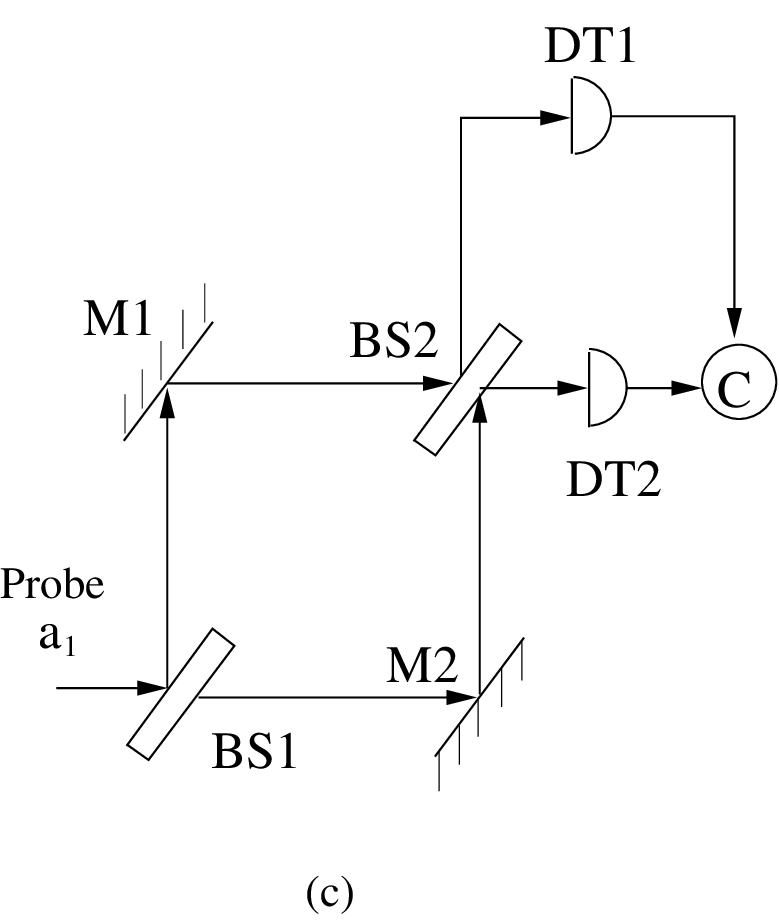}
\caption{ (a)  The classical pump with a Rabi frequency $
\Omega_{0} $ is shined on the 2 dimensional optical lattice at the
angle $ (\theta_0, \phi_0 ) $. A Ring cavity is used to enhance the
scattered light $ a $ at the angle $ (\theta_1, \phi_1 ) $. Dashed
ellipsis is a trap. Compare with the light scattering experiment Fig.1a.
(b) Mach-Zehnder Interferometer \cite{scully} to detect the single
photon correlation function Eqn.\ref{onephoton}. } \label{cavityqed}
\end{figure}

Because the cavity size in the experiment is $ L_{cav} ~ 175 \mu m $
is so small, it is very difficult to manipulate experimentally the
two cavities, so it is necessary to replace the pumping cavity in
\cite{off1} by a strong classical laser beam in
Fig.\ref{cavityqed}a. However, the probing cavity is crucial to
enhance the scattered photon mode $ a $ along a given direction,
then the experiments are much easier to implement. By using this
workable experimental set-up, we will develop systematically a
theory to detect the nature of quantum phases such as both the
ground state and the excitation spectrum above the ground state of
interacting bosons loaded in optical lattices. We explicitly show
that off-resonant photons not only couple to the density order
parameter, but also the valence bond order parameter due to the
hopping of the bosons on the lattice. By tuning the angles between
the classical laser beam and cavity photons, the photon
characteristics such as quadrature in Eqn.4 ( to be measured by
phase sensitive homedyne detection ) or one photon correlation
functions in Eqn.7 ( to be measured by the Mach-Zehnder
Interferometer (MZI) ) can detect not only the well known superfluid
and Mott insulating phases, but also other interesting phases such
as charge density wave (CDW), valence bond solid (VBS), CDW
supersolid and VBS supersolid.

  In the light scattering experiment Fig.1a, the light is scattered
  to any direction, so one has to use a small aperture to select the
  light scattered into a given direction $ \vec{k}_1 $. In this
  section, we propose to replace the aperture  by a ring cavity ( Fig.\ref{cavityqed}a ) to
  select a cavity photon mode, to enhance the scattered light and also
  select a given direction $ \vec{k}_{1} $. In this section, we will
  show that by measuring the characteristics of the leaking photon
  out of the cavity, one can also determine the ground state and
  excitation spectrum of the atoms loaded in the optical lattice.
  So the experimental set-up in Fig.\ref{cavityqed} could be used as
  alternative to the light scattering experiment in Fig.1a

   A non-destructive measurement is to probe the quantum
   phases formed by two level cold bosons  by shining a classical laser beam with a Rabi frequency $ \Omega $ and  with a
   frequency $ \omega_{0} $  far off the resonant frequency of the two level
   atoms $ \omega_{a} $ and then measure the characteristics of scattered light $ a $
   from these quantum phases in a cavity with frequency $ \omega_c $
   ( Fig.\ref{cavityqed}a ). The cavity will greatly enhance the scattering amplitude of the probe photons $ a $.
   The boson Hamiltonian is given by Eqn.\ref{boson}.
   Assuming the mode functions $ u_{l}( \vec{r} ) = e^{i \vec{k}_{l} \cdot \vec{r}_{i} + i \phi_{l} }, l=0,1 $
   for the pump with the frequency $ \omega_0 $  and the traveling wave with the frequency $ \omega_1= \omega_c $ in a ring cavity ( Fig.\ref{cavityqed}b)
   where the two in-plane  momenta of the two lights $ \vec{k}_{0}=k_{0} \sin \theta_{0} ( \cos
   \phi_{0}, \sin \phi_{0} ),  \vec{k}_{1}=k_{1} \sin \theta_{1} ( \cos
   \phi_{1}, \sin \phi_{1} ) $.   All the atoms are loaded in optical lattices formed by
   laser beams with wavevector $ \vec{k} $. The pumping laser beam and the cavity photon $ a $ are much
   weaker than the standing wave laser beams which form the optical lattices ( not shown in Fig.\ref{cavityqed} ).
   When it is far off the resonance, the light-atom detuning $ \Delta_{la}= \omega_{l}
  -\omega_{a} $ is much larger than the Rabi frequency $ \Omega_{0} $
  which, in turn, is larger than  $ \kappa $ and $ \gamma $, then  after adiabatically eliminating
  the upper level of the two level atoms ( Fig.\ref{raman} ), in the frame rotating with the pumping frequency $ \omega_0 $, the
  effective cavity QED Hamiltonian describing the interaction among the pumping laser, the off-resonant cavity photons and the ground level
  is:
\begin{eqnarray}
  H_{c}  & =  & (\omega_c-\omega_0 ) a^{\dagger} a + \int d^{2} \vec{x}
  \Psi^{\dagger}( \vec{x} ) [ \frac{ \Omega^{2}_{0} }{ \Delta_{0a}} + \frac{ g^{2} }{ \Delta_{0a}} a^{\dagger} a
     \nonumber  \\
   &  +  &  \frac{ g \Omega_{0} }{ \Delta_{0a} } ( a^{\dagger} e^{i (
    \vec{k}_0-\vec{k}_1 ) \cdot \vec{r} } + h.c. ) ]\Psi( \vec{x} )
\label{cavity1}
\end{eqnarray}

 Again, expanding the ground state atom field operator $ \psi_{g}(\vec{r}) =\sum _{i} b_{i} w( \vec{r} -\vec{r}_{i} )  $ in Eqn.\ref{bragg}
 where $ w( \vec{r} -\vec{r}_{i} ) $ is the localized Wannier functions of the lowest Bloch band corresponding to $ V_{OL}( \vec{r} ) $ and $  b_i $
 is the annihilation operator of an atom at the site $ i $ in the
 Eqn.\ref{boson}, one get:
\begin{eqnarray}
  H_{c} & = & (\omega_c-\omega_0 + \frac{ g^{2} }{ \Delta_{0a}} N_{at} ) a^{\dagger} a +  \frac{ \Omega^{2}_{0} }{ \Delta_{0a}} N_{at}
       \nonumber  \\
        & +  & \frac{ g \Omega_{0} }{ \Delta_{0a} } a^{\dagger} [ \sum^{K}_{i}  J_{i,i} n_{i}  +  \sum^{K}_{<ij>}  J_{i,j} b^{\dagger}_{i}b_{j}
        ] + h.c.
\label{cavityint}
\end{eqnarray}
  where the interacting matrix element is $
  J_{i,j}= \int d \vec{r} w( \vec{r}-\vec{r}_{i} )
  u^{*}_{0}(\vec{r} ) u_{1}(\vec{r} ) w( \vec{r}-\vec{r}_{j} ) $.

  It  is very instructive to note the similarity and difference between  Eqn.\ref{cavityint}
  and  Eqn.\ref{laserint}. The incident light in Fig.1a and
  Fig.\ref{cavityqed} are the same, both are classical lights, but the classical scattered light is Fig.1a
  was replaced by the quantum cavity photon $ a $ in Fig.\ref{cavityqed}. So the scattered light Rabi frequency  $ \Omega $ in
  Eqn.\ref{laserint} is replaced by $ g a^{\dagger} $ in
  Eqn.\ref{cavityint} ( correspondingly, the scattered light in Fig.\ref{raman} was replaced by the cavity photon $ a $ ),
  the $ e^{-i \omega t } $ in
  Eqn.\ref{laserint} was taken care of by the effective detuning  of the cavity photon from the incident light
  $ \Delta_c = \omega_c - \omega_0  + \frac{ g^{2} }{ \Delta_{0a}} N_{at} $ in Eqn.\ref{cavityint}.
  In view of these similarity and differences,  the $ \hat{D} =\sum^{K}_{i}  J_{i,i} n_{i} $ and $ \hat{K} = \sum^{K}_{<ij>}  J_{i,j} b^{\dagger}_{i}b_{j} $
  in  Eqn.\ref{cavityint} can be very similarly manipulated as in
  Eqn.\ref{d} and \ref{kxy} respectively.

   In this section, we focus on the global illumination $ K=N $ in Fig.\ref{cavityqed}a.
   If there is a wedding cake structure inside a trap \cite{manybody} ( see also Sec. XI ), then $ K=N_{P} $ is the
   number of atoms in a given phase $ P $ in the wedding cake. This can be more easily realized inside a flat
   trap \cite{flattrap} or inside a harmonic trap with only one shell structure such as in Fig.\ref{motttrap}.

   It is constructive to compare with the photon-exciton coupling $
   i \sum_{k} g(k) a^{\dagger}_{k} b_{\vec{k}} + h.c. $ in
   the electron-hole bilayer (EHBL) system \cite{short1,short2,excitonlong}
   where photons couple to the SF order parameter $
   b_{\vec{k}} $ directly.  Here the photons couple to both the density
   order parameter and also the valence bond order parameter
   instead of  coupling to the SF order parameter directly. However, as shown in
   Eqn.\ref{sf}, inside the SF state, the density-density correlation function can
   also reflect the nature of SF precisely. So the coupling in
   Eqn.\ref{cavityint} can reflect all the three orders: density order, valence bond  order and SF order
   precisely.

   The Heisenberg equation of motion for $ a $ is:
\begin{equation}
  \frac{ d a }{ d t }= -i  \Delta_c  a-i \frac{\Omega_{0} g }{ \Delta_{0a}} ( \hat{D} + \hat{K} )
  - \kappa a + F(t)
\label{hei}
\end{equation}
   where $ \Delta_c = \omega_c - \omega_0  + \frac{ g^{2} }{ \Delta_{0a}} N_{at} $
   stands for the effective detuning to the pumping frequency.
   The $ F(t) $ is the  noise operator satisfying:
  $\left\langle F^{\dagger }(t)F(t^{\prime })\right\rangle_R =
\kappa \bar{n}_{\omega_c } \delta (t-t^{\prime }), \left\langle
F(t)F^{\dagger}(t^{\prime })\right\rangle_R = \kappa (
\bar{n}_{\omega_c} + 1 ) \delta (t-t^{\prime }), \left\langle
F(t)F(t^{\prime })\right\rangle_R= \left\langle
F^{\dagger}(t)F^{\dagger}(t^{\prime })\right\rangle_R =0 $ where
the average $ R $ is taken with respect to the reservoir, the $\bar{n}_{\omega_c }=1/(e^{\omega_c /T}-1)$ and $ T $ is the
temperature of the photon reservoir outside of the cavity.

{\sl (a) One time average: photon expectation value }

   The stationary solution for the Heisenberg  equation of motion for $ a $  Eqn.\ref{hei} is:
\begin{equation}
 \langle a \rangle = C  N ( \langle \hat{D} \rangle + \langle \hat{K} \rangle )
\label{dk}
\end{equation}
   where the $  C= -\frac{  i \Omega_{0} g  }{ \Delta_{0a}( \kappa -i \Delta_c ) } $.
   The ensemble average $  \langle \cdots \rangle $ is taken with respect to the initial state $ | atom \rangle \times |0 \rangle_{ph} $
   where the $ | atom \rangle $
   stands for the ground state of interacting atoms in Eqn.\ref{boson} and the $ |0 \rangle_{ph} $ stands for the initial zero photon state.

   Substituting Eqn.\ref{d} and \ref{kxy} into the Eqn.\ref{dk} leads to
\begin{equation}
  \langle a (\vec{q}) \rangle = C N ( \langle n(\vec{q}) \rangle +
     f_{x}( \vec{q})  \langle K_{x}( \vec{q} )\rangle +
                f_{y}( \vec{q} ) \langle  K_{y}( \vec{q} ) \rangle )
\label{zerophoton}
\end{equation}
    which can be measured by phase sensitive homedyne measurement \cite{scully,short2}.

    Again, we first look at the SF to Mott transition at integer fillings as discussed in Sect.V.
    Inside the SF,  the $ K_{i,i+ \hat{x}}=  K_{i,i+ \hat{y}} $ is uniform which depends on $ t-t_{c} $ in Fig.\ref{sfmott}, so
    at a reciprocal lattice $ \vec{q}= \vec{K} $, Eqn.\ref{zerophoton} leads to:
\begin{equation}
    \langle a (\vec{q}) \rangle_{SF}- \langle a_{1}(\vec{q}) \rangle_{Mott}  =  2 C N f_x( \vec{q} ) K
\label{mottsfinc2}
\end{equation}
    which can be detected by phase sensitive homodyne detection \cite{scully,short2}. It is an effective measurement of
    the kinetic energy inside the SF. This measurement can be
    contrasted to the increase of the scattering cross section from
    the Mott to the SF shown in Eqn.\ref{mottsfinc} or from the CDW to  CDW-SS, or  from the VBS to  VB-SS discussed in Section
    V-a and VI-a respectively.

{\sl (b) Two time averages: single photon correlation functions}

   Now we need to calculate the two-time one photon correlation
   function. The solution of Eqn.\ref{hei} is \cite{eom}:
\begin{eqnarray}
   a (t) &  =  &e^{ -(\kappa +i  \Delta_c )( t-t_{0}) } a (
   t_0)  \nonumber  \\
   & + & \int^{t}_{t_0} d\tau
   e^{ -(\kappa +i  \Delta_c )( t-\tau ) } (\eta_{a} (\tau) + F(\tau))
\label{integral}
\end{eqnarray}
  where $ \eta_{a}(t)=-i \frac{\Omega_{0} g}{ \Delta_{0a}} ( \hat{D}(t) + \hat{K}(t) )
  $ stands for the "effective " pumping force on $ a $ from the atoms.

  We observe the following three important facts:
  (1) If $ t-t_0 \gg 1/\kappa $, the first term drops out in the steady state.
  (2) The second term
  contributes significantly only when $ t-\tau < 1/\kappa $.
  It was known that in an optical lattice \cite{bloch,boson}, the hopping energy scale is much smaller than
  the cavity decay energy scale: $ J \sim 10^{3} Hz \ll \kappa \sim 10^{7} Hz
  $, so we can approximate $ \eta_{a} (\tau) \sim \eta_{a} (t) $
  when $ t-\tau < 1/\kappa $.
  (3) In the  optical cavity frequency regime $ \hbar \omega_c \gg k_{B} T $, so $
  \bar{n}( \omega_c) \sim 0 $, then the noise term $ F(t) $ drops out for a
  normal ordered correlation functions.
  The three facts lead to the two time photon correlation function:
\begin{eqnarray}
 \langle a^{\dagger}( \vec{q}, t) a( \vec{q}, 0 ) \rangle  & = & | C |^{2}
 [ \langle D^{*}( \vec{q}, t)  D(  \vec{q}, 0 ) \rangle   \nonumber \\
 & + & \langle K^{*}( \vec{q}, t) K( \vec{q} , 0 ) \rangle ]
\label{onephoton}
\end{eqnarray}
    where $ \vec{q}=\vec{k}_{1}-\vec{k}_{0} $.  We expect the crossing correlator
    $ \langle D^{*}( \vec{q}, t) K(  \vec{q}, 0 ) \rangle $ is negligibly small in any phases.
    As shown in the Fig.\ref{cavityqed}b, using the Mach-Zehnder Interferometer (MZI) \cite{scully} and
    adjusting the difference between the two light paths, one can
    measure this two time one photon correlation function.

    From Eqn.\ref{onephoton}, one can see that at any given  momentum $ \vec{q} $, the Florescence spectrum of the probing
    photons  $ I_{out}(  \vec{q}, \omega ) = \int d \tau  e^{-i \omega \tau } \langle a^{\dagger} ( \vec{q}, t + \tau ) a ( \vec{q}, t )
    \rangle $ is :
\begin{eqnarray}
     I_{out}(  \vec{q}, \omega ) & = &  |C|^{2} N^{2} [ |f_{0}( \vec{q} ) |^{2} S_{n}( \vec{q}, \omega )
  \nonumber  \\
  & + &  \sum_{\alpha=\hat{x},\hat{y} } |f_{\alpha}( \vec{q} ) |^{2} S_{K_{\alpha} }( \vec{q}, \omega ) ]
\label{power}
\end{eqnarray}
    which is similar to Eqn.\ref{cross} or Eqn.\ref{cloud} after
    replacing $  ( \frac{ \Omega^{2} }{ \Delta } )^{2}  $ by $ |C|^{2} $.

  In Eqn.\ref{onephoton}, by setting $ t=0 $, one can see that the
  leaking photon number gives just the structure factor:
\begin{eqnarray}
 n_{ph}( \vec{q} )= \langle a^{\dagger}( \vec{q}) a( \vec{q} ) \rangle  & = & |C|^{2} N^{2} [  |f_{0}( \vec{q} ) |^{2} S_{n}( \vec{q}
    )     \nonumber  \\
      & +  & \sum_{\alpha=\hat{x},\hat{y} } |f_{\alpha}( \vec{q} ) |^{2} S_{K_{\alpha}}( \vec{q})   ]
\label{number}
\end{eqnarray}
   which is similar to Eqn.\ref{crossequal} or Eqn.\ref{cloudequal} after
   replacing $  ( \frac{ \Omega^{2} }{ \Delta } )^{2}  $ by $ |C|^{2} $.
   All the discussions in previous sections can also be applied
   here. For example, the scattering cross sections in
   a square lattice Fig.\ref{squareoptical} or in a triangular lattice Fig.\ref{trioptical} should just be replaced
   by the photon numbers at the corresponding wavevectors.
   So we show that the Florescence spectrum of the leaking cavity photons
   can directly reflect the ground state and the excitation spectrum of any quantum state.

\section{ Quantum phases, phase transitions, Local density approximation and in situ measurements inside a harmonic trap }

    So far, we have been discussing the detections of quantum phases and phase transitions inside a flat trap.
    But most of the traps used in cold atom experiments are harmonic traps. Here we will discuss the effects of
    a harmonic trap. Inside a harmonic trap $ V(r)= \frac{1}{2} \alpha r^{2} $ where the $ \alpha $ is the curvature, one can construct
    the length scale \cite{curvature}:
\begin{equation}
    L_c \sim ( \alpha/t )^{-1/2}
\label{lc}
\end{equation}
    where the $ t $ is the hopping in Eqn.\ref{boson}.  As shown in \cite{curvature}, it plays a similar role as a
    finite size $ L $ in the homogeneous system.

    In cold atom experiments inside a harmonic trap, it is more convenient to express the scaling functions
    in Eqn.\ref{densityscaling} in terms of the finite temperature $ T $ :
\begin{eqnarray}
        S_{n}( \vec{Q}_{N}, i \omega_{n} =0 ) & =  & T^{d-2+\eta} F_{ns} ( \frac{K-K_{c}}{T^{1/\nu z}}, \beta/L^{z}_c )    \nonumber   \\
        S_{n}( \vec{Q}_{N}, \tau =0 ) & = & T^{\frac{ 2 \beta }{ z \nu}}  F_{ne} (\frac{K-K_{c}}{T^{1/\nu z}}, \beta/L^{z}_c )
\label{densityscalingT}
\end{eqnarray}

   The Local density approximation (LDA) means that the system's properties at the local chemical potential
   $ \mu(r)= \mu-V(r) $ can reflect those of a homogeneous system at this local $ \mu(r) $.
   Then determining the bulk thermodynamic quantities as a function of the chemical potential corresponds to determining the
   $ r $ dependence in the harmonic trap \cite{trapthe0,trapthe1}. We expect that the LDA works in the $ 1/T < L_c $ limit, so
   the trapped system feels the  temperature effects before it feels the curvature effects of the trap.
   In all the present experiments, the $ T \sim 20 nK $, so $ 1/T $ is indeed smaller than the $ L_c $, so the system feels the
   temperature effects before it feels the curvature effects, the LDA is valid. Then the tuning parameter $ K $  in Eqn.\ref{densityscalingT}
   can be taken as the local chemical potential  $ \mu(r)= \mu-V(r) $.  Setting $ K= \mu(r) $ and  $ \beta/L^{z}_c  \rightarrow 0 $ in
   Eqn.\ref{densityscalingT} and lead to:
\begin{eqnarray}
        S^{LDA}_{n}( \vec{Q}_{N}, i \omega_{n} =0 ) & =  & T^{d-2+\eta} F_{ns} ( \frac{\mu-\mu_{c}}{T^{1/\nu z}} )    \nonumber   \\
        S^{LDA}_{n}( \vec{Q}_{N}, \tau =0 ) & = & T^{\frac{ 2 \beta }{ z \nu}}  F_{ne} (\frac{\mu-\mu_{c}}{T^{1/\nu z}} )
\label{densityscalingTc}
\end{eqnarray}
    Note that we expect that the scaling functions will eventually break down at lower temperatures where the system starts to
    feel the curvature effects  of the trap.

\begin{figure}
\includegraphics[width=7cm]{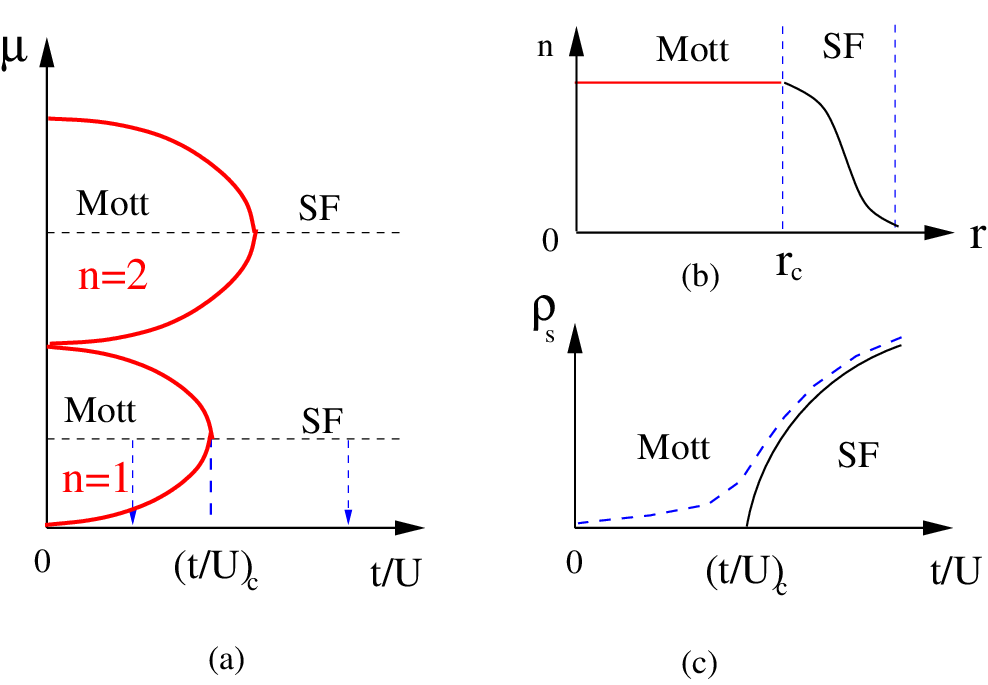}
\caption{ (a) The zero temperature phase diagram of a homogeneous short-ranged boson Hubbard model
\cite{boson}. See Fig.\ref{sfmott} for its finite temperature phase diagram.
Inside a trap, the local chemical potential $ \mu(r) =
\mu -\frac{1}{2} K r^{2} $ decreases from the center to the
boundary. Under the LDA, there is a Mott to SF transition at $ r=r_c $.
(b) The well-known shell structure of Mott state in the
center and the SF in the boundary (c) The red solid line is the
superfluid density in a homogeneous system with $ \rho_{s} \sim
(t-t_c)^{\nu} $. The blue dashed line is inside a trap }
\label{motttrap}
\end{figure}

   Another important effect of the trap is that there exist multiple phases inside an harmonic trap \cite{curvature}.
   As the local chemical  potential $ \mu(r)= \mu-V(r) $ decreasing from the center of the trap to the
   boundary, there is always a shell structure of phases inside a trap.
   For example, at filling $ n=1 $, there is a Mott phase at the
   center, then there is always a SF shell around the boundary ( Fig.\ref{motttrap} ).
   So there is a Mott gap in the center, gapless SF around the boundary. To some extent, this is similar to
   quantum Hall state where there is a gap in the bulk, but gapless edge state along the boundary.
   But the main difference is that here there is a harmonic trap,
   while in Quantum Hall system, there is a sharp sample edge within a few magnetic length.

   Under the LDA, there is a Mott to SF transition in Fig.\ref{motttrap}b at $ r=r_c $, one can apply the thermodynamic scaling
   Eqn.\ref{deltanz2} and Eqn.\ref{rhos} as:
\begin{eqnarray}
   \rho_s(r) & \sim & \delta n(r)=1-n(r)   \nonumber  \\
   & = & \frac{  m_{a} a^{2} \delta \mu(r) }{ 4 \pi \hbar^{2}} \ln [ \frac{\hbar^{2}}{ 2 m_{a} a^{2} \delta \mu(r) }]
\label{deltanz2r}
\end{eqnarray}
   where $ \delta \mu(r) = \mu_c- \mu(r)= \frac{1}{2} K ( r^{2}- r^{2}_{c} ) \sim r-r_c $.
   In principal, this scaling relation can be tested by the {\sl in situ} measurement in \cite{insitu0,insitu1}.
   However, it remains challenging to measure the dynamic density-density correlation function Eqn.\ref{scalingz2}
   by the {\sl in situ} meathod. It seems only the three scattering experiments can measure the dynamic correlations.
   So the {\sl in situ} measurement and the  three scattering measurements are complimentary to each other.

  Under the Local density approximation (LDA), a scaling analysis can be written down across
  the 2d Kosterlitz-Thouless (KT) superfluid to normal gas transition from the center of the trap to the boundary.
  The KT transition is  a finite temperature transition at $ T=T_c $, the correlation length $ \xi \sim e^{ 1/\sqrt{T-T_c}} $.
  Indeed, the recent {\sl in situ} measurements \cite{insitu1} on local density and local density fluctuations were
  performed to confirm the scaling functions at different temperatures and different interaction strengths.

\begin{figure}
\includegraphics[width=7cm]{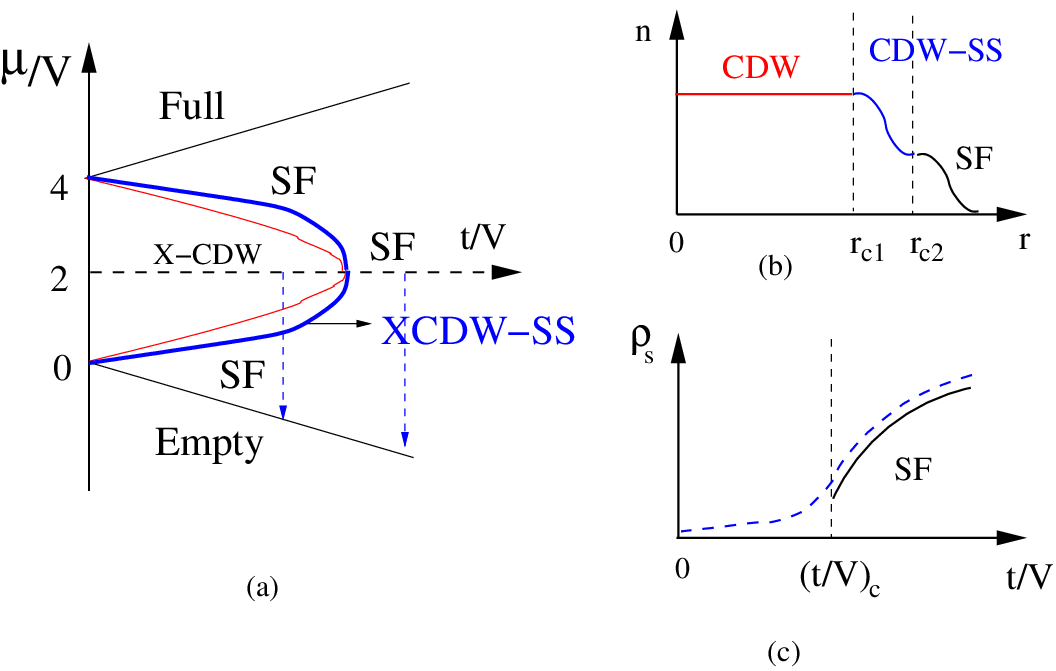}
\caption{ (a) The zero temperature phase diagram of a homogeneous boson Hubbard model
with dipole-dipole interaction strength $ V $ \cite{dipolarss,dipolarsstri}.  Comparing with the Fig.\ref{squarediagram}.
(b) The shell structure of X-CDW  state in the
center and the SF in the boundary with the CDW-SS intervening
between the two. The 2nd order transitions at $ r_{c1} $ and $ r_{c2} $ are discussed in the text.
(c) The transition from the X-CDW to the SF is not studied yet. It is likely to be a first order transition. Then
the red solid line is the superfluid density in
a homogeneous system with $ \rho_{s} $ jump discontinuously at the
CDW-SF transition. The blue dashed line is inside a trap }
\label{cdwtrap}
\end{figure}

  There are many possible ways to observe a stabilized supersolid in a cold atom experiment. One possible route is to using
  the shell structure inside a harmonic trap.
  For  hard core bosons, it was shown by the QMC in \cite{square}, the $ ( \pi, \pi ) $ X-CDW SS is not stable against
  phase separation with  $ V_1 > 0, V_2 = 0 $, but
  the $ ( \pi, 0 ) $ stripe SS may be stable with  $ V_1=0, V_2 > 0 $. The transition from the stripe SS to the SF is a first order transition.
  However, for hard core bosons with a dipole-dipole interaction, the $ (\pi,\pi) $ ( $ \vec{Q}_n= 2 \pi/3 (1,1) $ in Fig. \ref{triphase}a )  X-CDW supersolid was
  found to be stable in a square ( triangular ) lattice \cite{dipolarss,dipolarsstri} in a large parameter regimes near the half filling \cite{int}.
  Furthermore, it was found the CDW-SS to the SF transition is a second order transition
  in the $ 3d $ Ising universality class.
  Now if we take a half filling at the center of the trap, so it is
  a X-CDW at the center, a SF near the boundary, then there could be a  stable vacancy-like X-CDW supersolid \cite{bipart,frus}
  separating the X-CDW from the SF ( Fig.\ref{cdwtrap}b ).
  This is a ring structure with a periodic boundary condition around the center, so it
  may be favorable to stabilize this X-CDW SS in the quasi-1d ring
  structure. Indeed, a SS was found to be stable in 1d optical lattice in \cite{sca}.
  Under the LDA, the transition at $ r_{c1} $ is a second order
  in the universality class of the Mott to the SF transition with $ z=2, \nu=1/2, \eta=0 $ in Fig. \ref{motttrap}.
  Then Eqn.\ref{deltanz2r} still holds with $ \delta n(r)=1/2-n(r) $.
  The transition at $ r_{c2} $ is also a second order
  in the universality class of the 3d Ising class with $ z=1, 1/\nu=1.594, 2\beta/\nu= 1.037 $. Then we can write Eqn.\ref{densityscalingTc} as
\begin{equation}
        S^{LDA}_{n}( \vec{Q}_{N}, \tau =0 )  =  T^{1.037} F_{ne} (\frac{\mu-\mu_{c}}{T^{1.594}} )
\label{isingtc}
\end{equation}
     which can be tested by the combination of {\sl in situ } measurements and the scattering measurements.

\section{ Conclusions }

   Due to the dilutees and charge neutral of cold atoms,  the experimental ways to detect possible quantum phases and quantum phase
transitions of cold atoms in optical lattices in a minimum
destructive way remain very limited.
In this paper, we developed a systematic and unified theory to use
the three different experiments: optical Bragg scattering, atom
Bragg spectroscopy or off-resonant cavity enhanced scattering to
detect the ground states, the elementary excitation spectra and the
corresponding spectral weights of many quantum phases in both
bipartite and frustrated lattices.
   We show that the two photon Raman processes in all the three measurements not only couple to
   the density order parameter, but also the  {\sl valence bond order } parameter due to the hopping of the bosons
   on the lattice. This coupling to the VBS order is extremely
   sensitive to the superfluid order or VBS order at corresponding
   ordering wavevectors. It is this coupling which make the three
   experiments being able to detect not only the well known SF and
   Mott phases, but also many other important phases such as CDW,
   VBS, CDW-VBS and all kinds of supersolids.
   The first  experiment \cite{braggbog,braggeng} was well established, the second\cite{braggafm,blochlight} is being
  performed in several experimental groups, the third is a possible new
  experimental set-up complementary to the first two.  The combinations of the three can be
  used as powerful and complete tools to study the properties of many
  quantum phases of atoms in optical  lattices.

          The physical measurable quantities of the three
          experiments are the light scattering cross sections, the
          atom scattered clouds and the cavity leaking photons
          respectively. All these experimental measurable quantities
          are determined  by the density-density and bond-bond   correlation functions.
         From symmetry points of view, we analyzed several general properties of the density-density and bond-bond
         correlation functions in both bipartite and frustrated lattices.
         The CDW and VBS order parameters in bipartite lattices are Ising
         order parameter $ m $ and $ K $ respectively, while those in
         frustrated lattices are complex $ U(1) $ order parameter $ \phi_c $ and $ \phi_v $  respectively. This
         difference leads to several new features of light scattering
         cross section in frustrated lattices. Previous literatures in
         quantum phase transitions focused on computing order parameter  correlation
         functions. Here, motivated by the fact that all the three experimental inelastic scattering only couples to the
         density-density correlation function instead of coupling to the order parameter directly, we computed the
         density-density correlation function in the Mott phase, Superfluid phase and the  quantum critical regime
         in both $ z=1 $ and the $ z=2 $ universality class in Fig.\ref{sfmott} and Fig.\ref{sfmott23} respectively.

   At integer fillings, when $ \vec{q} $ matches a reciprocal lattice vector $ \vec{K} $ of the underlying OL,
   there is a increase in the optical scattering cross section as the system evolves from the Mott to the SF state.
   This increase may be used as an effective measure of the average   kinetic energy inside the SF.
   At half integer fillings, in the CDW state, when $  \vec{q} $ matches the CDW ordering
   wavevector $ \vec{Q}_n $ and $ \vec{K} $, there is a diffraction peak proportional to the CDW order parameter squared and the density squared
   respectively (Fig.\ref{squareoptical}a), the ratio of the two peaks are good measure of the CDW order parameter. In the VBS state, when
   $ \vec{q} $ matches the VBS ordering wavevector $ \vec{Q}_K $, there is a much smaller, but detectable diffraction peak
   proportional to the VBS order parameter squared,  when it matches $ \vec{K}
   $, there is also a diffraction peak proportional to the uniform
   density in the VBS state (Fig.\ref{squareoptical}b).  The ratio of the two peaks are good measure of the VBS order parameter.
   All the diffraction peaks scale as the square of the numbers of atoms inside the trap.
   All these characteristics can determine uniquely CDW and VBS state at commensurate fillings and
   the corresponding CDW supersolid and VBS supersolid slightly away from the commensurate fillings.
   In a frustrated lattice such as a triangular lattice, there are a
   new kind phase with both CDW and VBS orders called CDW+VBS phase in Fig.\ref{tricdwvbs}.
   There are corresponding several new features in the light scattering cross sections. We
   also describe how it can detect both the CDW order $ |\langle \phi_c \rangle |^{2} $ and the VBS order
   $ |\langle \phi_v \rangle |^{2} $ in the CDW-VBS phase in Fig.\ref{trioptical}.
  We also point out that due to the  smallness of the CDW gap  and even smaller VBS gap
  compared to the energy of the incident light, the tiny energy difference $ \delta E=E_f-E_i $ could be detected
  by quantum beats in phase sensitive homodyne interference experiments.
  The superfluid components in the CDW and VBS supersolids can be determined by
  the momentum \cite{braggbog} transfer Bragg spectroscopy. While
  the excitation gaps in the CDW and
  VBS and corresponding spectral weights can be detected by the energy
  transfer \cite{braggeng} Bragg spectroscopy.
  So the combinations of these photon and atom scatterings can be used to detect many conventional
  and exotic quantum phases and their excitation spectra of cold atoms in optical lattices as soon as these
  phases are within experimental reach.

  We also propose the cavity QED as another possible effective
  detection method. It is constructive to compare Fig.\ref{cavityqed} with the very recent
  experiments to realize the $ Z_2 $ super-radiant phase \cite{orbital} and
  to realize all kinds of bi-stabilities in BEC, BEC spinor, fermion or spin-orbit coupled systems\cite{qedbecoff,spinor,fermion,spinorbit}.
  In the super-radiant experiment \cite{orbital}, the pumping is a transverse pumping similar to that in Fig.\ref{cavityqed}.
  The system is in a good cavity limit, so the Hamiltonian dynamics dominates over any dissipation process.  As the transverse
  pumping power increases above a critical value, the system will
  evolve from a normal phase into a super-radiant phase. So the  pumping laser is used to induce the dramatic change of the ground state
  through the Raman transition.
  The change of the ground state and the collective excitation spectrum of the strongly coupled atom-photon system can be detected by the
  Florescence spectrum.
  In the  bi-stability experiments \cite{qedbecoff,spinor,fermion,spinorbit}, the pumping is a longitudinal pumping, so it pumps the cavity photon directly,
  no Raman process in Fig.\ref{raman} is involved.
  As the longitudinal pumping power increases above a critical value, the system will
  suffer bi-instabilities, even tri-stabilities \cite{spinorbit}. So the
  pumping laser is also used to change the properties of the strongly coupled photon-atom system by non-linear effects which can be
  detected by cavity transmission spectrum.
  The system is in  a bad cavity limit.
  In both cases, the pumping is strong which induces highly
  non-linear optical and matter effects. The approximation is a two
  level matter wave approximation which is justified in the strong pumping case.
  However, in the present cavity QED detection scheme in Fig. \ref{cavityqed}, the classical laser is used to just probe the quantum
  phases of the cold atoms so that it will not disturb the
  properties of the system itself.  The cavity is in the bad cavity limit. The approximation we made is a
  linear response theory which is justified in the weak pumping
  regime. But we treat all the matter modes exactly which are what one like to detect by the Florescence spectrum.
  Furthermore, the cavity QED detection does not involve the TOF
  measurements, so it is non-destructive. However, putting OL inside
  a ring cavity shown in Fig. \ref{cavityqed} may still present
  serious  experimental  challenges.






    The local density and local density fluctuations can be directly measured by recently advanced {\sl in situ} spatially resolved
    imagings. In the temperature regime where the LDA holds, the Mott to SF transition can be directly probed across the shell structure
    inside a harmonic trap. The CDW to the CDW-SS and then the CDW-SS to the SF transition can also be directly probed across the
    multi-shells structure inside a harmonic trap.
    The two kinds of measurement are complementary and dual to each other.
    The combination of both class of detection methods could be used to
    match the combination of STM, the ARPES and neutron scatterings in condensed matter systems, therefore achieve the putative goals of quantum simulations
    of quantum phases and quantum phase transition.

{\bf Acknowledgements }

We thank I. Bloch, Jason Ho, R.Hulet, Juan Pino, R. Scalettar and Han Pu for very helpful discussions.
 J. Ye also thanks  Jason Ho, A. V. Balatsky and Han Pu for their
hospitalities during his visit at Ohio state university, the LANL
and Rice university where part of this work is done. J. Ye's research is supported by
NSF-DMR-1161497, NSFC-11074173, Beijing Municipal Commission of Education under grant No.PHR201107121, at KITP is supported in part by the
NSF under grant No. PHY-1125915. YC was supported by NSFC-10874032 and
11074043, the State Key Programs of China (Grant no. 2009CB929204) and Shanghai
Municipal Government.
W.P. Zhang's research was supported by the National Basic Research Program of China (973 Program)
 under Grant No.2011CB921604, and NSFC under Grant Nos.10588402 and 10474055.

\appendix

\section{ Review of quantum phases in bipartite and frustrated lattices  }

  Quantum phases of the Extended Boson Hubbard Model (EBHM) with long range interactions in Eqn.1 are
  thoroughly studied by various analytic and numerical methods such as the spin wave expansion in Ref.\cite{gan},
  the dual vortex method (DVM) in Refs.\cite{pq1,mob,bipart,frus}  and quantum Monte-Carlo simulations in Refs.\cite{square,squaresoft,frusqmc,sandvik,honey}.
  In the following, we review some interesting quantum phases in bipartite and
  frustrated lattices respectively. This appendix is not new, but
  pave the way for the discussions  on the detections of these quantum
  phases in the main text.

\subsection{ Some Quantum phases in bipartite optical lattices  }

    Some of the important Mott insulating phases at commensurate fillings
    in a square lattice were already summarized in the Fig.\ref{squarephase}.
    Some quantum phases, especially supersolid phases and quantum phase transitions slightly away
    from commensurate fillings are studied by the DVM in \cite{mob,bipart,frus}. The zero temperature phase diagrams of the chemical potential $
    \mu $ against the quantum fluctuations $ t/V_1 $ slightly away from $ 1/2 $ fillings  are shown in the Fig.\ref{squarediagram}.
    The DVM is a symmetry based approach, so the results achieved from the DVM can be compared to any microscopic models.
    For example, it can be compared to $ V_1, V_2,....$ model, or to the dipole-dipole interaction model.
    If the supersolid phases are stable or not
    depend on the specific microscopic interactions. As said in Sect.II-D, the dipole-dipole interaction is particularly
    favorable to the formation of CDW supersolids slightly away $1/2 $ fillings.


In a honeycomb lattice, we can also find some CDW and VBS phases
\cite{bipart,frus}. Because of their similarity to the quantum
phases in a square lattice, we will not discuss the honeycomb
lattice specifically.

\subsection{ Some Quantum phases in frustrated optical lattices  }

In a triangular lattice at the filling factor $ f=1/3 $, it is easy
to see a CDW phase and a triangular Valence bond ( TVB ) phase
\cite{frus}.

\begin{figure}
\includegraphics[width=3cm]{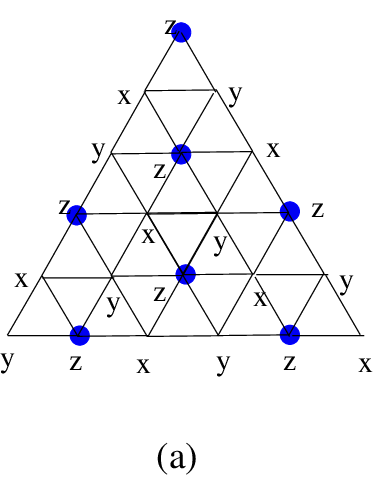}
\hspace{0.5cm}
\includegraphics[width=3cm]{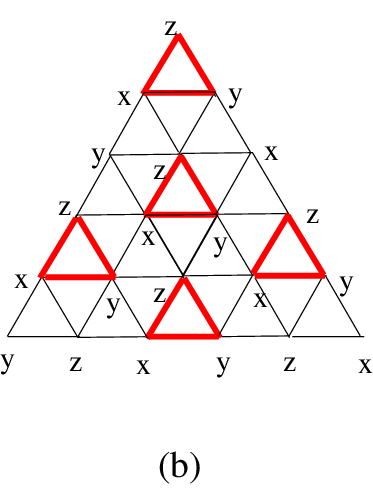}
\caption{ Several insulating states at filling factor $ f=1/3 $ in a
triangular lattice (a) the X-CDW  at  $ \vec{Q}_{n}=2\pi/3(1,1) $
(b) VBS at $ \vec{Q}_{K}= 2\pi/3(1,1) $. See \cite{gan,frus}.   }
\label{triphase}
\end{figure}

   The density in the X-CDW Fig.\ref{triphase}a is:
\begin{equation}
    \rho( \vec{r} )= A \cos \vec{Q}_{n} \cdot \vec{r} + 1/3
\label{xcdw}
\end{equation}
   where $  \vec{Q}_n= 2\pi/3(1, 1 ) $.  Putting $ \vec{r} \rightarrow \vec{r}-\vec{a}_1,  \vec{r} \rightarrow \vec{r}-\vec{a}_2 $
   will correspond to the other  two X-CDW located at the other two sublattices.

   If $ \vec{Q}_n= 2\pi/3(1, 0 ), 2\pi/3(0, 1 ),  2\pi/3(1, -1 ) $, then Eqn.\ref{xcdw} stands for a
   stripe phase along $ \vec{a}_1, \vec{a}_2, \vec{a}_3 $ respectively. For hard core bosons with a dipole-dipole interaction,
   this  X-CDW at $ f=1/3  $ and the X-CDW supersolid slightly away from $ f=1/3  $ was
   found to be stable in a large parameter regimes in a  triangular lattice \cite{dipolarss}.

    The triangular VB state in Fig.\ref{triphase}b is:
\begin{eqnarray}
    B_1( \vec{r} ) & = & K_1 \cos \vec{Q}_{K} \cdot \vec{r} + K_2  \nonumber  \\
    B_d( \vec{r} ) & = & K_1 \cos \vec{Q}_{K} \cdot \vec{r} + K_2  \nonumber  \\
    B_2( \vec{r} ) & = & K_1 \cos ( \vec{Q}_{K} \cdot \vec{r} -\frac{ 2 \pi }{3 } ) + K_2
\label{trivbs}
\end{eqnarray}
    where $  \vec{Q}_K= 2\pi/3(1, 1 ) $ and the $ B_1( \vec{r} ), B_d( \vec{r} ), B_2( \vec{r} ) $ are the
    three bonds along the 3 directions along $ \vec{a}_1, \vec{a}_d,
    \vec{a}_2 $. One can see that $ B_{2}( \vec{r}
    )= B_{1} ( \vec{r} - \vec{a}_1 ) $ as expected from the symmetry  breaking patterns in Fig.\ref{triphase}b.

\begin{figure}
\includegraphics[width=6cm]{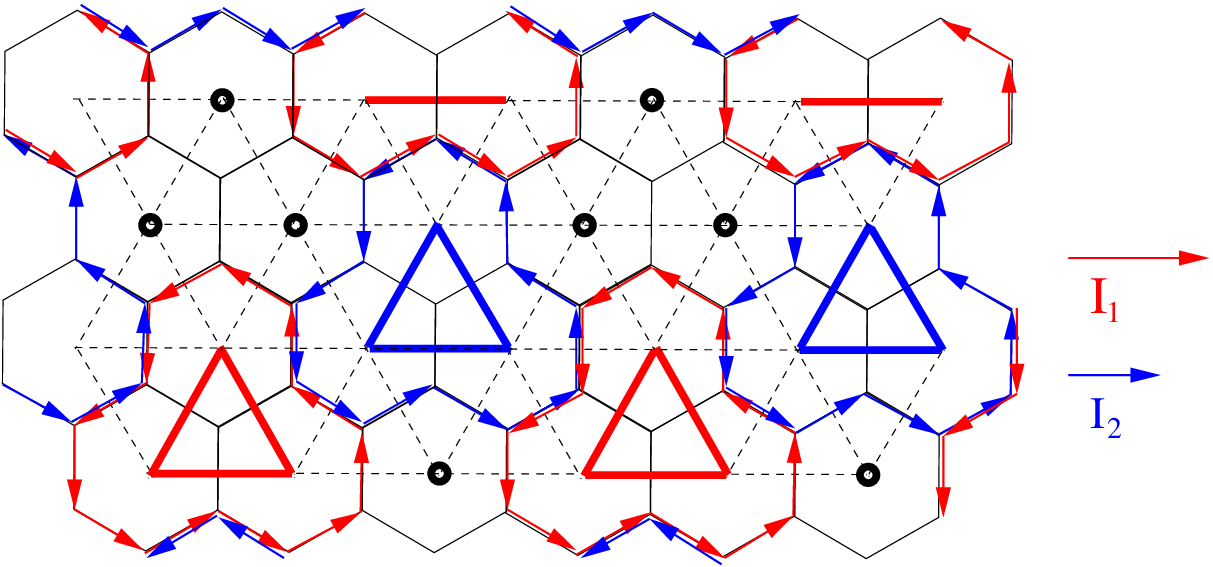}
\caption{ The CDW-VB phase in a triangular lattice at $ f=1/3 $ in
the easy plane limit $ v<0, w> 0 $. The $ (0,0) $ sets the origin of
the direct triangular lattice. There are two different vortex
currents flowing in the dual honeycomb lattice. The red current is $
I_{1}= \sin \frac{ 3 \pi}{9} + \sin \frac{ 2 \pi}{9} $, the blue
current $ I_{2}= \sin \frac{ \pi}{9} + \sin \frac{ 2 \pi}{9} $. The
two different currents indicate 3 boson different densities in the
direct triangular lattice. The density and bond distributions are
given in Eqn.\ref{directtricv} and Eqn.\ref{directtricvb}
respectively. See \cite{frus}. } \label{tricdwvbs}
\end{figure}

  In a triangular lattice, we also identify a phase with both CDW and
  VBS order in Fig.\ref{tricdwvbs} \cite{frus}.  The CDW order in the CDW+VBS phase is given by:
\begin{eqnarray}
 \rho_{CV}( \vec{x} )  &  =  &  1/3 + 4 \delta \sqrt{ \frac{I^{2}_1-I_{1} I_{2} + I^{2}_2
 }{3}} [ - \cos ( \frac{ 2 \pi }{3} x + \frac{ 5 \pi}{18} )   \nonumber \\
 & + & \cos ( \frac{ 2 \pi }{3} y  + \frac{ \pi}{18}
 ) + \cos ( \frac{ 2 \pi }{3} (x-y)  + \frac{ \pi}{18} )]
\label{directtricv}
\end{eqnarray}
  where $ \delta $ is the CDW order parameter and $ I_1, I_2 $ are
  the vortex currents in Fig.\ref{tricdwvbs}.

   The VBS order in the CDW+VBS phase is given by:
\begin{eqnarray}
 B_{1}( \vec{x} ) &  =  &  B_{d}( \vec{x} )= c \delta ( I_1 + I_2 ) ( 1 + 2 \cos  \frac{ 2 \pi }{3} (x+y)
 )   \nonumber   \\
 & + & 2 c \delta \sqrt{ I^{2}_1-I_{1} I_{2} + I^{2}_2 }[  \cos ( \frac{ 2 \pi }{3} x + \frac{ 2 \pi}{9} )   \nonumber \\
 & + & \cos ( \frac{ 2 \pi }{3} y  - \frac{ 2 \pi}{9}
 ) + \cos ( \frac{ 2 \pi }{3} (x-y)  - \frac{ 2 \pi}{9} )]     \nonumber  \\
  B_{2}( \vec{x} ) &  =  & c \delta ( I_1 + I_2 ) ( 1 + 2 \cos  [\frac{ 2 \pi }{3}
 (x+y)-\frac{ 2 \pi }{3} ])   \nonumber   \\
 & + & 2 c \delta \sqrt{ I^{2}_1-I_{1} I_{2} + I^{2}_2 } [  \cos ( \frac{ 2 \pi }{3} x - \frac{ 4 \pi}{9} )   \nonumber \\
 & + & \cos ( \frac{ 2 \pi }{3} y  - \frac{ 2 \pi}{9}
 ) + \cos ( \frac{ 2 \pi }{3} (x-y)  - \frac{ 8 \pi}{9} )]
\label{directtricvb}
\end{eqnarray}
   where the $ c $ is an unknown constant. One can see that $ B_{2}( \vec{x}
   )= B_{1} ( \vec{x} - \vec{a}_1 ) $ as expected from the symmetry
   breaking patterns in Fig.\ref{tricdwvbs}.

   When comparing with the CDW order in Eqn.\ref{directtricv}, we can see that in addition to the 3 ordering wave
   vectors $ \vec{Q}_{\alpha}= 2\pi/3(1,0), 2\pi/3(0,1), 2\pi/3(1,-1),\alpha=1,2,3
   $, there is also a new VBS ordering  wave vector $ \vec{Q}_{4}=2\pi/3(1,
   1) $. It is this new ordering wave vector which makes the detection of the
   VBS order inside the CDW-VB phase possible to be  discussed in Sec.VI.

  In a Kagome lattice, we also found some interesting CDW,  VBS  and CDW+VBS phase
  \cite{frus}.  Because of their similarity to the quantum
  phases in a triangular lattice, we will not discuss the Kagome
  lattice specifically.

\section{ Photon Bragg scattering experiments }

 The main difference of light scattering in cold atom systems from
that in condensed matter system is that the former is charge
neutral, so the off-resonant Ramon scattering processes in
Fig.\ref{raman} are involved, what are measured are the
density-density and bond-bond correlation functions listed in
Eqn.\ref{cross}. While in the latter, the ions carry electric
charges, so there is a direct scattering. So the latter usually has
a much bigger scattering cross sections. For example, in high $ T_c
$ superconductors \cite{hightc}, the ARPES directly measures the
single quasi-particle spectral weights.

In the previous sections, we showed that the elastic photon Bragg
scattering  can not only detect
  the CDW ordering at $ \vec{q}=\vec{Q}_{n} $ easily ( Fig.\ref{squareoptical}a ), but also be used to detect the VBS
  ordering at $ \vec{q}=\vec{Q}_{K} $( Fig.\ref{squareoptical}b ).  It is also quite sensitive to the small superfluid component in
  the CDW-SS and VB-SS. We also showed that the photon Bragg scattering cross section increases from the Mott to the SF phase due to the
  boson flow through the whole optical lattice inside the SF phase.
  So this detection method is especially powerful to detect the ground
  states of various quantum phases.  But it may not be
  easy to detect the excitation spectrum by using this method due to
  the following reasons:
  Due to the smallness of the CDW gap $ \Delta_{CDW} \sim U \sim 10 kHz \sim 1 \mu K $ and
  even smaller VBS gap $ \Delta_{VBS} \sim t^{2}/U \sim n K $
  compared to the energy of the incident light $ E_i \sim 10^{5}
  GHz $, the tiny energy difference $ \delta E=E_f-E_i $ is not easy be detected in the present inelastic Bragg scattering experiments.
  So far, all the previous Bragg scattering experiments \cite{braggsingle} only focused on the elastic
  scattering with $ \omega =0 $ which detect the static orders of OL
  itself. However, it was argued \cite{braggsingle} that the integrated  in-elastic scattering
  can be used to detect the temperature of cold atoms in OL with
  incident light intensity $ I_{in} \sim 500 \mu W/cm^{2} $ and the detuning in Fig.\ref{raman} $
  \Delta \sim 20 \gamma $ where $ \gamma $ is the linewidth of the
  excited state. We expect that the tiny energy difference is still detectable by measuring the quantum beats of
  phase sensitive homodyne interference experiments. So the possible
  future  combination of the light scattering in Fig.1a combined with the
  quantum beat measurement can be used to  detect not only the ground state, but also the excitation
  spectrum. Note that in contrast to the atom Bragg spectroscopy to be discussed in the following appendix,  the photon
  Bragg spectroscopy does not involve the  TOF measurements, so it  is a non-destructive measurement.

\section{ Atom Bragg scattering experiments }

 There are also two different Atom Bragg scattering experiments. The
 Momentum Bragg spectroscopy is more applicable in the superfluid side to
 detect its Bogoliubov excitations. While the
 energy Bragg spectroscopy is more applicable in the insulating side to
 detect its excitation gap. But it may not be as effective as the photon Bragg scattering to
 detect the ground states of the quantum phases.

 {\sl (a) Momentum Bragg spectroscopy }

  It measures the atom diffraction from the light gratings formed by the two laser beams in Fig.1b, so it is
  a complementary ( or dual ) to the light scattering which measures the light diffraction from the atoms
  gratings. It is used to measure the Bogoliubov excitation spectrum inside a superfluid.

\begin{figure}
\includegraphics[width=7.5cm]{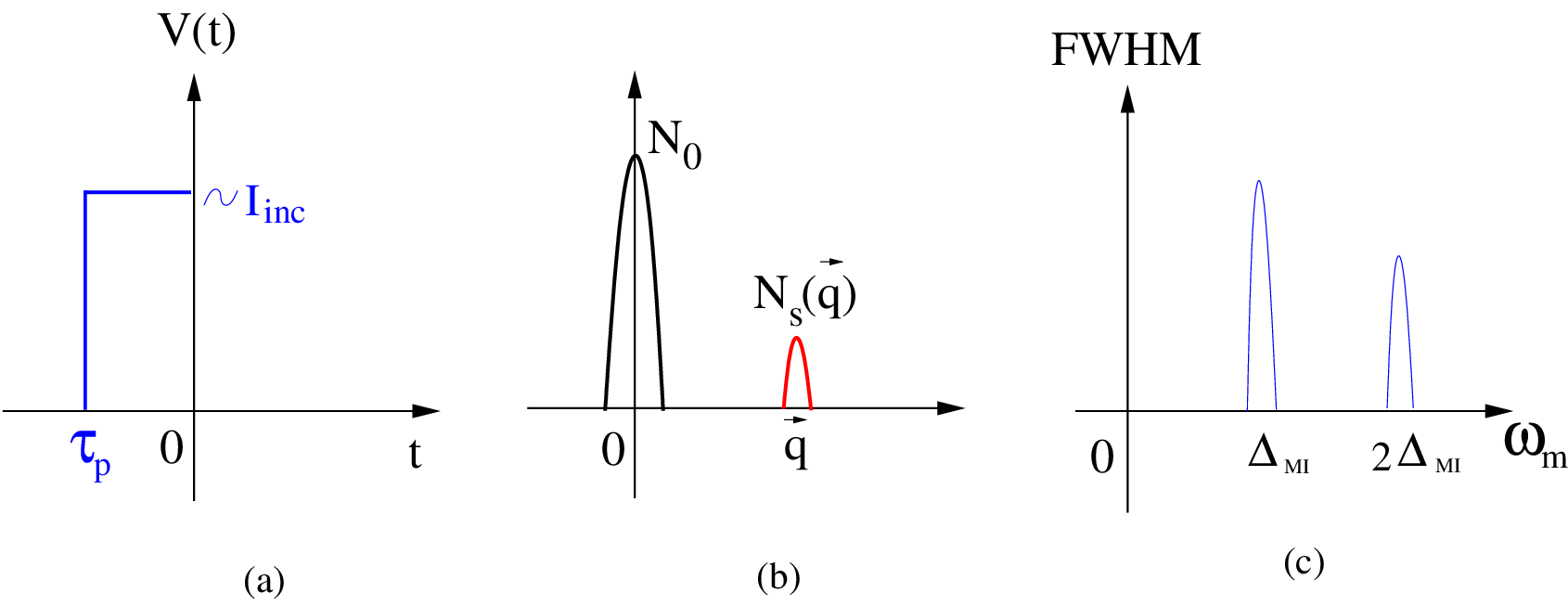}
\caption{ (a) The duration and strength of the two laser beams in
the atom Bragg spectroscopy in Fig.1b. (b) In the momentum transfer
Bragg spectroscopy, the scattered atom cloud at $ \vec{q} $ verses
the condensate cloud at $ \vec{q} =0 $. In a TOF image, the momentum
is mapped to the axial distance $ z $. So the two peaks are
separated by a distance in the axial direction. (c) In the energy
Bragg spectroscopy, the full width at half maximum (FWHM ) of the
central peak in the TOF image peaks when the modulation frequency $
\omega_m $ matches the Mott gap $ \Delta_{Mott} $ in the Mott
insulating phase. The momentum is fixed at $ \vec{q}=0 $ in the Mott
state, but can be tuned to $ \vec{Q}_n $ or  $ \vec{Q}_K $ in a CDW
or VBS state to measure $ \Delta_{CDW} $ and $ \Delta_{VBS} $
respectively. } \label{atombragg}
\end{figure}

  The main experimental procedures of the Momentum Bragg  spectroscopy \cite{braggbog} are:
  (1) The duration of the two incident light pulses in Fig.1b is about
  $ \tau_p \sim 400 \mu s $, then shut off ( Fig.\ref{atombragg} a). (2) Then the  optical lattices and also the trap in Fig.1b are released shortly after
  the turn-off of the light pulse.
  (3) Then after some expanding time $ \tau_{exp} \sim 200 \tau_p \sim 8 m s $, the time of flight images are taken to measure the momentum
  distribution of the expanding gas. During the expansion process,
  the energy and momentum of the elementary excitations are transferred to the free particles.

  Only when the imparted energy $ \omega $ and the imparted momentum $ \vec{k} $ match the
  dispersion relation of the elementary excitations in the system,
  one can observe a scattered cloud ( Fig.\ref{atombragg} b) whose number $ N_{sc}( \vec{q}, \omega ) $  is proportional to the
  dynamic structure factor:
\begin{eqnarray}
  N_{sc}( \vec{q}, \omega )  & \sim  & ( \frac{ \Omega^{2} }{ \Delta } )^{2}
  N^{2} [ |f_{0}( \vec{q} ) |^{2} S_{n}( \vec{q}, \omega )
  \nonumber  \\
  & + &  \sum_{\alpha=\hat{x},\hat{y} } |f_{\alpha}( \vec{q} ) |^{2} S_{K_{\alpha} }( \vec{q}, \omega ) ]
\label{cloud}
\end{eqnarray}
   where the prefactor is proportional to the duration $ \tau_p
   $, the incident density $ I_{in} \sim 1 mW/cm^{2} $ in Fig.\ref{atombragg}.
   When comparing with Eqn.\ref{cross}, one can see that the
   scattering cross section in the light scattering experiment is
   replaced by the scattered clouds in the atom Bragg spectroscopy.

   The integrated scattered atoms  $ N_{sc}( \vec{q} )  = \int d
   \omega N_{sc}( \vec{q}, \omega )  $ is
   proportional to the {\sl equal-time} response function
\begin{eqnarray}
     N_{sc}( \vec{q} )  & \sim   &  ( \frac{ \Omega^{2} }{ \Delta } )^{2} N^{2} [  |f_{0}( \vec{q} ) |^{2} S_{n}( \vec{q}
     )    \nonumber  \\
     &  +  & \sum_{\alpha=\hat{x},\hat{y} } |f_{\alpha}( \vec{q} ) |^{2} S_{K_{\alpha}}( \vec{q}) ]
\label{cloudequal}
\end{eqnarray}
     This can be measured by scanning the frequency $ \omega $ at
     fixed $ \vec{q} $.  For the current
   experimental pulse duration time and the
   incident density, the scattered atoms around $ \vec{q} $ are about $ 10-20 \% $ of the condensate atoms around $\vec{q}=0 $:
   $ N_{sc}(\vec{q})/N_0(\vec{q}=0) \sim 0.1-0.2 $. Then one may neglect nonlinear terms at the required laser powers.
Possible spontaneous scattering processes can also be suppressed. In
real atom Bragg spectroscopy experiments, there are several
technical issues such as the signal strength, the geometry
dependence, the suppressions of the multi-photon processes and
super-radiance... need to be considered, but the basic picture is
presented in this section and sketched in the Fig.\ref{atombragg}.

\vspace{0.2cm}

\vspace{0.2cm}

 {\sl (b) Energy Bragg spectroscopy }

  The excitation gaps in the Mott, CDW or
  VBS and corresponding spectral weights in Fig.\ref{excitations} can be detected by the energy
  transfer  Bragg spectroscopy \cite{braggeng}.
  In this experiment, the two laser beams in Fig.1b is
  counter-propagating, so simply super-imposed on the external laser
  beams forming the optical lattice: $ V_{OL}(x,t)=( V_{OL,0}(x,t)+
  A_m \sin \omega_m t )\cos^{2} k x $. The modulation $ A_m $ and
  the frequency difference $ \omega_m $ introduces two sidebands
  with frequencies $ \pm  \omega_m $ relative to the OL laser
  frequency. So when the imparted energy $ \omega_m $ matches the
  excitation gap $ \Delta $ at $ \vec{q}=0 $ or any reciprocal lattice vector $ \vec{K} $ inside the  Mott state, there is a maximum
  response of the atoms. After the excitation, the OL potential is
  ramping down linearly to the SF regime and kept in this shallow potential
  for some time. Then the trap in Fig.1b is suddenly switched off,
  the atoms' interference images are  taken after their $ 20 ms $ ballistic expansion.
  The full width at half maximum (FWHM) of the central peak is taken as the
  measure of the energy deposited in the atom clouds by the
  excitation. Namely, the FWHM represents the system's repones to
  the two laser beams in Fig.1b. Indeed, discrete peaks in the FWHM were observed  \cite{braggeng} at discrete spectrum $ \omega_m= n
  \Delta_{MI},n=0,\pm 1, \cdots $ in Fig.\ref{atombragg}c.

  Note that the gaps inside the CDW, VBS, CDW-VBS and the corresponding
  supersolids in Fig.\ref{excitations} are at momentum $ \vec{q} =\vec{Q}_{n} $ or $ \vec{q} =\vec{Q}_{K}
  $.  We expect similar energy  transfer  Bragg spectroscopy can be used to
  measure these gaps when one tune $ \vec{q}= \vec{Q}_{n} $ or $ \vec{q} =\vec{Q}_{K} $ in Fig.1b and Fig.\ref{atombragg}c.
  The corresponding spectral weights are just the areas under the
  peaks in Fig.\ref{atombragg}c.

  It maybe useful to make a brief comparison between the atom and photon Bragg scattering here,
  the  photon Bragg scattering is very good for detecting the orders,
  it also does not involve the  TOF measurements, so it
  is a non-destructive measurement. As explained in
  the previous section, so far, it is very challenging, but still possible  to measure
  the excitation spectrum by using very precise quantum beating interference measurements. However, as discussed in this section,
  the atom Bragg spectroscopy is not very suitable to detect the
  ground state orders, especially all the possible insulating states, but very effective to detect the excitation
  spectra. For example,  the superfluid  components in the CDW and VBS supersolids can be determined by
  the momentum \cite{braggbog} transfer Bragg spectroscopy. The gaps
  in the CDW and VBS supersolids can be measured by the energy \cite{braggeng} transfer Bragg spectroscopy.
  Because the atom Bragg spectroscopy involves the  TOF measurements, so it is a destructive measurement.
  So the photon and atom Bragg spectroscopy are two complementary detection methods, the combinations of both can be used to detect many conventional
   and exotic quantum phases and their excitation spectra of cold atoms in optical lattices as soon as these
   phases are within experimental reach.

\section{  Superfluid stiffness and QED detection at the 2 dimensional classical diffraction minima $
\vec{q}= ( \pi,\pi ) $.}


    The Cavity QED scattering from the SF and Mott phase at 1d near $ q= \pi $ was addressed in
    \cite{off1}. Particularly, they discussed the cavity QED scattering  from the Mott and the SF phase
    at the geometry  to suppress the classical diffraction. In this geometry,
    $ \theta_{0}= 0, \theta_{1} =\pi/2 $ in Fig.\ref{cavityqed}a, so $ \hat{D}= \sum_{i} (-1)^{i} n_{i} $, so $ \langle
    a^{\dagger}_{1} a_{1} \rangle_{MI} =0 $ in the $ t/U \rightarrow 0 $ limit in Fig.\ref{sfmott} ( see below
    Eqn.\ref{decayqcp} ).
    In Ref.\cite{off1}, the SF ground state was taken as a
    non-interacting BEC state $ | BEC \rangle_{non} \sim ( \sum_{i} b^{\dagger}_{i} )^{N} |0 \rangle $, so
    the results in \cite{off1} on the SF side are based on the fact that the non-local pair correlation functions between the
    atom numbers at different sites $ \langle n_{i} n_{j} \rangle $ in the  non-interacting BEC state are the same for any $ i \neq j $.
    This infinite long-range correlation is certainly not valid in any
    SF, so their photon spectrum on the SF side are in-correct. In this
    appendix, we will calculate the photon number spectrum inside
    the SF phase and show that it is an effective measure of the
    superfluid density $ \rho_s $ and the phonon velocity $ v $  of the SF.
    While the two fundamental physical quantities  for the SF are
    not even defined for the non-interacting BEC state.

          It is well known that a non-interacting BEC
          state is a pathological state. A repulsive interaction is needed to
          transform the BEC state to a SF state and leads to the gapless
          Goldstone mode above the SF ground state shown in Eqn.\ref{sf} and in
          Fig.\ref{excitations}. The Fourier transform of the equal density-density correlation function in Eqn.\ref{sfequal} at $ d=2 $ is:
\begin{equation}
 \langle \delta n_{i} \delta n_{j} \rangle_{SF} =  \langle n_{i} n_{j} \rangle -
     n^{2} \sim \rho_{s}/v |i-j|^{3}
\label{decaysf}
\end{equation}
     which decays as a power law  as
     the distance between the two sites $ |i-j | \gg a  $
     due to the gapless Goldstone mode  in the SF and also depends on the
     superfluid density $ \rho_{s} $ and the Goldstone mode velocity $ v $.
     Strictly speaking, Eqn.\ref{decaysf} only holds at very long distance $ |i-j| \gg a $, for example, it certainly breaks down at $ i=j $,
     but we expect $ \langle n^{2}_{i} \rangle_{SF} - n^{2} \sim \rho_{s}/v $ still holds.
     In contrast, from Eqn.\ref{cdwexp}, inside the Mott state
\begin{equation}
      \langle \delta n_{i} \delta n_{j} \rangle_{MI} =  \langle n_{i} n_{j} \rangle -  n^{2} \sim e^{- |i-j|/\xi_{MI}
      }
\label{decaymott}
\end{equation}
     where $ \xi_{MI} \sim 1/\Delta_{MI} $ is the correlation length inside the Mott
     state.  This equation is consistent with
     Eqn.\ref{mott} in the momentum space in the Mott phase.

     While at the QCP between the SF to Mott transition in
     Fig.\ref{sfmott}, because $ \delta n $ is a conserved quantity,
     so no anomalous dimension, then we have:
 \begin{equation}
 \langle \delta n_{i} \delta n_{j} \rangle_{QC} =  \langle n_{i} n_{j} \rangle -
     n^{2} \sim 1/ |i-j|^{4}
\label{decayqcp}
\end{equation}
     for $  |i-j | \gg a  $. This equation is consistent with
     Eqn.\ref{qc} in the momentum space at the QC.

     Then from Eqn.\ref{onephoton}, we can see that at $ \vec{q}= ( \pi,\pi ) $,
     the photon number is $
     \langle a^{\dagger}( \vec{q} , 0) a( \vec{q}, 0 ) \rangle   =  | C |^{2}
     \langle D^{*}( \vec{q}, 0)  D(  \vec{q}, 0 ) \rangle  = | C |^{2}
     \sum_{i,j} (-1)^{i-j} \langle n_i n_j \rangle = | C |^{2} N
     \sum_j (-1)^{i-j} [ \langle n_i n_j \rangle -n^{2} ] $. By using Eqn.\ref{decaysf} and Eqn.\ref{decaymott}, we can see that $
     \langle a^{\dagger}( \vec{q} , 0) a( \vec{q}, 0 ) \rangle_{SF} \sim | C
     |^{2} N \rho_{s}/v $ and $ \langle a^{\dagger}( \vec{q} , 0) a( \vec{q}, 0 )
     \rangle_{MI} \sim 0 $ in the $ t/U \rightarrow 0 $ limit in Fig.\ref{sfmott}. So the photon number at $ \vec{q}= ( \pi,\pi )
     $ maybe an effective measurement of the superfluid density $
     \rho_s $  inside the SF. The same conclusions apply to the
     light scattering cross section after replacing  $ |C|^{2} $ by
     $  ( \frac{ \Omega^{2} }{ \Delta } )^{2}  $.



\begin{thebibliography}{99}
\bibitem{finiteqmc} In a QMC simulations on a finite $ L \times L $ lattice, becasue there is no real symmetry breaking at any finite system \cite{phased},
 the order parameter $ m $ in Eqn.\ref{nnstatic} ( the order parameter $ K $ in Eqn.\ref{kkstatic} )
 vanishes in both the superfluid and the CDW phase ( VBS ) phase, so one need only focus on Eqn.\ref{nnqn} ( \ref{kkqk} ),
 see Ref. \cite{curvature,dipolarss,dipolarsstri} for QMC in CDW  and Ref.\cite{sandvik} for QMC on Antiferromagnetic order and the VBS.
\bibitem{int} Note that  the EBHM Eqn.\ref{boson} with  the dipole-dipole interaction may be in a different universality class than that with  the
  $ V_1 > 0, V_2 > 0 $ interaction.
\bibitem{justify} The CDW and VBS gaps are within the lowest Bloch
band, so the approximation of neglecting the higher Bloch bands such as the $ p $ band is
justified.
\bibitem{afm} This is similar to non-collinear order ( or spiral
order ) in a frustrated Anti-ferromagnet where the order parameter
can be written as $ \vec{S} (\vec{r} )= \vec{n}_{1} \cos
\vec{Q}\cdot \vec{r} + \vec{n}_{2} \sin \vec{Q}\cdot \vec{r} = (
\vec{n}_{1} -i \vec{n}_2 ) e^{ i \vec{Q}\cdot \vec{r} } + h.c. $
where $ \vec{n}^{2}_{1}= \vec{n}^{2}_{2}=1, \vec{n}_{1} \cdot
\vec{n}_{2} =0 $.
\end{thebibliography}
\end{document}